\documentclass[%
 reprint,
superscriptaddress,
 amsmath,amssymb,
 aps,
prb,
]{revtex4-2}

\usepackage{graphicx}
\usepackage{dcolumn}
\usepackage{bm}
\usepackage{siunitx}
\usepackage{amsmath}
\usepackage{mathtools}
\usepackage{hyperref}

\usepackage{xcolor}
\usepackage{verbatim}
\usepackage{lipsum}
\usepackage[english]{babel}
\usepackage[T1]{fontenc}
\usepackage{upgreek}

\begin{document}

\preprint{APS/123-QED}

\title{Reconfigurable spin-wave platform based on interplay between nanodots and waveguide in hybrid magnonic crystal}


 
\author{Krzysztof~Szulc}
\email{szulc@ifmpan.poznan.pl}
\affiliation{%
Institute of Molecular Physics, Polish Academy of Sciences, M. Smoluchowskiego 17, 60-179 Pozna\'{n}, Poland
}%
\affiliation{%
Institute of Spintronics and Quantum Information, Faculty of Physics and Astronomy, Adam Mickiewicz University, Pozna\'{n}, Uniwersytetu Pozna\'{n}skiego 2, 61-614 Pozna\'{n}, Poland 
}%
\author{Mateusz~Zelent}%
\affiliation{%
Institute of Spintronics and Quantum Information, Faculty of Physics and Astronomy, Adam Mickiewicz University, Pozna\'{n}, Uniwersytetu Pozna\'{n}skiego 2, 61-614 Pozna\'{n}, Poland 
}%
\author{Maciej~Krawczyk}%
\affiliation{%
Institute of Spintronics and Quantum Information, Faculty of Physics and Astronomy, Adam Mickiewicz University, Pozna\'{n}, Uniwersytetu Pozna\'{n}skiego 2, 61-614 Pozna\'{n}, Poland 
}%

\date{\today}

\begin{abstract}

We present a hybrid magnonic crystal composed of a chain of nanodots with strong perpendicular magnetic anisotropy and Dzyaloshinskii--Moriya interaction, positioned above a permalloy waveguide. The micromagnetic study examines two different magnetization states in the nanodots: a single-domain state and an egg-shaped skyrmion state. Due to the dipolar coupling between the dot and the waveguide, a strongly bound hybrid magnetization texture is formed in the system. Our results show complex spin-wave spectra, combining the effects of periodicity, magnetization texture, and hybridization of the propagating waves in the waveguide with the dot/skyrmion modes. The dynamics of the systems are characterized by several key features which include differences in band-gap sizes, the presence of flat bands in the skyrmion state that can form both bound and hybridized states, the latter sometimes leading to the presence of additional non-Bragg band gaps, and a broad frequency range of only waveguide-dominated modes in the single-domain state. Thus, the study shows that the proposed hybrid magnonic crystals have many distinct functionalities, highlighting their reconfigurable potential, magnon--magnon couplings, mode localization, and bound states overlapping with the propagating waves. This opens up potential applications in analog and quantum magnonics, spin-wave filtering, and the establishment of magnonic neural networks.

\end{abstract}

\maketitle


\section{Introduction}

Over the past decade, spin-wave (SW) computing has been extensively researched as a potential candidate to complement and surpass CMOS-based technologies~\cite{Chumak2022AdvancesComputing,Wang2023PerspectiveNetworks} for digital~\cite{Chumak2014} or analog signal processing~\cite{Klinger2014,Fischer2017} and neural network implementation~\cite{Lee2022ReservoirCrystal,Lee2024}. 
This is because SWs offer high-frequency operation, even at tens of GHz, miniaturization well below 100 nm, and most importantly, ultralow power consumption, as low as 1 aJ per operation. Moreover, they can locally interact with magnetic solitons, i.e. domain walls in 1D and magnetic vortices or skyrmions in 2D, and thus can hybridize with, be excited by, and be controlled by soliton dynamics~\cite{Wintz2016,Yu2021,Petti2022}.  

Magnetic skyrmions are topologically protected 2D magnetization textures, known for their stability and very small size, especially N{\'e}el skyrmions in thin ferromagnetic films, which are stabilized by Dzyaloshinskii--Moriya interaction (DMI)~\cite{Fert2013}. Their dynamics can be driven by external forces such as magnetic field, electric current, structural stress, thermal fluctuations, or laser pulses~\cite{Lonsky2020DynamicTextures}, which expands their potential applications for information storage and processing~\cite{Vakili2021Skyrmionics-computingMagnets, Everschor-Sitte2018Perspective:Field,AFert2017MagneticApplications,Marrows2021PerspectiveSpintronics} as well as in magnonics \cite{Li2022}, e.g., to control SW propagation~\cite{Moon2016}, to scatter SWs~\cite{Lan2021,Kotus2022}, to form SW frequency combs~\cite{Wang2021}, or to excite propagating SWs in thin films~\cite{Diaz2020}. Furthermore, 1D and 2D arrays of skyrmions in thin films form a periodic potential for SWs, and these structures can be considered as reprogrammable magnonic crystals with transmission bands separated by band gaps \cite{Ma2015b,Kim2018,Chen2021,Bassotti2022}. In the presented results, the first bands are associated with skyrmion excitations, and interestingly, the bandwidth, i.e., the coupling strength between skyrmion oscillations, can be controlled not only by the separation between the skyrmions but also by the bias magnetic field \cite{Kim2018,ZHANG2019,Bassotti2022}.
However, due to the high damping of SWs in multilayers possessing DMI~\cite{Dhiman2021,Azzawi2017}, and high frequencies considered (tens of GHz) with a few exceptions, these effects remain mainly numerical demonstrations~\cite{Tang2023}.

In a skyrmion within confined geometry, three types of eigenmodes have been observed~\cite{Jin2022Spin-waveNanodots, Zelent2022SpinAnisotropy}: gyroscopic, breathing, and azimuthal modes. The gyroscopic mode refers to the rotational motion of the skyrmion core~\cite{guslienko2016gyrotropic}. The breathing mode involves the radial oscillation of the size of the skyrmion~\cite{Kim2014BreathingDotsb}. Azimuthal modes are SWs propagating along the skyrmion circumference~\cite{mruczkiewicz2016collective, Garst2017CollectiveMagnets,Mruczkiewicz2018}. Their quantization is described by an azimuthal wave number, with clockwise (CW) and counterclockwise (CCW) degeneracy lifted by the asymmetric exchange interaction. When the dots are arranged in a chain or array, bands of collective skyrmion excitations can be formed~\cite{Chen2021}. However, the dynamic dipolar coupling between the skyrmions is rather weak, especially between the azimuthal modes, and the non-zero bandwidths have been numerically demonstrated only for gyrotropic or breathing modes and only for dots being in direct contact, which indicates an important role of the exchange coupling \cite{mruczkiewicz2016collective}. Thus, the bandwidths of these collective skyrmion modes are narrower than in thin films with immersed skyrmions, stabilized purely by DMI \cite{Kim2018,ZHANG2019,Bassotti2022} or with the help of the stray field from the array of dots in the vortex state placed above the out-of-plane magnetized thin film, i.e., a hybrid structure \cite{Wang2020b}.

Hybrid structures are commonly used to obtain systems that combine two, usually mutually exclusive, material properties, such as ferromagnetism and superconductivity~\cite{Golovchanskiy2019,Petrovic2021} or SWs and enhanced programmability provided by the artificial spin ice system \cite{Negrello2022}. This is also true for magnonics and skyrmions. The former requires long propagation distances \cite{Wang2023PerspectiveNetworks} and coherence times \cite{Pan2024}, and thus low damping, while the latter requires DMI resulting from spin--orbit interactions and neighboring heavy metals, which is associated with increased damping. Such a hybrid structure is also the system composed of yttrium iron garnet and the Co/Pt multilayer with the skyrmion, where skyrmion acts as a point source to excite short SWs in yttrium iron garnet with tens of nm wavelength \cite{Chen2021ChiralSkyrmions}. 

Following these ideas, we propose a hybrid structure consisting of a SW conduit made of a low-damping material (Py) and a chain of (Ir/Co/Pt) nanodots of 300 nm diameter, forming a hybrid magnonic crystal (HMC) which serves as a multifunctional platform for SW applications. Such an HMC structure has already been shown an extended range of DMI values, which facilitates N\'eel-type skyrmion stabilization at comparatively lower DMI values~\cite{Zelent2023}. Using micromagnetic simulations, we show that this HMC also exhibits several interesting properties that arise from the coupling between the subsystems relevant to the control of SWs and have potential for their practical exploitation. These include flat magnonic bands both below and above the bottom of the SW spectrum in Py, and a reprogrammable magnonic band structure, where the width of the band gaps is modified by the magnetization texture in the dots at the remanence: skyrmion or single-domain state. The former property provides a system for the realization of bound states in the continuum in a magnonic domain~\cite{Hsu2016}, the latter for the SW filtering or enhanced nonlinear dynamics \cite{Gallardo2019}. Moreover, we observe enhanced coupling between skyrmion modes when coupled to the Py waveguide with respect to the isolated dot chain, and the dynamic coupling between SWs propagating in the waveguide and SWs confined to the domain wall of the skyrmion leads to band anticrossing, thus additional band gaps, making this system suitable for exploiting the magnon--magnon coupling and thus useful for quantum magnonics applications~\cite{Awschalom2021,Li2020}. Considering this wide range of interesting properties and their multifunctionality, the proposed HMCs represent a promising platform for magnonic artificial neural networks, as proposed in Ref.~\cite{Papp2021}, or where the waveguide serves as synapses connected by propagating SWs, and interacting resonant neurons, i.e. nanodots on the waveguide with the rich spectra of SW modes~\cite{Fripp2023NonlinearNeurons}. 

\section{Methods}

\begin{figure}[t]
    \includegraphics[width=\linewidth]{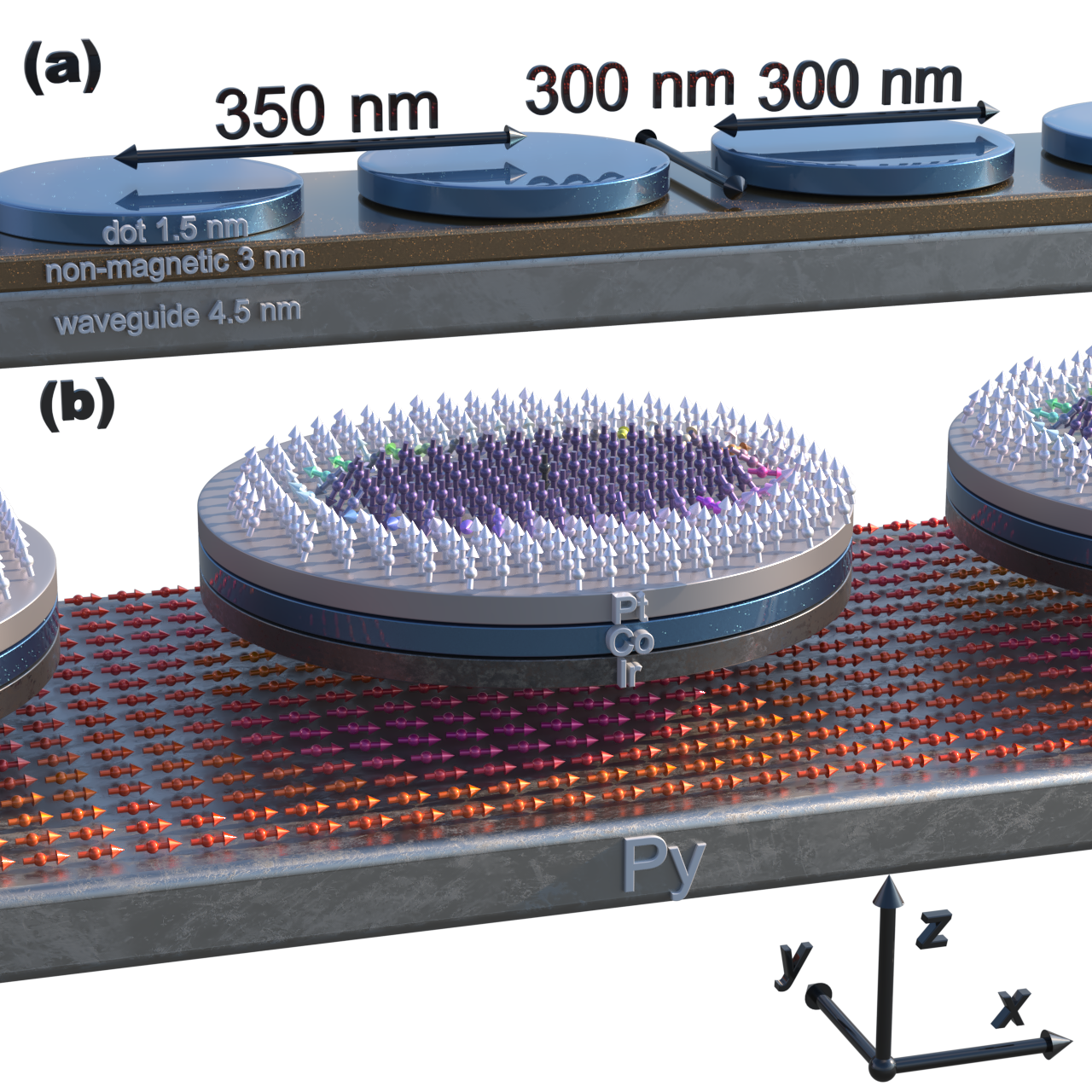}
    \caption{The visual representation of the system under consideration. (a) A schematic geometry of the Ir/Co/Pt multilayer dot, separated from the 4.5~nm-thick Py strip by a 3~nm-thick nonmagnetic layer. (b) An artist's rendering of the simulated magnetization configuration, showing an egg-shaped N\'eel-type skyrmion in the dot stabilized by the magnetostatic coupling to the skyrmion imprint on the in-plane magnetized strip. The arrows and their color (according to the HSL-cone color scale) indicate the direction of magnetization. Note that the figure is not to scale.}
    \label{fig:structure}
\end{figure}

The system under investigation is presented in Fig.~\ref{fig:structure}. It consists of the infinitely long waveguide made of permalloy (Py, Ni$_{80}$Fe$_{20}$) with a width of 300~nm and a thickness of 4.5~nm and a chain of Co dots with a 300~nm diameter and 1.5~nm thickness. The dots are laying centrally above the waveguide with a relative separation of 50~nm, resulting in the periodic structure with a lattice constant of 350~nm. The waveguide and the dots are separated by a 3~nm-thick nonmagnetic layer.

The magnetization dynamics of the system are described by the Landau--Lifshitz--Gilbert equation:
\begin{equation}\label{eq:llg}
    \frac{\partial \mathbf{M}}{\partial t} = -|\gamma| \mu_0 \mathbf{M} \times \mathbf{H}_\mathrm{eff} + \frac{\alpha}{M_\mathrm{S}} \mathbf{M} \times \frac{\partial \mathbf{M}}{\partial t},
\end{equation}
where $\gamma$ is the gyromagnetic ratio, $\mu_0$ is the permeability of vacuum, $\mathbf{H}_\mathrm{eff}$ is the effective magnetic field, $\alpha$ is the damping constant, and $M_\mathrm{S}$ is the saturation magnetization. The effective magnetic field $\mathbf{H}_\mathrm{eff}$ is described as follows:
\begin{equation}
\begin{split}
    \mathbf{H}_\mathrm{eff} = \mathbf{H}_0 + \frac{2A_\mathrm{ex}}{\mu_0 M_\mathrm{S}^2}\nabla^2 \mathbf{M} + \frac{2K_\mathrm{PMA}}{\mu_0 M_\mathrm{S}^2} M_z \hat{\bf z} - \nabla \varphi + \\ + \frac{2D}{\mu_0 M_\mathrm{S}^2} \left( \frac{\partial M_z}{\partial x}\hat{\bf x} + \frac{\partial M_z}{\partial y}\hat{\bf y} - \left( \frac{\partial M_x}{\partial x}+\frac{\partial M_y}{\partial y} \right) \hat{\bf z} \right),
\end{split}
\end{equation}
where $\mathbf{H}_0$ is the external magnetic field, $A_\mathrm{ex}$ is the exchange stiffness constant, $K_\mathrm{PMA}$ is the perpendicular magnetic anisotropy constant, $D$ is the Dzyaloshinskii--Moriya constant, and $\varphi$ is the magnetic scalar potential, which can be determined from the formula
\begin{equation}\label{eq:pot}
    \nabla^2 \varphi = \nabla \cdot \mathbf{M},
\end{equation}
which is derived from Maxwell equations in the magnetostatic approximation.

The system was studied using the finite-element method simulations in COMSOL Multiphysics \cite{Szulc2022}. The simulations were performed in the 3D model with the implementation of Eqs.~(\ref{eq:llg}) and~(\ref{eq:pot}). The static magnetization configuration was stabilized in the time-domain simulation with periodic boundary conditions placed at the ends of the unit cell perpendicular to the $x$-axis to introduce the periodicity into the system. For the proper calculation of the stray magnetic field, the condition $\varphi=0$ is applied at a distance of $\SI{10}{\micro\meter}$ from the system. As an initial magnetization configuration, the waveguide is uniformly magnetized along the $x$-axis while the dots are uniformly magnetized along the $z$-axis (for the study of a single-domain state configuration) or have a skyrmion inside (for the study of skyrmion state) \cite{Guslienko2017}. The magnetic state relaxation lasts \SI{1}{\micro\second}. The dispersion relation was calculated using the eigenfrequency solver. For this purpose, the Landau--Lifshitz--Gilbert equation is solved in its linearized form, where the total magnetization vector $\mathbf{M}=\mathbf{M}_0+\mathbf{m}\, e^{i\omega t}$ is split to a static component $\mathbf{M}_0=(M_{0x},M_{0y},M_{0z})$ and a dynamic component $\mathbf{m}=(m_x,m_y,m_z)$. The equation takes the form of an eigenvalue equation, where the complex eigenvalues give the frequencies, the dynamic magnetization $\mathbf{m}$ and the dynamic magnetic scalar potential are the eigenvectors, and the wavevector is a sweep parameter. Here, the periodic boundary conditions are replaced by Bloch boundary conditions. The tetrahedral mesh is used with a maximum size of 5 nm in the dot and 7 nm in the waveguide. Outside the magnetic material, the mesh grows with ratio 1.4. On the sides where Bloch boundary condition is applied, we prepared identical triangular meshes.

The material parameters of Py are $M_\mathrm{S} = \SI{800}{\kilo\ampere/\meter}$, $A_\mathrm{ex} = \SI{13}{\pico\joule/\meter}$, $D = 0$, $K_\mathrm{PMA} = 0$, $\alpha = 0.005$. The magnetic dot is defined with an effective-medium approach~\cite{PSSR:PSSR201700259, lemesh2018TwistedMultilayers,woo2016observation,Suna1986PerpendicularFilm} as a structure with DMI and PMA, where the three repetitions of the 0.5 nm-thick Ir/Co/Pt multilayer are simulated as a single Co layer with an effective thickness. The effective parameters of the dot are $M_\mathrm{S} = \SI{956}{\kilo\ampere/\meter}$, $A_\mathrm{ex} = \SI{10}{\pico\joule/\meter}$, $D = \SI{-1.6}{\milli\joule/\meter^2}$, $K_\mathrm{PMA} = \SI{717}{\kilo\joule/\meter^3}$, $\alpha = 0.02$. In all calculations, the external magnetic field $\mathbf{H}_0=0$.

\section{Results and discussion}

\begin{figure}[t]
    \includegraphics{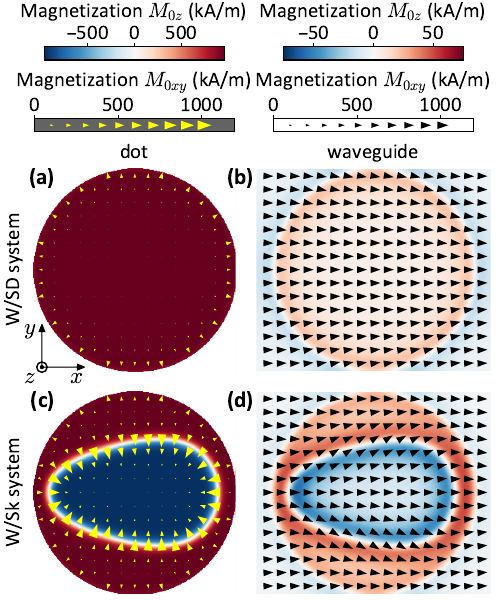}
    \caption{Configuration of the magnetization in the unit cell of the coupled system of waveguide with the chain of dots in (a,b) the single-domain state and (c,d) skyrmion state. The magnetization is shown in the $xy$-planes crossing (a,c) the center of the dot and (b,d) the center of the waveguide. The color map shows the $M_{0z}$ component of the magnetization, and the in-plane component $M_{0xy}$ is presented with the arrows.}
    \label{fig:config}
\end{figure}

\subsection{Static magnetization texture}

In the multilayer nanodot, characterized by strong PMA and DMI, various magnetization states can be stabilized, including an out-of-plane single-domain state, a Néel-type skyrmion, a double-domain structure, and a worm-like domains~\cite{Vetrova2021InvestigationDot}. In this paper, we are focusing on two of the above-mentioned configurations---a single-domain state (SD) and a skyrmion state (Sk). While the literature is well-versed in the static and dynamic behavior of these structures in isolation~\cite{Riveros2021Field-dependentNanodots, Mruczkiewicz2018}, their static and dynamic properties become complex in a compact hybrid system~\cite{Zelent2023}.

Figure~\ref{fig:config} presents the static magnetic configuration of a single unit cell of the HMC. Here, we consider hybrid systems with two different magnetization configurations---the waveguide (W) with a chain of dots with the single-domain out-of-plane magnetization state (W/SD) and the waveguide with a chain of dots with skyrmions (W/Sk). The magnetization texture is shown on the $xy$-planes crossing the center of the dot (a,c) and the center of the waveguide (b,d), respectively. In the HMC, the magnetization configuration differs from that of isolated subsystems. This change, induced by the dipolar coupling, is mainly caused by the competition between the strong PMA in the dots, which favors magnetization along the $z$-axis, and the shape anisotropy inherent in the waveguide, which induces a preference for magnetization along the $x$-axis.

In the W/SD system, the most pronounced effect of dipolar interaction between the subsystems manifests just beneath the edges of the dot, as illustrated in Fig.~\ref{fig:config}(b). Here, the peak deviation in magnetization reaches $\max{|M_{0y}|}=216$~kA/m (i.e., 27\% of $M_\text{S}$ in Py) along the $y$-axis and $\max{|M_{0z}|}=23$~kA/m along the $z$-axis. Notably, within the nanodot itself, the magnetization deviation of approximately 2\% is present close to the dot edge.

The static magnetic texture in the W/Sk system undergoes more significant modification. Unlike the configurations observed in the individual subsystems, the skyrmion is not only imprinted in the waveguide, but also takes on an egg-like shape instead of being circular. This static effect has already been demonstrated in the system with a single dot and finite strip in Ref.~\cite{Zelent2023}. Please note that the imprint intensity is stronger in the W/Sk system than in the W/SD system. The average net magnetization along the easy axis in the W/Sk system decreases by 20~kA/m in comparison with only 5~kA/m for W/SD system. Also, the maximum deviation reaches $\max{|M_{0y}|}=475$~kA/m and $\max{|M_{0z}|}=59$~kA/m.

\subsection{Spin-wave dynamics}

\begin{figure*}[t]
    \includegraphics{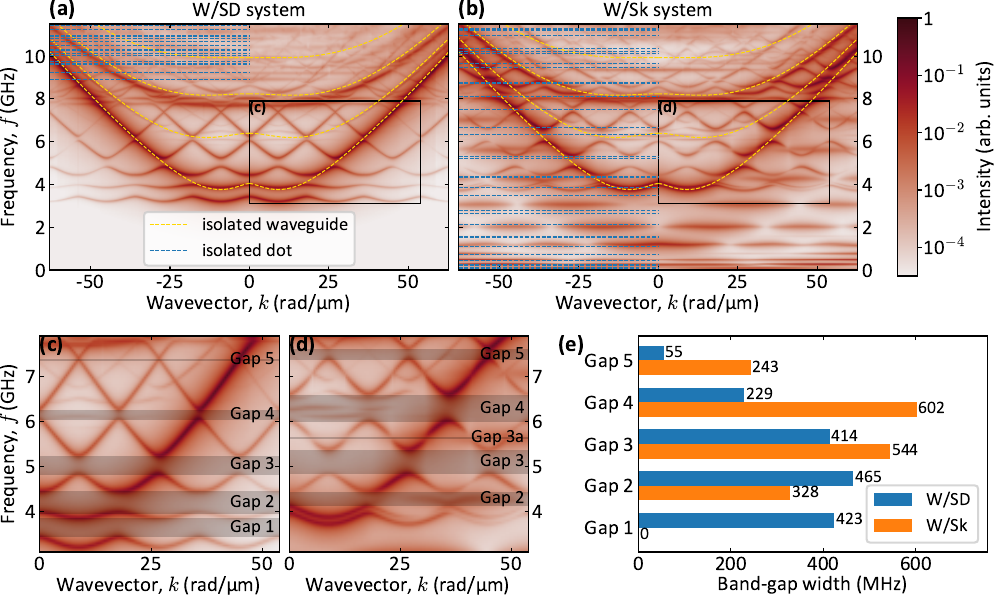}
    \caption{Dispersion relation of (a) the W/SD system and (b) the W/Sk system. Color maps show the intensity measured for the out-of-plane component of the magnetization $m_z$. Intensity is scaled logarithmically. Dashed yellow lines show the dispersion relation of the waveguide itself without dots and horizontal dashed blue lines mark the resonance frequencies of a dot in the SD state (a) and Sk state (b). Subfigures (c) and (d) are zoom-ins of the W/SD and W/Sk systems' dispersion relations as marked by black rectangles in subfigures (a) and (b), respectively. Gray rectangles mark the Bragg gaps in the dispersion relation. (e) Bar chart collecting band-gap widths for both W/SD and W/Sk systems.}
    \label{fig:disp}
\end{figure*}

Following the analysis of the system's static magnetization configurations, we performed numerical simulations of the SW dynamics (see Methods section for details). The dispersion relations of W/SD and W/Sk systems are depicted in Fig.~\ref{fig:disp}(a) and Fig.~\ref{fig:disp}(b), respectively, where the color map indicates the intensity $I$ of the out-of-plane dynamic magnetization component $m_z$ across the entire system. The intensity of each mode is quantified as follows:
\begin{equation}
    I_\mathrm{mode}(k,f_n(k)) = \left| \iiint_{V} m_z(f_n(k)) e^{ikx}\,\mathrm{d}x\,\mathrm{d}y\,\mathrm{d}z \right|^2,
\end{equation}
where $V$ denotes the volume of magnetic material within a single unit cell and $f_n(k)$ is the frequency of the $n$th mode at wavevector $k$, being a complex function. Next, the intensities of all modes are converted into the Lorentzian function and then summed to give the total intensity $I$
\begin{equation}
    I(k,f) = \sum_n \frac{I_\mathrm{mode}(k,f_n(k))}{\mathrm{Im}[f_n(k)]\left(1+\left(\frac{f-\mathrm{Re}[f_n(k)]}{\mathrm{Im}[f_n(k)])}\right)^2\right)}
\end{equation}
at wavevector $k$ and frequency $f$. This method of quantifying intensity makes these results comparable to the Brillouin light scattering measurement results~\cite{Gubbiotti2018SpinStudy}. For comparison, the dispersion relation of an isolated waveguide without a chain of dots is illustrated with dashed yellow lines and the resonance frequencies of the dot with the horizontal blue lines. The SW modes start at 8.887 GHz and 54 MHz in the first and next systems, respectively. The more detailed comparison between the dispersion relation of the isolated waveguide and the frequencies of the SW modes in the isolated dot in the single-domain and skyrmion states is presented in Section S2 of the Supplementary Materials.

The dispersion relations of both systems contain complex mode dependencies, caused by the interaction between the dots and their imprints in the waveguide. The highest-intensity mode follows the fundamental mode of an isolated waveguide. The antisymmetric waveguide modes are barely visible in the plots due to the nature of the intensity calculation. Above the third waveguide mode, the intensity distribution is unclear and only the fundamental modes are recognizable. The reflected branches and band gaps are present as a result of the periodicity induced by the arrangement of the dots.

However, there are significant differences in the band-gap width of the Bragg gaps among the systems. The zoom-ins of the dispersions of W/SD and W/Sk systems are shown in Fig.~\ref{fig:disp}(c) and Fig.~\ref{fig:disp}(d), respectively, with gray strips marking the positions of the band gaps. The widths of the first five band gaps are listed in Fig.~\ref{fig:disp}(e). The W/SD system is characterized by larger low-order gaps, with the size exceeding 400 MHz. The size of higher-order gaps is much smaller, with gap 5 already being similar in size to the linewidth of the modes (which is 77 MHz), making it barely noticeable. In contrast, the W/Sk system is characterized by larger sizes of higher-order band gaps, i.e., third and higher. Interestingly, the first band gap is completely absent. Due to the backward wave character of the mode at low wavevectors, the edge of the first Brillouin zone lies close to the frequency minimum. As a result, the first and second bands share the same character. In the W/SD system, the stronger interaction between the modes pushes the first band much below the frequency of the isolated waveguide (see Fig.~\ref{fig:disp}(a)). In the W/Sk system, this interaction is weaker, which causes the first band maximum to be at a higher frequency than the second band minimum, leading to the absence of the band gap, as visible in Fig.~\ref{fig:disp}(d). These properties clearly demonstrate the reprogrammable nature of the proposed HMC system. By preserving the frequency positions of the band gaps, we can modify their width and even close or open the first band gap simply by changing the magnetization state in the dots.

An important question is about the mechanisms of the band-gap formation in the W/SD and W/Sk systems. There are two main factors which can be involved in the formation. One is a periodic static configuration. The magnetization imprint in the waveguide coming from the dipolar interaction with the dot enables the Bragg scattering in the system. However, the dynamic coupling between the propagating SWs in the waveguide and the magnetization dynamics confined to the nanodots can also provide such an effect. It is more intuitive when the resonator eigenfrequencies overlap with the continuous bands of the waveguide \cite{Mruczkiewicz2017,Graczyk2018,Negrello2022}. Nevertheless, the coupling can also occur when the resonant frequencies of the nanoelements are much higher than the lowest propagation bands of the waveguide \cite{Qin2021,Wang2024c}. In our study, the former situation is realized in the W/Sk system and the latter in the W/SD system.

Another difference between the spectra in both systems is the presence of numerous flat bands in the dispersion of the W/Sk system, which lie below the waveguide modes and begin at frequencies below 1 GHz. These modes are directly connected to the dynamics of the skyrmion domain wall in the dots which starts at the level of tens of MHz (see, Fig.~S2 in the Supplementary Materials). At higher frequencies, some of the skyrmion modes hybridize with the waveguide modes. Interestingly, one of these modes leads to the generation of an additional band gap with a width of 53 MHz, marked as gap 3a in Fig.~\ref{fig:disp}(d). The modified spectra indicate that the presence of skyrmions in dots can directly affect the dynamics of SWs propagating in the waveguide. Obviously, such modes are not present in the W/SD system since the lowest resonant mode of the dot in a single-domain state is at the frequency of 8.887 GHz, which is above the third waveguide mode. These results show that the change of a magnetization configuration of the dots can induce additional flat bands in the SW spectrum and also magnon--magnon coupling, effects which are currently under intense investigation and also important from an application point of view~\cite{Gallardo2019,Tacchi2023,Centala2023,Wang2024b,Chen2018,Wang2024c,Moalic2024,Kumar2023,Bo2024}. In the next section, we will explore these properties, focusing on understanding of the physical mechanisms involved. 

\subsection{Mode localization}

\begin{figure*}[!t]
    \includegraphics{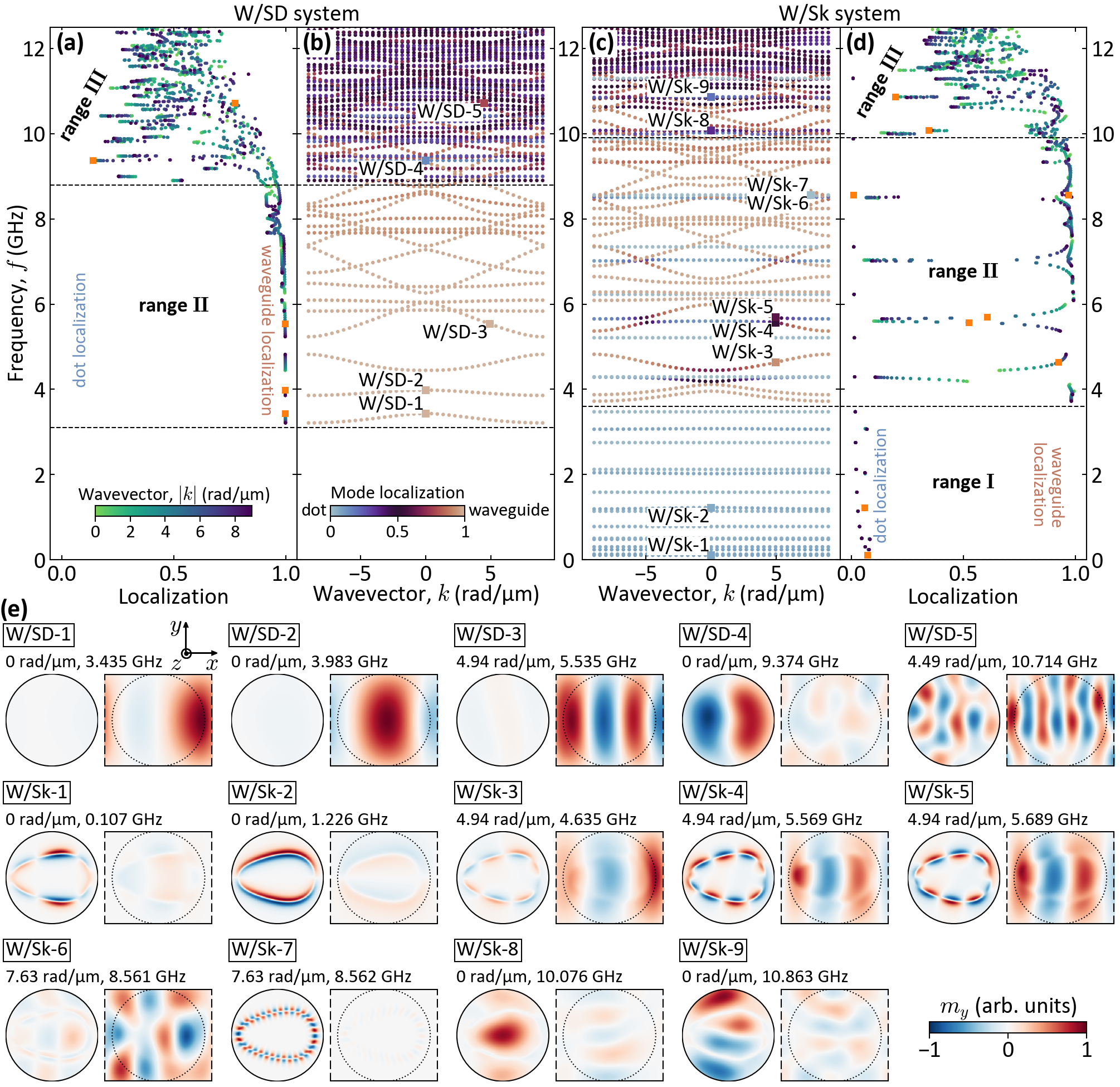}
    \caption{The dispersion relation in the first Brillouin zone presents the localization of modes in both (b) W/SD and (c) W/Sk systems. Each mode localization value is indicated by the color of the point on the dispersion. The corresponding plots with the localization value are shown in (a) for the W/SD system and in (d) for the W/Sk system. Here, the color of the point marks the absolute value of the wavevector. Dashed black vertical lines mark the limits of ranges. (e) SW mode profiles for 5 modes in the W/SD system and 9 modes in the W/Sk system. The modes are marked on the dispersion relations with a square point and a label. In each mode profile, the left color map displays the $m_y$ magnetization component in the $xy$-plane at the center of the dot, while the right color map displays $m_y$ in the $xy$-plane at the center of the waveguide. The intensity is normalized so that the maximum value of $|m_y|$ is 1 for each of the mode profiles. All profiles are labeled and their wavevector and frequency are given. The animated version of this figure is available in Supplementary Materials.}
    \label{fig:localization}
\end{figure*}

In order to deepen the analysis of the SW dynamics, we calculated the localization of the modes and plotted it on the dispersion relation folded to the first Brillouin zone as shown in Fig.~\ref{fig:localization}(b) and (c). We define the mode localization as a measure of how much of the intensity of a given mode comes from a given subsystem (in this case -- a waveguide). It is calculated as
\begin{equation}
    L(k,f_n) = \frac{\mathcal{I}_\mathrm{w}}{\mathcal{I}_\mathrm{w}+\mathcal{I}_\mathrm{d}},
\end{equation}
where $\mathcal{I}_\mathrm{w}$ and $\mathcal{I}_\mathrm{d}$ are the intensity of the mode in the waveguide and dot, respectively. Intensity is measured as
\begin{equation}
    \mathcal{I}_\mathrm{w(d)}(k,f_n) = \iiint_{V_\mathrm{w(d)}} \left|\mathbf{m}(k,f_n)\right|^2\,\mathrm{d}x\,\mathrm{d}y\,\mathrm{d}z 
\end{equation}
where $V_\mathrm{w(d)}$ is the volume of the waveguide (dot) in a unit cell.
The mode fully localized in the dot has a value of $L=0$ and is marked in Figs.~\ref{fig:localization}(b,c) with light blue color, while the mode fully localized in the waveguide has a value of $L=1$ and is marked with light brown color. A more extensive description of the mode localization is presented in Section S1 of the Supplementary Materials.

The analysis of the SW spectrum will be conducted in three different frequency ranges: range I, marking the frequencies below the possible SW excitation in the waveguide but available skyrmion excitation in the dot; range II, where the waveguide can be excited but still only the skyrmion in the dot; range III, where the excitation of the whole dot is possible.

The mode localization for the W/SD system is shown in Fig.~\ref{fig:localization}(b). The SW profiles of 5 selected modes (W/SD-1--W/SD-5) are presented in the first row of Fig.~\ref{fig:localization}(e). Additionally, we plot the frequency as a function of localization in Fig.~\ref{fig:localization}(a), with $k$ dependence coded in the color: green for $k=0$ and navy blue for the Brillouin zone boundary. There are no modes existing in the range I. It is because the excitation in the isolated subsystems are also not possible here. The W/Sk system is presented in Fig.~\ref{fig:localization}(c). The SW profiles of 9 selected modes are presented in two bottom rows of Fig.~\ref{fig:localization}(e). Additionally, we plot the frequency as a function of localization in Fig.~\ref{fig:localization}(d).

\subsubsection{W/SD system -- waveguide modes}

In the range II of the W/SD system, the lowest mode has a frequency above 3 GHz, and all modes up to 8.9 GHz are predominantly confined to the waveguide with minimal amplitude in the dot. The lowest localization value in this region is $L=0.89$. This result is consistent with the simulations of eigenstates of an isolated dot, where the lowest mode was observed at 8.887 GHz (see Section S2 in the Supplementary Materials). Therefore, any excitation in the dot below this frequency is only a forced oscillation, and the Bragg scattering and band-gap formation [Fig.~\ref{fig:disp}(c)] can be attributed to the static magnetization imprint created in the waveguide by the nanodot [see Fig.~\ref{fig:config}(b)]. Most of the modes in this area resemble typical waveguide modes, e.g. mode W/SD-3. However, other modes, such as W/SD-1 and W/SD-2, are significantly distorted. Despite the Bragg gap being present only at the edge of the Brillouin zone, their branches are coupled throughout the entire range of the zone. This results in a non-uniform SW amplitude distribution, even at zero $k$.

\subsubsection{W/Sk system -- narrow-bandwidth skyrmion modes}

In the W/Sk system, the range I spans the frequencies from 100 MHz to 3.5~GHz, which is below the frequency of the lowest waveguide band. It contains 15 flat bands related to the modes localized in dots, which are clockwise and counterclockwise azimuthal modes of the skyrmion domain wall, similar to the skyrmion in an isolated dot (see Fig.~S1 in Supplementary Materials) \cite{Mruczkiewicz2018}. The lowest frequency mode, W/Sk-1, is a 3rd order counterclockwise mode. The skyrmion breathing mode, W/Sk-2, has a frequency of 1.226 GHz and is only the ninth lowest mode. The largest localization in this range is $L=0.08$. The modes in this range have very small bandwidths ranging from 9 kHz to 34 MHz (breathing mode, W/Sk-2). This is because these modes can only interact through dipolar interactions or evanescent SWs, which are the only ones that can exist in the waveguide at such low frequencies. The simulation of the dot chain in the skyrmion state but without the waveguide (see Table~1 in Section S3 of the Supplementary Materials) shows that the bandwidths of most of the bands are significantly smaller in the absence of the waveguide, except for the breathing mode (of comparable bandwidth, 33~MHz) and the fourth counterclockwise mode, which is larger than in the system with the waveguide (further details on this comparison can be found in Section S3 of the Supplementary Materials). This suggests the coupling of the skyrmion modes from the range I through the evanescent waves in the waveguide. Such an effect is similar to the enhanced SW transmission in bi-component 1D MC at frequencies below the FMR frequency of one of the constituent materials~\cite{Qin2018}. However, here it is theoretically predicted for 1D HMC consisting of a homogeneous film and a chain of dots in the skyrmion state, the system easily extendable to a 2D array. The effects described above could be exploited for the design of frequency-selective magnonic devices, enabling precise control over signal modulation and processing at the nanoscale. The distinct localization and narrow bandwidths of these modes offer opportunities to create highly-efficient filters or oscillators that operate within a precisely-defined frequency range. Furthermore, these weakly dispersive bands can be used to exploit the flat-band physics recently demonstrated in 1D MCs with periodic modulation of a DMI~\cite{Gallardo2019,Tacchi2023}.

\subsubsection{W/Sk system -- hybridized and bound states}

Starting from 3.5~GHz, similarly to W/SD system, waveguide modes start to appear and they coexist with skyrmion modes in this range (marked as range II). Interestingly, the skyrmion is always slightly excited even if the mode is strongly localized in the waveguide (see mode W/Sk-3 with localization 0.925). Such a strong waveguide localization is also found at the edges of the Bragg band gaps, indicating that their origin is mainly in the imprinted magnetization in the waveguide, similar to the W/SD system. On the one hand, the clockwise skyrmion modes hybridize with the waveguide modes, resulting in mixed modes that are confined to both the waveguide and the dot. In Fig.~\ref{fig:localization}(d), these modes form characteristic horizontally aligned points with localization between 0.05 and about 0.55. As it was shown before, such a hybridization leads to the presence of an additional band gap marked as gap 3a in Fig.~\ref{fig:disp}(d), whose modes are labeled W/Sk-4 and W/Sk-5. Thus, it is possible to excite propagating modes in the waveguide with a specific wavelength by excitation of specific skyrmion modes, and study the recently intensively explored physics of the dynamically coupled systems, in particular magnon--magnon coupling~\cite{Adhikari_2021,Gartside2021,Wang2024}. Moreover, the resonant coupling offers a possibility for the realization of artificial neural networks~\cite{Papp2021,Fripp2023NonlinearNeurons}, where the propagating SWs act as synapses connecting neurons, playing through the nanodot in skyrmion or single-domain state. However, this requires the activation of neurons by propagating SWs. Such a nonlinear property can be achieved by slightly increasing the SW amplitude so that the static magnetic component decreases, resulting in a change of the resonance frequency (e.g. around the modes W/Sk-4 and W/Sk-5 or around 4~GHz, with a change of just about 10~MHz)~\cite{Krivosik2010,Merbouche2022,Qi2023} and, depending on the realization, activating or deactivating of the resonance effect.

On the other hand, counterclockwise skyrmion modes form a vertical line of points in Fig.~\ref{fig:localization}(d) and have a strong localization in dots with $L$ not exceeding 0.02. Thus, we can conclude that the interaction between counterclockwise modes and waveguide modes is negligible. Interesting is the pair of modes W/Sk-6 and W/Sk-7, which differs in frequency only by 1 MHz but their localization values are 0.97 and 0.01, respectively. As expected, the small amplitude in the dot for mode W/Sk-6 is not connected with the skyrmion mode W/Sk-7, further confirming a lack of coupling between them. It points to the possibility of exploiting these modes, which are strongly localized in the dot or waveguide but are uncoupled, as bound states in the continuum. This effect has been extensively studied in photonics but has yet to be explored in magnonics~\cite{Hsu2016}.

The hybridization of the modes is influenced by the magnetic damping,especially the hybridization between the waveguide and the skyrmion modes. The presence of high damping of $\alpha=0.02$ in the Co/Pt dots and relatively low damping of $\alpha=0.005$ in Py leads to the decrease of the anticrossing gap size, in some cases only slightly (e.g., around 5.5~GHz), in other cases even leading to a closure of the gap and crossing of the modes (e.g., around 8.5~GHz). It also leads to a significant change in the lifetime of the hybridized modes. Interestingly, high damping in the dots in SD state has much less influence on the collective dynamics in HMC. These effects are discussed in detail in Section S4 of the Supplementary Materials.

\subsubsection{Range III -- mixed regime}

In the frequency range marked as range III, the modes have a mixed character. It starts at 8.9 GHz for W/SD system and at 10 GHz for W/Sk system. Higher frequency for the W/Sk system is due to the presence of skyrmion in the dot, which induces specific confinement of the resonant modes in the dot, leading to an increase in their frequency. The localization value in the range vary between 0.13 and 0.96 (after excluding the weakly interacting counterclockwise skyrmion modes), and the range further reduces to 0.14--0.83 above 10.5 GHz in W/SD system and to 0.23--0.82 above 11.5 GHz in the W/Sk system. In this range there are modes localized predominantly in the dot (e.g. mode W/SD-4, W/Sk-8, and W/Sk-9), which originate from the resonant modes in the dot. However, there is always significant energy leakage to the waveguide. On the other hand, the propagating waveguide modes, such as mode W/SD-5, also have significant amplitude in the dot. This indicates the coupled nature of most nanodot modes with the waveguide SWs, also resulting in the smeared intensity observed in Fig.~\ref{fig:disp}(a), above the dispersion of the fundamental mode. Also, a large number of hybridized modes in the range II does not support the formation of band gaps.

The presence of the dot also strongly modifies the wavefront of the SW propagating in the waveguide (see mode W/SD-5). This effect can be used to excite propagating SWs in the waveguide by exciting the dots themselves, similar to the excitation of short-wavelength SWs with 2D diffraction couplers~\cite{Yu2016}.

\section{Conclusions}

We have studied a one-dimensional HMC consisting of an infinitely long Py waveguide and a chain of nanodots with PMA and DMI (Ir/Co/Pt), in which we consider two different magnetic states: a skyrmion and a single-domain state. The static magnetization configuration in the HMC differs from that of its isolated subsystems. This is because the configuration of the dot imprints the magnetization texture upon the waveguide, at the same time, the skyrmion shape becomes strongly distorted, taking on an egg-like shape. This makes a SW dynamic in an HMC complex while increasing the skyrmion stability and offering multifunctional properties for advancing magnonics. 

The dispersion relations of both systems exhibit characteristic features of magnonic crystals, such as folded branches and band gaps that open at the border or center of the Brillouin zone, resulting from the Bragg scattering on the static magnetization imprints in the waveguide. However, there is a substantial difference in the sizes of the band gaps, demonstrating the programmability of the proposed HMC. Additionally, the W/Sk system has a large number of flat low-frequency skyrmion modes below the resonance frequency of the waveguide. These modes are azimuthal rotating modes, both clockwise and counterclockwise, localized in the skyrmion domain wall and are characterized by very narrow bandwidths ranging from single kHz to single MHz. Interestingly, the bandwidths are significantly larger than those of the dots with the skyrmions forming a chain but without a waveguide, indicating evanescent-wave coupling between the skyrmions in W/Sk system. The flat bands may also overlap with the waveguide modes at higher frequencies and, interestingly, depending on their sense of rotation, can hybridize with them, sometimes even leading to additional band gaps in the spectrum, or be uncoupled. In the same frequency range in the W/SD system, all modes are almost exclusively localized in the waveguide. Interestingly, some of the anticrossing band gaps in the W/Sk system are preserved even at high damping values in Co/Pt dots.

At frequencies above 9 GHz, the resonant modes of the dots, also in a single-domain state, begin to appear and strongly hybridize with the waveguide modes, causing the localization of the modes to become mixed. However, in the W/Sk system, some of the modes (localized in the skyrmion domain wall) still can not interact with the waveguide at high frequencies, which promises the realization of the bound state in the continuum in magnonics.

The above-mentioned properties offer several useful functionalities for magnonics, including reconfigurability, filtering, evanescent coupling, magnon--magnon hybridizations, flat bands, uncoupled SW modes in the band structure, as well as SW-skyrmion bands together with their evanescence coupling. These functionalities are suitable for applications, in particular the realization of magnonic artificial neural networks.

\section*{Data availability}

The raw data files that support this study are available via
the Zenodo repository \cite{data}.

\begin{acknowledgments}
K. Szulc and M. Krawczyk acknowledge the financial support from National Science Centre, Poland, grants no. UMO-2020/39/I/ST3/02413 and UMO-2021/41/N/ST3/04478. K. Szulc acknowledges the financial support from the Foundation for Polish Science. This work was supported by the EU Research and Innovation Programme Horizon Europe (HORIZON-CL4-2021-DIGITAL-EMERGING-01) under grant agreement no. 101070347 (MANNGA).
\end{acknowledgments}


\begin{thebibliography}{79}%
    \makeatletter
    \providecommand \@ifxundefined [1]{%
     \@ifx{#1\undefined}
    }%
    \providecommand \@ifnum [1]{%
     \ifnum #1\expandafter \@firstoftwo
     \else \expandafter \@secondoftwo
     \fi
    }%
    \providecommand \@ifx [1]{%
     \ifx #1\expandafter \@firstoftwo
     \else \expandafter \@secondoftwo
     \fi
    }%
    \providecommand \natexlab [1]{#1}%
    \providecommand \enquote  [1]{``#1''}%
    \providecommand \bibnamefont  [1]{#1}%
    \providecommand \bibfnamefont [1]{#1}%
    \providecommand \citenamefont [1]{#1}%
    \providecommand \href@noop [0]{\@secondoftwo}%
    \providecommand \href [0]{\begingroup \@sanitize@url \@href}%
    \providecommand \@href[1]{\@@startlink{#1}\@@href}%
    \providecommand \@@href[1]{\endgroup#1\@@endlink}%
    \providecommand \@sanitize@url [0]{\catcode `\\12\catcode `\$12\catcode `\&12\catcode `\#12\catcode `\^12\catcode `\_12\catcode `\%12\relax}%
    \providecommand \@@startlink[1]{}%
    \providecommand \@@endlink[0]{}%
    \providecommand \url  [0]{\begingroup\@sanitize@url \@url }%
    \providecommand \@url [1]{\endgroup\@href {#1}{\urlprefix }}%
    \providecommand \urlprefix  [0]{URL }%
    \providecommand \Eprint [0]{\href }%
    \providecommand \doibase [0]{https://doi.org/}%
    \providecommand \selectlanguage [0]{\@gobble}%
    \providecommand \bibinfo  [0]{\@secondoftwo}%
    \providecommand \bibfield  [0]{\@secondoftwo}%
    \providecommand \translation [1]{[#1]}%
    \providecommand \BibitemOpen [0]{}%
    \providecommand \bibitemStop [0]{}%
    \providecommand \bibitemNoStop [0]{.\EOS\space}%
    \providecommand \EOS [0]{\spacefactor3000\relax}%
    \providecommand \BibitemShut  [1]{\csname bibitem#1\endcsname}%
    \let\auto@bib@innerbib\@empty
    \bibitem [{\citenamefont {Chumak}\ \emph {et~al.}(2022)\citenamefont {Chumak}, \citenamefont {Kabos}, \citenamefont {Wu}, \citenamefont {Abert}, \citenamefont {Adelmann}, \citenamefont {Adeyeye}, \citenamefont {Akerman}, \citenamefont {Aliev}, \citenamefont {Anane}, \citenamefont {Awad}, \citenamefont {Back}, \citenamefont {Barman}, \citenamefont {Bauer}, \citenamefont {Becherer}, \citenamefont {Beginin}, \citenamefont {Bittencourt}, \citenamefont {Blanter}, \citenamefont {Bortolotti}, \citenamefont {Boventer}, \citenamefont {Bozhko}, \citenamefont {Bunyaev}, \citenamefont {Carmiggelt}, \citenamefont {Cheenikundil}, \citenamefont {Ciubotaru}, \citenamefont {Cotofana}, \citenamefont {Csaba}, \citenamefont {Dobrovolskiy}, \citenamefont {Dubs}, \citenamefont {Elyasi}, \citenamefont {Fripp}, \citenamefont {Fulara}, \citenamefont {Golovchanskiy}, \citenamefont {Gonzalez-Ballestero}, \citenamefont {Graczyk}, \citenamefont {Grundler}, \citenamefont {Gruszecki}, \citenamefont {Gubbiotti}, \citenamefont {Guslienko},
      \citenamefont {Haldar}, \citenamefont {Hamdioui}, \citenamefont {Hertel}, \citenamefont {Hillebrands}, \citenamefont {Hioki}, \citenamefont {Houshang}, \citenamefont {Hu}, \citenamefont {Huebl}, \citenamefont {Huth}, \citenamefont {Iacocca}, \citenamefont {Jungfleisch}, \citenamefont {Kakazei}, \citenamefont {Khitun}, \citenamefont {Khymyn}, \citenamefont {Kikkawa}, \citenamefont {Klaui}, \citenamefont {Klein}, \citenamefont {Klos}, \citenamefont {Knauer}, \citenamefont {Koraltan}, \citenamefont {Kostylev}, \citenamefont {Krawczyk}, \citenamefont {Krivorotov}, \citenamefont {Kruglyak}, \citenamefont {Lachance-Quirion}, \citenamefont {Ladak}, \citenamefont {Lebrun}, \citenamefont {Li}, \citenamefont {Lindner}, \citenamefont {Macedo}, \citenamefont {Mayr}, \citenamefont {Melkov}, \citenamefont {Mieszczak}, \citenamefont {Nakamura}, \citenamefont {Nembach}, \citenamefont {Nikitin}, \citenamefont {Nikitov}, \citenamefont {Novosad}, \citenamefont {Otalora}, \citenamefont {Otani}, \citenamefont {Papp},
      \citenamefont {Pigeau}, \citenamefont {Pirro}, \citenamefont {Porod}, \citenamefont {Porrati}, \citenamefont {Qin}, \citenamefont {Rana}, \citenamefont {Reimann}, \citenamefont {Riente}, \citenamefont {Romero-Isart}, \citenamefont {Ross}, \citenamefont {Sadovnikov}, \citenamefont {Safin}, \citenamefont {Saitoh}, \citenamefont {Schmidt}, \citenamefont {Schultheiss}, \citenamefont {Schultheiss}, \citenamefont {Serga}, \citenamefont {Sharma}, \citenamefont {Shaw}, \citenamefont {Suess}, \citenamefont {Surzhenko}, \citenamefont {Szulc}, \citenamefont {Taniguchi}, \citenamefont {Urbanek}, \citenamefont {Usami}, \citenamefont {Ustinov}, \citenamefont {van~der Sar}, \citenamefont {van Dijken}, \citenamefont {Vasyuchka}, \citenamefont {Verba}, \citenamefont {Kusminskiy}, \citenamefont {Wang}, \citenamefont {Weides}, \citenamefont {Weiler}, \citenamefont {Wintz}, \citenamefont {Wolski},\ and\ \citenamefont {Zhang}}]{Chumak2022AdvancesComputing}%
      \BibitemOpen
      \bibfield  {author} {\bibinfo {author} {\bibfnamefont {A.~V.}\ \bibnamefont {Chumak}}, \bibinfo {author} {\bibfnamefont {P.}~\bibnamefont {Kabos}}, \bibinfo {author} {\bibfnamefont {M.}~\bibnamefont {Wu}}, \bibinfo {author} {\bibfnamefont {C.}~\bibnamefont {Abert}}, \bibinfo {author} {\bibfnamefont {C.}~\bibnamefont {Adelmann}}, \bibinfo {author} {\bibfnamefont {A.~O.}\ \bibnamefont {Adeyeye}}, \bibinfo {author} {\bibfnamefont {J.}~\bibnamefont {Akerman}}, \bibinfo {author} {\bibfnamefont {F.~G.}\ \bibnamefont {Aliev}}, \bibinfo {author} {\bibfnamefont {A.}~\bibnamefont {Anane}}, \bibinfo {author} {\bibfnamefont {A.}~\bibnamefont {Awad}}, \bibinfo {author} {\bibfnamefont {C.~H.}\ \bibnamefont {Back}}, \bibinfo {author} {\bibfnamefont {A.}~\bibnamefont {Barman}}, \bibinfo {author} {\bibfnamefont {G.~E.~W.}\ \bibnamefont {Bauer}}, \bibinfo {author} {\bibfnamefont {M.}~\bibnamefont {Becherer}}, \bibinfo {author} {\bibfnamefont {E.~N.}\ \bibnamefont {Beginin}}, \bibinfo {author} {\bibfnamefont {V.~A. S.~V.}\
      \bibnamefont {Bittencourt}}, \bibinfo {author} {\bibfnamefont {Y.~M.}\ \bibnamefont {Blanter}}, \bibinfo {author} {\bibfnamefont {P.}~\bibnamefont {Bortolotti}}, \bibinfo {author} {\bibfnamefont {I.}~\bibnamefont {Boventer}}, \bibinfo {author} {\bibfnamefont {D.~A.}\ \bibnamefont {Bozhko}}, \bibinfo {author} {\bibfnamefont {S.~A.}\ \bibnamefont {Bunyaev}}, \bibinfo {author} {\bibfnamefont {J.~J.}\ \bibnamefont {Carmiggelt}}, \bibinfo {author} {\bibfnamefont {R.~R.}\ \bibnamefont {Cheenikundil}}, \bibinfo {author} {\bibfnamefont {F.}~\bibnamefont {Ciubotaru}}, \bibinfo {author} {\bibfnamefont {S.}~\bibnamefont {Cotofana}}, \bibinfo {author} {\bibfnamefont {G.}~\bibnamefont {Csaba}}, \bibinfo {author} {\bibfnamefont {O.~V.}\ \bibnamefont {Dobrovolskiy}}, \bibinfo {author} {\bibfnamefont {C.}~\bibnamefont {Dubs}}, \bibinfo {author} {\bibfnamefont {M.}~\bibnamefont {Elyasi}}, \bibinfo {author} {\bibfnamefont {K.~G.}\ \bibnamefont {Fripp}}, \bibinfo {author} {\bibfnamefont {H.}~\bibnamefont {Fulara}}, \bibinfo
      {author} {\bibfnamefont {I.~A.}\ \bibnamefont {Golovchanskiy}}, \bibinfo {author} {\bibfnamefont {C.}~\bibnamefont {Gonzalez-Ballestero}}, \bibinfo {author} {\bibfnamefont {P.}~\bibnamefont {Graczyk}}, \bibinfo {author} {\bibfnamefont {D.}~\bibnamefont {Grundler}}, \bibinfo {author} {\bibfnamefont {P.}~\bibnamefont {Gruszecki}}, \bibinfo {author} {\bibfnamefont {G.}~\bibnamefont {Gubbiotti}}, \bibinfo {author} {\bibfnamefont {K.}~\bibnamefont {Guslienko}}, \bibinfo {author} {\bibfnamefont {A.}~\bibnamefont {Haldar}}, \bibinfo {author} {\bibfnamefont {S.}~\bibnamefont {Hamdioui}}, \bibinfo {author} {\bibfnamefont {R.}~\bibnamefont {Hertel}}, \bibinfo {author} {\bibfnamefont {B.}~\bibnamefont {Hillebrands}}, \bibinfo {author} {\bibfnamefont {T.}~\bibnamefont {Hioki}}, \bibinfo {author} {\bibfnamefont {A.}~\bibnamefont {Houshang}}, \bibinfo {author} {\bibfnamefont {C.-M.}\ \bibnamefont {Hu}}, \bibinfo {author} {\bibfnamefont {H.}~\bibnamefont {Huebl}}, \bibinfo {author} {\bibfnamefont {M.}~\bibnamefont
      {Huth}}, \bibinfo {author} {\bibfnamefont {E.}~\bibnamefont {Iacocca}}, \bibinfo {author} {\bibfnamefont {M.~B.}\ \bibnamefont {Jungfleisch}}, \bibinfo {author} {\bibfnamefont {G.~N.}\ \bibnamefont {Kakazei}}, \bibinfo {author} {\bibfnamefont {A.}~\bibnamefont {Khitun}}, \bibinfo {author} {\bibfnamefont {R.}~\bibnamefont {Khymyn}}, \bibinfo {author} {\bibfnamefont {T.}~\bibnamefont {Kikkawa}}, \bibinfo {author} {\bibfnamefont {M.}~\bibnamefont {Klaui}}, \bibinfo {author} {\bibfnamefont {O.}~\bibnamefont {Klein}}, \bibinfo {author} {\bibfnamefont {J.~W.}\ \bibnamefont {Klos}}, \bibinfo {author} {\bibfnamefont {S.}~\bibnamefont {Knauer}}, \bibinfo {author} {\bibfnamefont {S.}~\bibnamefont {Koraltan}}, \bibinfo {author} {\bibfnamefont {M.}~\bibnamefont {Kostylev}}, \bibinfo {author} {\bibfnamefont {M.}~\bibnamefont {Krawczyk}}, \bibinfo {author} {\bibfnamefont {I.~N.}\ \bibnamefont {Krivorotov}}, \bibinfo {author} {\bibfnamefont {V.~V.}\ \bibnamefont {Kruglyak}}, \bibinfo {author} {\bibfnamefont
      {D.}~\bibnamefont {Lachance-Quirion}}, \bibinfo {author} {\bibfnamefont {S.}~\bibnamefont {Ladak}}, \bibinfo {author} {\bibfnamefont {R.}~\bibnamefont {Lebrun}}, \bibinfo {author} {\bibfnamefont {Y.}~\bibnamefont {Li}}, \bibinfo {author} {\bibfnamefont {M.}~\bibnamefont {Lindner}}, \bibinfo {author} {\bibfnamefont {R.}~\bibnamefont {Macedo}}, \bibinfo {author} {\bibfnamefont {S.}~\bibnamefont {Mayr}}, \bibinfo {author} {\bibfnamefont {G.~A.}\ \bibnamefont {Melkov}}, \bibinfo {author} {\bibfnamefont {S.}~\bibnamefont {Mieszczak}}, \bibinfo {author} {\bibfnamefont {Y.}~\bibnamefont {Nakamura}}, \bibinfo {author} {\bibfnamefont {H.~T.}\ \bibnamefont {Nembach}}, \bibinfo {author} {\bibfnamefont {A.~A.}\ \bibnamefont {Nikitin}}, \bibinfo {author} {\bibfnamefont {S.~A.}\ \bibnamefont {Nikitov}}, \bibinfo {author} {\bibfnamefont {V.}~\bibnamefont {Novosad}}, \bibinfo {author} {\bibfnamefont {J.~A.}\ \bibnamefont {Otalora}}, \bibinfo {author} {\bibfnamefont {Y.}~\bibnamefont {Otani}}, \bibinfo {author}
      {\bibfnamefont {A.}~\bibnamefont {Papp}}, \bibinfo {author} {\bibfnamefont {B.}~\bibnamefont {Pigeau}}, \bibinfo {author} {\bibfnamefont {P.}~\bibnamefont {Pirro}}, \bibinfo {author} {\bibfnamefont {W.}~\bibnamefont {Porod}}, \bibinfo {author} {\bibfnamefont {F.}~\bibnamefont {Porrati}}, \bibinfo {author} {\bibfnamefont {H.}~\bibnamefont {Qin}}, \bibinfo {author} {\bibfnamefont {B.}~\bibnamefont {Rana}}, \bibinfo {author} {\bibfnamefont {T.}~\bibnamefont {Reimann}}, \bibinfo {author} {\bibfnamefont {F.}~\bibnamefont {Riente}}, \bibinfo {author} {\bibfnamefont {O.}~\bibnamefont {Romero-Isart}}, \bibinfo {author} {\bibfnamefont {A.}~\bibnamefont {Ross}}, \bibinfo {author} {\bibfnamefont {A.~V.}\ \bibnamefont {Sadovnikov}}, \bibinfo {author} {\bibfnamefont {A.~R.}\ \bibnamefont {Safin}}, \bibinfo {author} {\bibfnamefont {E.}~\bibnamefont {Saitoh}}, \bibinfo {author} {\bibfnamefont {G.}~\bibnamefont {Schmidt}}, \bibinfo {author} {\bibfnamefont {H.}~\bibnamefont {Schultheiss}}, \bibinfo {author} {\bibfnamefont
      {K.}~\bibnamefont {Schultheiss}}, \bibinfo {author} {\bibfnamefont {A.~A.}\ \bibnamefont {Serga}}, \bibinfo {author} {\bibfnamefont {S.}~\bibnamefont {Sharma}}, \bibinfo {author} {\bibfnamefont {J.~M.}\ \bibnamefont {Shaw}}, \bibinfo {author} {\bibfnamefont {D.}~\bibnamefont {Suess}}, \bibinfo {author} {\bibfnamefont {O.}~\bibnamefont {Surzhenko}}, \bibinfo {author} {\bibfnamefont {K.}~\bibnamefont {Szulc}}, \bibinfo {author} {\bibfnamefont {T.}~\bibnamefont {Taniguchi}}, \bibinfo {author} {\bibfnamefont {M.}~\bibnamefont {Urbanek}}, \bibinfo {author} {\bibfnamefont {K.}~\bibnamefont {Usami}}, \bibinfo {author} {\bibfnamefont {A.~B.}\ \bibnamefont {Ustinov}}, \bibinfo {author} {\bibfnamefont {T.}~\bibnamefont {van~der Sar}}, \bibinfo {author} {\bibfnamefont {S.}~\bibnamefont {van Dijken}}, \bibinfo {author} {\bibfnamefont {V.~I.}\ \bibnamefont {Vasyuchka}}, \bibinfo {author} {\bibfnamefont {R.}~\bibnamefont {Verba}}, \bibinfo {author} {\bibfnamefont {S.~V.}\ \bibnamefont {Kusminskiy}}, \bibinfo {author}
      {\bibfnamefont {Q.}~\bibnamefont {Wang}}, \bibinfo {author} {\bibfnamefont {M.}~\bibnamefont {Weides}}, \bibinfo {author} {\bibfnamefont {M.}~\bibnamefont {Weiler}}, \bibinfo {author} {\bibfnamefont {S.}~\bibnamefont {Wintz}}, \bibinfo {author} {\bibfnamefont {S.~P.}\ \bibnamefont {Wolski}},\ and\ \bibinfo {author} {\bibfnamefont {X.}~\bibnamefont {Zhang}},\ }\bibfield  {title} {\bibinfo {title} {{Advances in Magnetics Roadmap on Spin-Wave Computing}},\ }\href {https://doi.org/10.1109/TMAG.2022.3149664} {\bibfield  {journal} {\bibinfo  {journal} {IEEE Trans. Magn.}\ }\textbf {\bibinfo {volume} {58}},\ \bibinfo {pages} {1} (\bibinfo {year} {2022})}\BibitemShut {NoStop}%
    \bibitem [{\citenamefont {Wang}\ \emph {et~al.}(2023{\natexlab{a}})\citenamefont {Wang}, \citenamefont {Csaba}, \citenamefont {Verba}, \citenamefont {Chumak},\ and\ \citenamefont {Pirro}}]{Wang2023PerspectiveNetworks}%
      \BibitemOpen
      \bibfield  {author} {\bibinfo {author} {\bibfnamefont {Q.}~\bibnamefont {Wang}}, \bibinfo {author} {\bibfnamefont {G.}~\bibnamefont {Csaba}}, \bibinfo {author} {\bibfnamefont {R.}~\bibnamefont {Verba}}, \bibinfo {author} {\bibfnamefont {A.~V.}\ \bibnamefont {Chumak}},\ and\ \bibinfo {author} {\bibfnamefont {P.}~\bibnamefont {Pirro}},\ }\href@noop {} {\bibinfo {title} {Perspective on nanoscaled magnonic networks}} (\bibinfo {year} {2023}{\natexlab{a}}),\ \Eprint {https://arxiv.org/abs/2311.06129} {arXiv:2311.06129 [physics.app-ph]} \BibitemShut {NoStop}%
    \bibitem [{\citenamefont {Chumak}\ \emph {et~al.}(2014)\citenamefont {Chumak}, \citenamefont {Serga},\ and\ \citenamefont {Hillebrands}}]{Chumak2014}%
      \BibitemOpen
      \bibfield  {author} {\bibinfo {author} {\bibfnamefont {A.~V.}\ \bibnamefont {Chumak}}, \bibinfo {author} {\bibfnamefont {A.~A.}\ \bibnamefont {Serga}},\ and\ \bibinfo {author} {\bibfnamefont {B.}~\bibnamefont {Hillebrands}},\ }\bibfield  {title} {\bibinfo {title} {Magnon transistor for all-magnon data processing},\ }\href {https://doi.org/10.1038/ncomms5700} {\bibfield  {journal} {\bibinfo  {journal} {Nat. Commun.}\ }\textbf {\bibinfo {volume} {5}},\ \bibinfo {pages} {4700} (\bibinfo {year} {2014})}\BibitemShut {NoStop}%
    \bibitem [{\citenamefont {Klingler}\ \emph {et~al.}(2014)\citenamefont {Klingler}, \citenamefont {Pirro}, \citenamefont {Brächer}, \citenamefont {Leven}, \citenamefont {Hillebrands},\ and\ \citenamefont {Chumak}}]{Klinger2014}%
      \BibitemOpen
      \bibfield  {author} {\bibinfo {author} {\bibfnamefont {S.}~\bibnamefont {Klingler}}, \bibinfo {author} {\bibfnamefont {P.}~\bibnamefont {Pirro}}, \bibinfo {author} {\bibfnamefont {T.}~\bibnamefont {Brächer}}, \bibinfo {author} {\bibfnamefont {B.}~\bibnamefont {Leven}}, \bibinfo {author} {\bibfnamefont {B.}~\bibnamefont {Hillebrands}},\ and\ \bibinfo {author} {\bibfnamefont {A.~V.}\ \bibnamefont {Chumak}},\ }\bibfield  {title} {\bibinfo {title} {{Design of a spin-wave majority gate employing mode selection}},\ }\href {https://doi.org/10.1063/1.4898042} {\bibfield  {journal} {\bibinfo  {journal} {Appl. Phys. Lett.}\ }\textbf {\bibinfo {volume} {105}},\ \bibinfo {pages} {152410} (\bibinfo {year} {2014})}\BibitemShut {NoStop}%
    \bibitem [{\citenamefont {Fischer}\ \emph {et~al.}(2017)\citenamefont {Fischer}, \citenamefont {Kewenig}, \citenamefont {Bozhko}, \citenamefont {Serga}, \citenamefont {Syvorotka}, \citenamefont {Ciubotaru}, \citenamefont {Adelmann}, \citenamefont {Hillebrands},\ and\ \citenamefont {Chumak}}]{Fischer2017}%
      \BibitemOpen
      \bibfield  {author} {\bibinfo {author} {\bibfnamefont {T.}~\bibnamefont {Fischer}}, \bibinfo {author} {\bibfnamefont {M.}~\bibnamefont {Kewenig}}, \bibinfo {author} {\bibfnamefont {D.~A.}\ \bibnamefont {Bozhko}}, \bibinfo {author} {\bibfnamefont {A.~A.}\ \bibnamefont {Serga}}, \bibinfo {author} {\bibfnamefont {I.~I.}\ \bibnamefont {Syvorotka}}, \bibinfo {author} {\bibfnamefont {F.}~\bibnamefont {Ciubotaru}}, \bibinfo {author} {\bibfnamefont {C.}~\bibnamefont {Adelmann}}, \bibinfo {author} {\bibfnamefont {B.}~\bibnamefont {Hillebrands}},\ and\ \bibinfo {author} {\bibfnamefont {A.~V.}\ \bibnamefont {Chumak}},\ }\bibfield  {title} {\bibinfo {title} {{Experimental prototype of a spin-wave majority gate}},\ }\href {https://doi.org/10.1063/1.4979840} {\bibfield  {journal} {\bibinfo  {journal} {Appl. Phys. Lett.}\ }\textbf {\bibinfo {volume} {110}},\ \bibinfo {pages} {152401} (\bibinfo {year} {2017})}\BibitemShut {NoStop}%
    \bibitem [{\citenamefont {Lee}\ and\ \citenamefont {Mochizuki}(2022)}]{Lee2022ReservoirCrystal}%
      \BibitemOpen
      \bibfield  {author} {\bibinfo {author} {\bibfnamefont {M.-K.}\ \bibnamefont {Lee}}\ and\ \bibinfo {author} {\bibfnamefont {M.}~\bibnamefont {Mochizuki}},\ }\bibfield  {title} {\bibinfo {title} {Reservoir computing with spin waves in a skyrmion crystal},\ }\href {https://doi.org/10.1103/PhysRevApplied.18.014074} {\bibfield  {journal} {\bibinfo  {journal} {Phys. Rev. Appl.}\ }\textbf {\bibinfo {volume} {18}},\ \bibinfo {pages} {014074} (\bibinfo {year} {2022})}\BibitemShut {NoStop}%
    \bibitem [{\citenamefont {Lee}\ \emph {et~al.}(2024)\citenamefont {Lee}, \citenamefont {Wei}, \citenamefont {Stenning}, \citenamefont {Gartside}, \citenamefont {Prestwood}, \citenamefont {Seki}, \citenamefont {Aqeel}, \citenamefont {Karube}, \citenamefont {Kanazawa}, \citenamefont {Taguchi}, \citenamefont {Back}, \citenamefont {Tokura}, \citenamefont {Branford},\ and\ \citenamefont {Kurebayashi}}]{Lee2024}%
      \BibitemOpen
      \bibfield  {author} {\bibinfo {author} {\bibfnamefont {O.}~\bibnamefont {Lee}}, \bibinfo {author} {\bibfnamefont {T.}~\bibnamefont {Wei}}, \bibinfo {author} {\bibfnamefont {K.~D.}\ \bibnamefont {Stenning}}, \bibinfo {author} {\bibfnamefont {J.~C.}\ \bibnamefont {Gartside}}, \bibinfo {author} {\bibfnamefont {D.}~\bibnamefont {Prestwood}}, \bibinfo {author} {\bibfnamefont {S.}~\bibnamefont {Seki}}, \bibinfo {author} {\bibfnamefont {A.}~\bibnamefont {Aqeel}}, \bibinfo {author} {\bibfnamefont {K.}~\bibnamefont {Karube}}, \bibinfo {author} {\bibfnamefont {N.}~\bibnamefont {Kanazawa}}, \bibinfo {author} {\bibfnamefont {Y.}~\bibnamefont {Taguchi}}, \bibinfo {author} {\bibfnamefont {C.}~\bibnamefont {Back}}, \bibinfo {author} {\bibfnamefont {Y.}~\bibnamefont {Tokura}}, \bibinfo {author} {\bibfnamefont {W.~R.}\ \bibnamefont {Branford}},\ and\ \bibinfo {author} {\bibfnamefont {H.}~\bibnamefont {Kurebayashi}},\ }\bibfield  {title} {\bibinfo {title} {Task-adaptive physical reservoir computing},\ }\href
      {https://doi.org/10.1038/s41563-023-01698-8} {\bibfield  {journal} {\bibinfo  {journal} {Nat. Mater.}\ }\textbf {\bibinfo {volume} {23}},\ \bibinfo {pages} {79} (\bibinfo {year} {2024})}\BibitemShut {NoStop}%
    \bibitem [{\citenamefont {Wintz}\ \emph {et~al.}(2016)\citenamefont {Wintz}, \citenamefont {Tiberkevich}, \citenamefont {Weigand}, \citenamefont {Raabe}, \citenamefont {Lindner}, \citenamefont {Erbe}, \citenamefont {Slavin},\ and\ \citenamefont {Fassbender}}]{Wintz2016}%
      \BibitemOpen
      \bibfield  {author} {\bibinfo {author} {\bibfnamefont {S.}~\bibnamefont {Wintz}}, \bibinfo {author} {\bibfnamefont {V.}~\bibnamefont {Tiberkevich}}, \bibinfo {author} {\bibfnamefont {M.}~\bibnamefont {Weigand}}, \bibinfo {author} {\bibfnamefont {J.}~\bibnamefont {Raabe}}, \bibinfo {author} {\bibfnamefont {J.}~\bibnamefont {Lindner}}, \bibinfo {author} {\bibfnamefont {A.}~\bibnamefont {Erbe}}, \bibinfo {author} {\bibfnamefont {A.}~\bibnamefont {Slavin}},\ and\ \bibinfo {author} {\bibfnamefont {J.}~\bibnamefont {Fassbender}},\ }\bibfield  {title} {\bibinfo {title} {Magnetic vortex cores as tunable spin-wave emitters},\ }\href {https://doi.org/10.1038/nnano.2016.117} {\bibfield  {journal} {\bibinfo  {journal} {Nat. Nanotechnol.}\ }\textbf {\bibinfo {volume} {11}},\ \bibinfo {pages} {948} (\bibinfo {year} {2016})}\BibitemShut {NoStop}%
    \bibitem [{\citenamefont {Yu}\ \emph {et~al.}(2021)\citenamefont {Yu}, \citenamefont {Xiao},\ and\ \citenamefont {Schultheiss}}]{Yu2021}%
      \BibitemOpen
      \bibfield  {author} {\bibinfo {author} {\bibfnamefont {H.}~\bibnamefont {Yu}}, \bibinfo {author} {\bibfnamefont {J.}~\bibnamefont {Xiao}},\ and\ \bibinfo {author} {\bibfnamefont {H.}~\bibnamefont {Schultheiss}},\ }\bibfield  {title} {\bibinfo {title} {{Magnetic texture based magnonics}},\ }\href {https://doi.org/10.1016/J.PHYSREP.2020.12.004} {\bibfield  {journal} {\bibinfo  {journal} {Phys. Rep.}\ }\textbf {\bibinfo {volume} {905}},\ \bibinfo {pages} {1} (\bibinfo {year} {2021})}\BibitemShut {NoStop}%
    \bibitem [{\citenamefont {Petti}\ \emph {et~al.}(2022)\citenamefont {Petti}, \citenamefont {Tacchi},\ and\ \citenamefont {Albisetti}}]{Petti2022}%
      \BibitemOpen
      \bibfield  {author} {\bibinfo {author} {\bibfnamefont {D.}~\bibnamefont {Petti}}, \bibinfo {author} {\bibfnamefont {S.}~\bibnamefont {Tacchi}},\ and\ \bibinfo {author} {\bibfnamefont {E.}~\bibnamefont {Albisetti}},\ }\bibfield  {title} {\bibinfo {title} {Review on magnonics with engineered spin textures},\ }\href {https://doi.org/10.1088/1361-6463/ac6465} {\bibfield  {journal} {\bibinfo  {journal} {J. Phys. D: Appl. Phys.}\ }\textbf {\bibinfo {volume} {55}},\ \bibinfo {pages} {293003} (\bibinfo {year} {2022})}\BibitemShut {NoStop}%
    \bibitem [{\citenamefont {Fert}\ \emph {et~al.}(2013)\citenamefont {Fert}, \citenamefont {Cros},\ and\ \citenamefont {Sampaio}}]{Fert2013}%
      \BibitemOpen
      \bibfield  {author} {\bibinfo {author} {\bibfnamefont {A.}~\bibnamefont {Fert}}, \bibinfo {author} {\bibfnamefont {V.}~\bibnamefont {Cros}},\ and\ \bibinfo {author} {\bibfnamefont {J.}~\bibnamefont {Sampaio}},\ }\bibfield  {title} {\bibinfo {title} {Skyrmions on the track},\ }\href {https://doi.org/10.1038/nnano.2013.29} {\bibfield  {journal} {\bibinfo  {journal} {Nat. Nanotechnol.}\ }\textbf {\bibinfo {volume} {8}},\ \bibinfo {pages} {152} (\bibinfo {year} {2013})}\BibitemShut {NoStop}%
    \bibitem [{\citenamefont {Lonsky}\ and\ \citenamefont {Hoffmann}(2020)}]{Lonsky2020DynamicTextures}%
      \BibitemOpen
      \bibfield  {author} {\bibinfo {author} {\bibfnamefont {M.}~\bibnamefont {Lonsky}}\ and\ \bibinfo {author} {\bibfnamefont {A.}~\bibnamefont {Hoffmann}},\ }\bibfield  {title} {\bibinfo {title} {{Dynamic excitations of chiral magnetic textures}},\ }\href {https://doi.org/10.1063/5.0027042} {\bibfield  {journal} {\bibinfo  {journal} {APL Mater.}\ }\textbf {\bibinfo {volume} {8}},\ \bibinfo {pages} {100903} (\bibinfo {year} {2020})}\BibitemShut {NoStop}%
    \bibitem [{\citenamefont {Vakili}\ \emph {et~al.}(2021)\citenamefont {Vakili}, \citenamefont {Xu}, \citenamefont {Zhou}, \citenamefont {Sakib}, \citenamefont {Morshed}, \citenamefont {Hartnett}, \citenamefont {Quessab}, \citenamefont {Litzius}, \citenamefont {Ma}, \citenamefont {Ganguly}, \citenamefont {Stan}, \citenamefont {Balachandran}, \citenamefont {Beach}, \citenamefont {Poon}, \citenamefont {Kent},\ and\ \citenamefont {Ghosh}}]{Vakili2021Skyrmionics-computingMagnets}%
      \BibitemOpen
      \bibfield  {author} {\bibinfo {author} {\bibfnamefont {H.}~\bibnamefont {Vakili}}, \bibinfo {author} {\bibfnamefont {J.-W.}\ \bibnamefont {Xu}}, \bibinfo {author} {\bibfnamefont {W.}~\bibnamefont {Zhou}}, \bibinfo {author} {\bibfnamefont {M.~N.}\ \bibnamefont {Sakib}}, \bibinfo {author} {\bibfnamefont {M.~G.}\ \bibnamefont {Morshed}}, \bibinfo {author} {\bibfnamefont {T.}~\bibnamefont {Hartnett}}, \bibinfo {author} {\bibfnamefont {Y.}~\bibnamefont {Quessab}}, \bibinfo {author} {\bibfnamefont {K.}~\bibnamefont {Litzius}}, \bibinfo {author} {\bibfnamefont {C.~T.}\ \bibnamefont {Ma}}, \bibinfo {author} {\bibfnamefont {S.}~\bibnamefont {Ganguly}}, \bibinfo {author} {\bibfnamefont {M.~R.}\ \bibnamefont {Stan}}, \bibinfo {author} {\bibfnamefont {P.~V.}\ \bibnamefont {Balachandran}}, \bibinfo {author} {\bibfnamefont {G.~S.~D.}\ \bibnamefont {Beach}}, \bibinfo {author} {\bibfnamefont {S.~J.}\ \bibnamefont {Poon}}, \bibinfo {author} {\bibfnamefont {A.~D.}\ \bibnamefont {Kent}},\ and\ \bibinfo {author} {\bibfnamefont
      {A.~W.}\ \bibnamefont {Ghosh}},\ }\bibfield  {title} {\bibinfo {title} {{{Skyrmionics—Computing and memory technologies based on topological excitations in magnets}}},\ }\href {https://doi.org/10.1063/5.0046950} {\bibfield  {journal} {\bibinfo  {journal} {J. Appl. Phys.}\ }\textbf {\bibinfo {volume} {130}},\ \bibinfo {pages} {070908} (\bibinfo {year} {2021})}\BibitemShut {NoStop}%
    \bibitem [{\citenamefont {Everschor-Sitte}\ \emph {et~al.}(2018)\citenamefont {Everschor-Sitte}, \citenamefont {Masell}, \citenamefont {Reeve},\ and\ \citenamefont {Kläui}}]{Everschor-Sitte2018Perspective:Field}%
      \BibitemOpen
      \bibfield  {author} {\bibinfo {author} {\bibfnamefont {K.}~\bibnamefont {Everschor-Sitte}}, \bibinfo {author} {\bibfnamefont {J.}~\bibnamefont {Masell}}, \bibinfo {author} {\bibfnamefont {R.~M.}\ \bibnamefont {Reeve}},\ and\ \bibinfo {author} {\bibfnamefont {M.}~\bibnamefont {Kläui}},\ }\bibfield  {title} {\bibinfo {title} {{{Perspective: Magnetic skyrmions—Overview of recent progress in an active research field}}},\ }\href {https://doi.org/10.1063/1.5048972} {\bibfield  {journal} {\bibinfo  {journal} {J. Appl. Phys.}\ }\textbf {\bibinfo {volume} {124}},\ \bibinfo {pages} {240901} (\bibinfo {year} {2018})}\BibitemShut {NoStop}%
    \bibitem [{\citenamefont {Fert}\ \emph {et~al.}(2017)\citenamefont {Fert}, \citenamefont {Reyren},\ and\ \citenamefont {Cros}}]{AFert2017MagneticApplications}%
      \BibitemOpen
      \bibfield  {author} {\bibinfo {author} {\bibfnamefont {A.}~\bibnamefont {Fert}}, \bibinfo {author} {\bibfnamefont {N.}~\bibnamefont {Reyren}},\ and\ \bibinfo {author} {\bibfnamefont {V.}~\bibnamefont {Cros}},\ }\bibfield  {title} {\bibinfo {title} {{Magnetic skyrmions: advances in physics and potential applications}},\ }\href@noop {} {\bibfield  {journal} {\bibinfo  {journal} {Nat. Rev. Mater.}\ }\textbf {\bibinfo {volume} {2}},\ \bibinfo {pages} {17031} (\bibinfo {year} {2017})}\BibitemShut {NoStop}%
    \bibitem [{\citenamefont {Marrows}\ and\ \citenamefont {Zeissler}(2021)}]{Marrows2021PerspectiveSpintronics}%
      \BibitemOpen
      \bibfield  {author} {\bibinfo {author} {\bibfnamefont {C.~H.}\ \bibnamefont {Marrows}}\ and\ \bibinfo {author} {\bibfnamefont {K.}~\bibnamefont {Zeissler}},\ }\bibfield  {title} {\bibinfo {title} {{Perspective on skyrmion spintronics}},\ }\href {https://doi.org/10.1063/5.0072735} {\bibfield  {journal} {\bibinfo  {journal} {Appl. Phys. Lett.}\ }\textbf {\bibinfo {volume} {119}},\ \bibinfo {pages} {250502} (\bibinfo {year} {2021})}\BibitemShut {NoStop}%
    \bibitem [{\citenamefont {Li}\ \emph {et~al.}(2022)\citenamefont {Li}, \citenamefont {Ma}, \citenamefont {Chen}, \citenamefont {Xie},\ and\ \citenamefont {Ma}}]{Li2022}%
      \BibitemOpen
      \bibfield  {author} {\bibinfo {author} {\bibfnamefont {Z.}~\bibnamefont {Li}}, \bibinfo {author} {\bibfnamefont {M.}~\bibnamefont {Ma}}, \bibinfo {author} {\bibfnamefont {Z.}~\bibnamefont {Chen}}, \bibinfo {author} {\bibfnamefont {K.}~\bibnamefont {Xie}},\ and\ \bibinfo {author} {\bibfnamefont {F.}~\bibnamefont {Ma}},\ }\bibfield  {title} {\bibinfo {title} {{Interaction between magnon and skyrmion: Toward quantum magnonics}},\ }\href {https://doi.org/10.1063/5.0121314} {\bibfield  {journal} {\bibinfo  {journal} {J. Appl. Phys.}\ }\textbf {\bibinfo {volume} {132}},\ \bibinfo {pages} {210702} (\bibinfo {year} {2022})}\BibitemShut {NoStop}%
    \bibitem [{\citenamefont {Moon}\ \emph {et~al.}(2016)\citenamefont {Moon}, \citenamefont {Chun}, \citenamefont {Kim},\ and\ \citenamefont {Hwang}}]{Moon2016}%
      \BibitemOpen
      \bibfield  {author} {\bibinfo {author} {\bibfnamefont {K.-W.}\ \bibnamefont {Moon}}, \bibinfo {author} {\bibfnamefont {B.~S.}\ \bibnamefont {Chun}}, \bibinfo {author} {\bibfnamefont {W.}~\bibnamefont {Kim}},\ and\ \bibinfo {author} {\bibfnamefont {C.}~\bibnamefont {Hwang}},\ }\bibfield  {title} {\bibinfo {title} {Control of spin-wave refraction using arrays of skyrmions},\ }\href {https://doi.org/10.1103/PhysRevApplied.6.064027} {\bibfield  {journal} {\bibinfo  {journal} {Phys. Rev. Appl.}\ }\textbf {\bibinfo {volume} {6}},\ \bibinfo {pages} {064027} (\bibinfo {year} {2016})}\BibitemShut {NoStop}%
    \bibitem [{\citenamefont {Lan}\ and\ \citenamefont {Xiao}(2021)}]{Lan2021}%
      \BibitemOpen
      \bibfield  {author} {\bibinfo {author} {\bibfnamefont {J.}~\bibnamefont {Lan}}\ and\ \bibinfo {author} {\bibfnamefont {J.}~\bibnamefont {Xiao}},\ }\bibfield  {title} {\bibinfo {title} {Skew scattering and side jump of spin wave across magnetic texture},\ }\href {https://doi.org/10.1103/PhysRevB.103.054428} {\bibfield  {journal} {\bibinfo  {journal} {Phys. Rev. B}\ }\textbf {\bibinfo {volume} {103}},\ \bibinfo {pages} {054428} (\bibinfo {year} {2021})}\BibitemShut {NoStop}%
    \bibitem [{\citenamefont {Kotus}\ \emph {et~al.}(2022)\citenamefont {Kotus}, \citenamefont {Moalic}, \citenamefont {Zelent}, \citenamefont {Krawczyk},\ and\ \citenamefont {Gruszecki}}]{Kotus2022}%
      \BibitemOpen
      \bibfield  {author} {\bibinfo {author} {\bibfnamefont {K.~A.}\ \bibnamefont {Kotus}}, \bibinfo {author} {\bibfnamefont {M.}~\bibnamefont {Moalic}}, \bibinfo {author} {\bibfnamefont {M.}~\bibnamefont {Zelent}}, \bibinfo {author} {\bibfnamefont {M.}~\bibnamefont {Krawczyk}},\ and\ \bibinfo {author} {\bibfnamefont {P.}~\bibnamefont {Gruszecki}},\ }\bibfield  {title} {\bibinfo {title} {{Scattering of spin waves in a multimode waveguide under the influence of confined magnetic skyrmion}},\ }\href {https://doi.org/10.1063/5.0100594} {\bibfield  {journal} {\bibinfo  {journal} {APL Mater.}\ }\textbf {\bibinfo {volume} {10}},\ \bibinfo {pages} {091101} (\bibinfo {year} {2022})}\BibitemShut {NoStop}%
    \bibitem [{\citenamefont {Wang}\ \emph {et~al.}(2021)\citenamefont {Wang}, \citenamefont {Yuan}, \citenamefont {Cao}, \citenamefont {Li}, \citenamefont {Duine},\ and\ \citenamefont {Yan}}]{Wang2021}%
      \BibitemOpen
      \bibfield  {author} {\bibinfo {author} {\bibfnamefont {Z.}~\bibnamefont {Wang}}, \bibinfo {author} {\bibfnamefont {H.~Y.}\ \bibnamefont {Yuan}}, \bibinfo {author} {\bibfnamefont {Y.}~\bibnamefont {Cao}}, \bibinfo {author} {\bibfnamefont {Z.-X.}\ \bibnamefont {Li}}, \bibinfo {author} {\bibfnamefont {R.~A.}\ \bibnamefont {Duine}},\ and\ \bibinfo {author} {\bibfnamefont {P.}~\bibnamefont {Yan}},\ }\bibfield  {title} {\bibinfo {title} {Magnonic frequency comb through nonlinear magnon-skyrmion scattering},\ }\href {https://doi.org/10.1103/PhysRevLett.127.037202} {\bibfield  {journal} {\bibinfo  {journal} {Phys. Rev. Lett.}\ }\textbf {\bibinfo {volume} {127}},\ \bibinfo {pages} {037202} (\bibinfo {year} {2021})}\BibitemShut {NoStop}%
    \bibitem [{\citenamefont {D{\'i}az}\ \emph {et~al.}(2020)\citenamefont {D{\'i}az}, \citenamefont {Hirosawa}, \citenamefont {Loss},\ and\ \citenamefont {Psaroudaki}}]{Diaz2020}%
      \BibitemOpen
      \bibfield  {author} {\bibinfo {author} {\bibfnamefont {S.~A.}\ \bibnamefont {D{\'i}az}}, \bibinfo {author} {\bibfnamefont {T.}~\bibnamefont {Hirosawa}}, \bibinfo {author} {\bibfnamefont {D.}~\bibnamefont {Loss}},\ and\ \bibinfo {author} {\bibfnamefont {C.}~\bibnamefont {Psaroudaki}},\ }\bibfield  {title} {\bibinfo {title} {Spin wave radiation by a topological charge dipole},\ }\href {https://doi.org/10.1021/acs.nanolett.0c02192} {\bibfield  {journal} {\bibinfo  {journal} {Nano Lett.}\ }\textbf {\bibinfo {volume} {20}},\ \bibinfo {pages} {6556} (\bibinfo {year} {2020})}\BibitemShut {NoStop}%
    \bibitem [{\citenamefont {Ma}\ \emph {et~al.}(2015)\citenamefont {Ma}, \citenamefont {Zhou}, \citenamefont {Braun},\ and\ \citenamefont {Lew}}]{Ma2015b}%
      \BibitemOpen
      \bibfield  {author} {\bibinfo {author} {\bibfnamefont {F.}~\bibnamefont {Ma}}, \bibinfo {author} {\bibfnamefont {Y.}~\bibnamefont {Zhou}}, \bibinfo {author} {\bibfnamefont {H.~B.}\ \bibnamefont {Braun}},\ and\ \bibinfo {author} {\bibfnamefont {W.~S.}\ \bibnamefont {Lew}},\ }\bibfield  {title} {\bibinfo {title} {Skyrmion-based dynamic magnonic crystal},\ }\href {https://doi.org/10.1021/acs.nanolett.5b00996} {\bibfield  {journal} {\bibinfo  {journal} {Nano Lett.}\ }\textbf {\bibinfo {volume} {15}},\ \bibinfo {pages} {4029} (\bibinfo {year} {2015})}\BibitemShut {NoStop}%
    \bibitem [{\citenamefont {Kim}\ \emph {et~al.}(2018)\citenamefont {Kim}, \citenamefont {Yang}, \citenamefont {Cho}, \citenamefont {Kim},\ and\ \citenamefont {Kim}}]{Kim2018}%
      \BibitemOpen
      \bibfield  {author} {\bibinfo {author} {\bibfnamefont {J.}~\bibnamefont {Kim}}, \bibinfo {author} {\bibfnamefont {J.}~\bibnamefont {Yang}}, \bibinfo {author} {\bibfnamefont {Y.-J.}\ \bibnamefont {Cho}}, \bibinfo {author} {\bibfnamefont {B.}~\bibnamefont {Kim}},\ and\ \bibinfo {author} {\bibfnamefont {S.-K.}\ \bibnamefont {Kim}},\ }\bibfield  {title} {\bibinfo {title} {Coupled breathing modes in one-dimensional skyrmion lattices},\ }\href {https://doi.org/10.1063/1.5010948} {\bibfield  {journal} {\bibinfo  {journal} {J. Appl. Phys.}\ }\textbf {\bibinfo {volume} {123}},\ \bibinfo {pages} {053903} (\bibinfo {year} {2018})}\BibitemShut {NoStop}%
    \bibitem [{\citenamefont {Chen}\ and\ \citenamefont {Ma}(2021)}]{Chen2021}%
      \BibitemOpen
      \bibfield  {author} {\bibinfo {author} {\bibfnamefont {Z.}~\bibnamefont {Chen}}\ and\ \bibinfo {author} {\bibfnamefont {F.}~\bibnamefont {Ma}},\ }\bibfield  {title} {\bibinfo {title} {{Skyrmion based magnonic crystals}},\ }\href {https://doi.org/10.1063/5.0061832} {\bibfield  {journal} {\bibinfo  {journal} {J. Appl. Phys.}\ }\textbf {\bibinfo {volume} {130}},\ \bibinfo {pages} {090901} (\bibinfo {year} {2021})}\BibitemShut {NoStop}%
    \bibitem [{\citenamefont {Bassotti}\ \emph {et~al.}(2022)\citenamefont {Bassotti}, \citenamefont {Silvani},\ and\ \citenamefont {Carlotti}}]{Bassotti2022}%
      \BibitemOpen
      \bibfield  {author} {\bibinfo {author} {\bibfnamefont {M.}~\bibnamefont {Bassotti}}, \bibinfo {author} {\bibfnamefont {R.}~\bibnamefont {Silvani}},\ and\ \bibinfo {author} {\bibfnamefont {G.}~\bibnamefont {Carlotti}},\ }\bibfield  {title} {\bibinfo {title} {From the spin eigenmodes of isolated {N{\'e}el} skyrmions to the magnonic bands of skyrmionic crystals: {A} micromagnetic study as a function of the interfacial {Dzyaloshinskii-Moriya} interaction and the exchange constants},\ }\href {https://doi.org/10.1109/LMAG.2021.3136152} {\bibfield  {journal} {\bibinfo  {journal} {IEEE Magn. Lett.}\ }\textbf {\bibinfo {volume} {13}},\ \bibinfo {pages} {1} (\bibinfo {year} {2022})}\BibitemShut {NoStop}%
    \bibitem [{\citenamefont {Zhang}\ \emph {et~al.}(2019)\citenamefont {Zhang}, \citenamefont {Jin}, \citenamefont {Wang}, \citenamefont {Xia}, \citenamefont {Wang}, \citenamefont {Wang},\ and\ \citenamefont {Liu}}]{ZHANG2019}%
      \BibitemOpen
      \bibfield  {author} {\bibinfo {author} {\bibfnamefont {C.}~\bibnamefont {Zhang}}, \bibinfo {author} {\bibfnamefont {C.}~\bibnamefont {Jin}}, \bibinfo {author} {\bibfnamefont {J.}~\bibnamefont {Wang}}, \bibinfo {author} {\bibfnamefont {H.}~\bibnamefont {Xia}}, \bibinfo {author} {\bibfnamefont {J.}~\bibnamefont {Wang}}, \bibinfo {author} {\bibfnamefont {J.}~\bibnamefont {Wang}},\ and\ \bibinfo {author} {\bibfnamefont {Q.}~\bibnamefont {Liu}},\ }\bibfield  {title} {\bibinfo {title} {Directional spin-wave propagation in the skyrmion chain},\ }\href {https://doi.org/https://doi.org/10.1016/j.jmmm.2019.165542} {\bibfield  {journal} {\bibinfo  {journal} {J. Magn. Magn. Mater.}\ }\textbf {\bibinfo {volume} {490}},\ \bibinfo {pages} {165542} (\bibinfo {year} {2019})}\BibitemShut {NoStop}%
    \bibitem [{\citenamefont {Dhiman}\ \emph {et~al.}(2021)\citenamefont {Dhiman}, \citenamefont {Matczak}, \citenamefont {Gieniusz}, \citenamefont {Sveklo}, \citenamefont {Kurant}, \citenamefont {Guzowska}, \citenamefont {Stobiecki},\ and\ \citenamefont {Maziewski}}]{Dhiman2021}%
      \BibitemOpen
      \bibfield  {author} {\bibinfo {author} {\bibfnamefont {A.}~\bibnamefont {Dhiman}}, \bibinfo {author} {\bibfnamefont {M.}~\bibnamefont {Matczak}}, \bibinfo {author} {\bibfnamefont {R.}~\bibnamefont {Gieniusz}}, \bibinfo {author} {\bibfnamefont {I.}~\bibnamefont {Sveklo}}, \bibinfo {author} {\bibfnamefont {Z.}~\bibnamefont {Kurant}}, \bibinfo {author} {\bibfnamefont {U.}~\bibnamefont {Guzowska}}, \bibinfo {author} {\bibfnamefont {F.}~\bibnamefont {Stobiecki}},\ and\ \bibinfo {author} {\bibfnamefont {A.}~\bibnamefont {Maziewski}},\ }\bibfield  {title} {\bibinfo {title} {{Thickness dependence of interfacial Dzyaloshinskii-Moriya interaction, magnetic anisotropy and spin waves damping in Pt/Co/Ir and Ir/Co/Pt trilayers}},\ }\href {https://doi.org/https://doi.org/10.1016/j.jmmm.2020.167485} {\bibfield  {journal} {\bibinfo  {journal} {J. Magn. Magn. Mater.}\ }\textbf {\bibinfo {volume} {519}},\ \bibinfo {pages} {167485} (\bibinfo {year} {2021})}\BibitemShut {NoStop}%
    \bibitem [{\citenamefont {Azzawi}\ \emph {et~al.}(2017)\citenamefont {Azzawi}, \citenamefont {Hindmarch},\ and\ \citenamefont {Atkinson}}]{Azzawi2017}%
      \BibitemOpen
      \bibfield  {author} {\bibinfo {author} {\bibfnamefont {S.}~\bibnamefont {Azzawi}}, \bibinfo {author} {\bibfnamefont {A.~T.}\ \bibnamefont {Hindmarch}},\ and\ \bibinfo {author} {\bibfnamefont {D.}~\bibnamefont {Atkinson}},\ }\bibfield  {title} {\bibinfo {title} {Magnetic damping phenomena in ferromagnetic thin-films and multilayers},\ }\href {https://doi.org/10.1088/1361-6463/aa8dad} {\bibfield  {journal} {\bibinfo  {journal} {J. Phys. D: Appl. Phys.}\ }\textbf {\bibinfo {volume} {50}},\ \bibinfo {pages} {473001} (\bibinfo {year} {2017})}\BibitemShut {NoStop}%
    \bibitem [{\citenamefont {Tang}\ \emph {et~al.}(2023)\citenamefont {Tang}, \citenamefont {Liyanage}, \citenamefont {Montoya}, \citenamefont {Patel}, \citenamefont {Quigley}, \citenamefont {Grutter}, \citenamefont {Fitzsimmons}, \citenamefont {Sinha}, \citenamefont {Borchers}, \citenamefont {Fullerton}, \citenamefont {DeBeer-Schmitt},\ and\ \citenamefont {Gilbert}}]{Tang2023}%
      \BibitemOpen
      \bibfield  {author} {\bibinfo {author} {\bibfnamefont {N.}~\bibnamefont {Tang}}, \bibinfo {author} {\bibfnamefont {W.~L. N.~C.}\ \bibnamefont {Liyanage}}, \bibinfo {author} {\bibfnamefont {S.~A.}\ \bibnamefont {Montoya}}, \bibinfo {author} {\bibfnamefont {S.}~\bibnamefont {Patel}}, \bibinfo {author} {\bibfnamefont {L.~J.}\ \bibnamefont {Quigley}}, \bibinfo {author} {\bibfnamefont {A.~J.}\ \bibnamefont {Grutter}}, \bibinfo {author} {\bibfnamefont {M.~R.}\ \bibnamefont {Fitzsimmons}}, \bibinfo {author} {\bibfnamefont {S.}~\bibnamefont {Sinha}}, \bibinfo {author} {\bibfnamefont {J.~A.}\ \bibnamefont {Borchers}}, \bibinfo {author} {\bibfnamefont {E.~E.}\ \bibnamefont {Fullerton}}, \bibinfo {author} {\bibfnamefont {L.}~\bibnamefont {DeBeer-Schmitt}},\ and\ \bibinfo {author} {\bibfnamefont {D.~A.}\ \bibnamefont {Gilbert}},\ }\bibfield  {title} {\bibinfo {title} {Skyrmion-excited spin-wave fractal networks},\ }\href {https://doi.org/https://doi.org/10.1002/adma.202300416} {\bibfield  {journal} {\bibinfo  {journal}
      {Adv. Mater.}\ }\textbf {\bibinfo {volume} {35}},\ \bibinfo {pages} {2300416} (\bibinfo {year} {2023})}\BibitemShut {NoStop}%
    \bibitem [{\citenamefont {Jin}\ \emph {et~al.}(2022)\citenamefont {Jin}, \citenamefont {Li}, \citenamefont {Zhang}, \citenamefont {Wang}, \citenamefont {Wang}, \citenamefont {Lian}, \citenamefont {Gong},\ and\ \citenamefont {Shi}}]{Jin2022Spin-waveNanodots}%
      \BibitemOpen
      \bibfield  {author} {\bibinfo {author} {\bibfnamefont {C.}~\bibnamefont {Jin}}, \bibinfo {author} {\bibfnamefont {S.}~\bibnamefont {Li}}, \bibinfo {author} {\bibfnamefont {H.}~\bibnamefont {Zhang}}, \bibinfo {author} {\bibfnamefont {R.}~\bibnamefont {Wang}}, \bibinfo {author} {\bibfnamefont {J.}~\bibnamefont {Wang}}, \bibinfo {author} {\bibfnamefont {R.}~\bibnamefont {Lian}}, \bibinfo {author} {\bibfnamefont {P.}~\bibnamefont {Gong}},\ and\ \bibinfo {author} {\bibfnamefont {X.}~\bibnamefont {Shi}},\ }\bibfield  {title} {\bibinfo {title} {Spin-wave modes of elliptical skyrmions in magnetic nanodots},\ }\href {https://doi.org/10.1088/1367-2630/ac5df9} {\bibfield  {journal} {\bibinfo  {journal} {New J. Phys.}\ }\textbf {\bibinfo {volume} {24}},\ \bibinfo {pages} {043005} (\bibinfo {year} {2022})}\BibitemShut {NoStop}%
    \bibitem [{\citenamefont {Zelent}\ \emph {et~al.}(2022)\citenamefont {Zelent}, \citenamefont {Gruszecki}, \citenamefont {Moalic}, \citenamefont {Hellwig}, \citenamefont {Barman},\ and\ \citenamefont {Krawczyk}}]{Zelent2022SpinAnisotropy}%
      \BibitemOpen
      \bibfield  {author} {\bibinfo {author} {\bibfnamefont {M.}~\bibnamefont {Zelent}}, \bibinfo {author} {\bibfnamefont {P.}~\bibnamefont {Gruszecki}}, \bibinfo {author} {\bibfnamefont {M.}~\bibnamefont {Moalic}}, \bibinfo {author} {\bibfnamefont {O.}~\bibnamefont {Hellwig}}, \bibinfo {author} {\bibfnamefont {A.}~\bibnamefont {Barman}},\ and\ \bibinfo {author} {\bibfnamefont {M.}~\bibnamefont {Krawczyk}},\ }\bibfield  {title} {\bibinfo {title} {{Chapter One - Spin dynamics in patterned magnetic multilayers with perpendicular magnetic anisotropy}}\ }(\bibinfo  {publisher} {Academic Press},\ \bibinfo {year} {2022})\ pp.\ \bibinfo {pages} {1--51}\BibitemShut {NoStop}%
    \bibitem [{\citenamefont {Guslienko}\ and\ \citenamefont {Gareeva}(2017{\natexlab{a}})}]{guslienko2016gyrotropic}%
      \BibitemOpen
      \bibfield  {author} {\bibinfo {author} {\bibfnamefont {K.~Y.}\ \bibnamefont {Guslienko}}\ and\ \bibinfo {author} {\bibfnamefont {Z.~V.}\ \bibnamefont {Gareeva}},\ }\bibfield  {title} {\bibinfo {title} {Gyrotropic skyrmion modes in ultrathin magnetic circular dots},\ }\href {https://doi.org/10.1109/LMAG.2016.2616333} {\bibfield  {journal} {\bibinfo  {journal} {IEEE Magn. Lett.}\ }\textbf {\bibinfo {volume} {8}},\ \bibinfo {pages} {1} (\bibinfo {year} {2017}{\natexlab{a}})}\BibitemShut {NoStop}%
    \bibitem [{\citenamefont {Kim}\ \emph {et~al.}(2014)\citenamefont {Kim}, \citenamefont {Garcia-Sanchez}, \citenamefont {Sampaio}, \citenamefont {Moreau-Luchaire}, \citenamefont {Cros},\ and\ \citenamefont {Fert}}]{Kim2014BreathingDotsb}%
      \BibitemOpen
      \bibfield  {author} {\bibinfo {author} {\bibfnamefont {J.-V.}\ \bibnamefont {Kim}}, \bibinfo {author} {\bibfnamefont {F.}~\bibnamefont {Garcia-Sanchez}}, \bibinfo {author} {\bibfnamefont {J.}~\bibnamefont {Sampaio}}, \bibinfo {author} {\bibfnamefont {C.}~\bibnamefont {Moreau-Luchaire}}, \bibinfo {author} {\bibfnamefont {V.}~\bibnamefont {Cros}},\ and\ \bibinfo {author} {\bibfnamefont {A.}~\bibnamefont {Fert}},\ }\bibfield  {title} {\bibinfo {title} {Breathing modes of confined skyrmions in ultrathin magnetic dots},\ }\href {https://doi.org/10.1103/PhysRevB.90.064410} {\bibfield  {journal} {\bibinfo  {journal} {Phys. Rev. B}\ }\textbf {\bibinfo {volume} {90}},\ \bibinfo {pages} {064410} (\bibinfo {year} {2014})}\BibitemShut {NoStop}%
    \bibitem [{\citenamefont {Mruczkiewicz}\ \emph {et~al.}(2016)\citenamefont {Mruczkiewicz}, \citenamefont {Gruszecki}, \citenamefont {Zelent},\ and\ \citenamefont {Krawczyk}}]{mruczkiewicz2016collective}%
      \BibitemOpen
      \bibfield  {author} {\bibinfo {author} {\bibfnamefont {M.}~\bibnamefont {Mruczkiewicz}}, \bibinfo {author} {\bibfnamefont {P.}~\bibnamefont {Gruszecki}}, \bibinfo {author} {\bibfnamefont {M.}~\bibnamefont {Zelent}},\ and\ \bibinfo {author} {\bibfnamefont {M.}~\bibnamefont {Krawczyk}},\ }\bibfield  {title} {\bibinfo {title} {Collective dynamical skyrmion excitations in a magnonic crystal},\ }\href {https://doi.org/10.1103/PhysRevB.93.174429} {\bibfield  {journal} {\bibinfo  {journal} {Phys. Rev. B}\ }\textbf {\bibinfo {volume} {93}},\ \bibinfo {pages} {174429} (\bibinfo {year} {2016})}\BibitemShut {NoStop}%
    \bibitem [{\citenamefont {Garst}\ \emph {et~al.}(2017)\citenamefont {Garst}, \citenamefont {Waizner},\ and\ \citenamefont {Grundler}}]{Garst2017CollectiveMagnets}%
      \BibitemOpen
      \bibfield  {author} {\bibinfo {author} {\bibfnamefont {M.}~\bibnamefont {Garst}}, \bibinfo {author} {\bibfnamefont {J.}~\bibnamefont {Waizner}},\ and\ \bibinfo {author} {\bibfnamefont {D.}~\bibnamefont {Grundler}},\ }\bibfield  {title} {\bibinfo {title} {Collective spin excitations of helices and magnetic skyrmions: review and perspectives of magnonics in non-centrosymmetric magnets},\ }\href {https://doi.org/10.1088/1361-6463/aa7573} {\bibfield  {journal} {\bibinfo  {journal} {J. Phys. D: Appl. Phys.}\ }\textbf {\bibinfo {volume} {50}},\ \bibinfo {pages} {293002} (\bibinfo {year} {2017})}\BibitemShut {NoStop}%
    \bibitem [{\citenamefont {Mruczkiewicz}\ \emph {et~al.}(2018)\citenamefont {Mruczkiewicz}, \citenamefont {Gruszecki}, \citenamefont {Krawczyk},\ and\ \citenamefont {Guslienko}}]{Mruczkiewicz2018}%
      \BibitemOpen
      \bibfield  {author} {\bibinfo {author} {\bibfnamefont {M.}~\bibnamefont {Mruczkiewicz}}, \bibinfo {author} {\bibfnamefont {P.}~\bibnamefont {Gruszecki}}, \bibinfo {author} {\bibfnamefont {M.}~\bibnamefont {Krawczyk}},\ and\ \bibinfo {author} {\bibfnamefont {K.~Y.}\ \bibnamefont {Guslienko}},\ }\bibfield  {title} {\bibinfo {title} {{Azimuthal spin-wave excitations in magnetic nanodots over the soliton background: Vortex, Bloch, and N\'eel-like skyrmions}},\ }\href {https://doi.org/10.1103/PhysRevB.97.064418} {\bibfield  {journal} {\bibinfo  {journal} {Phys. Rev. B}\ }\textbf {\bibinfo {volume} {97}},\ \bibinfo {pages} {064418} (\bibinfo {year} {2018})}\BibitemShut {NoStop}%
    \bibitem [{\citenamefont {Wang}\ \emph {et~al.}(2020)\citenamefont {Wang}, \citenamefont {Nie}, \citenamefont {Xia},\ and\ \citenamefont {Guo}}]{Wang2020b}%
      \BibitemOpen
      \bibfield  {author} {\bibinfo {author} {\bibfnamefont {X.-g.}\ \bibnamefont {Wang}}, \bibinfo {author} {\bibfnamefont {Y.-Z.}\ \bibnamefont {Nie}}, \bibinfo {author} {\bibfnamefont {Q.-l.}\ \bibnamefont {Xia}},\ and\ \bibinfo {author} {\bibfnamefont {G.-h.}\ \bibnamefont {Guo}},\ }\bibfield  {title} {\bibinfo {title} {{Dynamically reconfigurable magnonic crystal composed of artificial magnetic skyrmion lattice}},\ }\href {https://doi.org/10.1063/5.0012791} {\bibfield  {journal} {\bibinfo  {journal} {J. Appl. Phys.}\ }\textbf {\bibinfo {volume} {128}},\ \bibinfo {pages} {063901} (\bibinfo {year} {2020})}\BibitemShut {NoStop}%
    \bibitem [{\citenamefont {Golovchanskiy}\ \emph {et~al.}(2019)\citenamefont {Golovchanskiy}, \citenamefont {Abramov}, \citenamefont {Stolyarov}, \citenamefont {Dzhumaev}, \citenamefont {Emelyanova}, \citenamefont {Golubov}, \citenamefont {Ryazanov},\ and\ \citenamefont {Ustinov}}]{Golovchanskiy2019}%
      \BibitemOpen
      \bibfield  {author} {\bibinfo {author} {\bibfnamefont {I.~A.}\ \bibnamefont {Golovchanskiy}}, \bibinfo {author} {\bibfnamefont {N.~N.}\ \bibnamefont {Abramov}}, \bibinfo {author} {\bibfnamefont {V.~S.}\ \bibnamefont {Stolyarov}}, \bibinfo {author} {\bibfnamefont {P.~S.}\ \bibnamefont {Dzhumaev}}, \bibinfo {author} {\bibfnamefont {O.~V.}\ \bibnamefont {Emelyanova}}, \bibinfo {author} {\bibfnamefont {A.~A.}\ \bibnamefont {Golubov}}, \bibinfo {author} {\bibfnamefont {V.~V.}\ \bibnamefont {Ryazanov}},\ and\ \bibinfo {author} {\bibfnamefont {A.~V.}\ \bibnamefont {Ustinov}},\ }\bibfield  {title} {\bibinfo {title} {Ferromagnet/superconductor hybrid magnonic metamaterials},\ }\href {https://doi.org/https://doi.org/10.1002/advs.201900435} {\bibfield  {journal} {\bibinfo  {journal} {Adv. Sci.}\ }\textbf {\bibinfo {volume} {6}},\ \bibinfo {pages} {1900435} (\bibinfo {year} {2019})}\BibitemShut {NoStop}%
    \bibitem [{\citenamefont {Petrovi{\'{c}}}\ \emph {et~al.}(2021)\citenamefont {Petrovi{\'{c}}}, \citenamefont {Raju}, \citenamefont {Tee}, \citenamefont {Louat}, \citenamefont {Maggio-Aprile}, \citenamefont {Menezes}, \citenamefont {Wyszy{\'n}ski}, \citenamefont {Duong}, \citenamefont {Reznikov}, \citenamefont {Renner}, \citenamefont {Milosevi{\'c}},\ and\ \citenamefont {Panagopoulos}}]{Petrovic2021}%
      \BibitemOpen
      \bibfield  {author} {\bibinfo {author} {\bibfnamefont {A.~P.}\ \bibnamefont {Petrovi{\'{c}}}}, \bibinfo {author} {\bibfnamefont {M.}~\bibnamefont {Raju}}, \bibinfo {author} {\bibfnamefont {X.~Y.}\ \bibnamefont {Tee}}, \bibinfo {author} {\bibfnamefont {A.}~\bibnamefont {Louat}}, \bibinfo {author} {\bibfnamefont {I.}~\bibnamefont {Maggio-Aprile}}, \bibinfo {author} {\bibfnamefont {R.~M.}\ \bibnamefont {Menezes}}, \bibinfo {author} {\bibfnamefont {M.~J.}\ \bibnamefont {Wyszy{\'n}ski}}, \bibinfo {author} {\bibfnamefont {N.~K.}\ \bibnamefont {Duong}}, \bibinfo {author} {\bibfnamefont {M.}~\bibnamefont {Reznikov}}, \bibinfo {author} {\bibfnamefont {C.}~\bibnamefont {Renner}}, \bibinfo {author} {\bibfnamefont {M.~V.}\ \bibnamefont {Milosevi{\'c}}},\ and\ \bibinfo {author} {\bibfnamefont {C.}~\bibnamefont {Panagopoulos}},\ }\bibfield  {title} {\bibinfo {title} {Skyrmion-(anti)vortex coupling in a chiral magnet-superconductor heterostructure},\ }\href {https://doi.org/10.1103/PhysRevLett.126.117205} {\bibfield
      {journal} {\bibinfo  {journal} {Phys. Rev. Lett.}\ }\textbf {\bibinfo {volume} {126}},\ \bibinfo {pages} {117205} (\bibinfo {year} {2021})}\BibitemShut {NoStop}%
    \bibitem [{\citenamefont {Negrello}\ \emph {et~al.}(2022)\citenamefont {Negrello}, \citenamefont {Montoncello}, \citenamefont {Kaffash}, \citenamefont {Jungfleisch},\ and\ \citenamefont {Gubbiotti}}]{Negrello2022}%
      \BibitemOpen
      \bibfield  {author} {\bibinfo {author} {\bibfnamefont {R.}~\bibnamefont {Negrello}}, \bibinfo {author} {\bibfnamefont {F.}~\bibnamefont {Montoncello}}, \bibinfo {author} {\bibfnamefont {M.~T.}\ \bibnamefont {Kaffash}}, \bibinfo {author} {\bibfnamefont {M.~B.}\ \bibnamefont {Jungfleisch}},\ and\ \bibinfo {author} {\bibfnamefont {G.}~\bibnamefont {Gubbiotti}},\ }\bibfield  {title} {\bibinfo {title} {{Dynamic coupling and spin-wave dispersions in a magnetic hybrid system made of an artificial spin-ice structure and an extended NiFe underlayer}},\ }\href {https://doi.org/10.1063/5.0102571} {\bibfield  {journal} {\bibinfo  {journal} {APL Mater.}\ }\textbf {\bibinfo {volume} {10}},\ \bibinfo {pages} {091115} (\bibinfo {year} {2022})}\BibitemShut {NoStop}%
    \bibitem [{\citenamefont {Pan}\ \emph {et~al.}(2024)\citenamefont {Pan}, \citenamefont {Li}, \citenamefont {Hei}, \citenamefont {Zhang}, \citenamefont {Mochizuki}, \citenamefont {Li},\ and\ \citenamefont {Nori}}]{Pan2024}%
      \BibitemOpen
      \bibfield  {author} {\bibinfo {author} {\bibfnamefont {X.-F.}\ \bibnamefont {Pan}}, \bibinfo {author} {\bibfnamefont {P.-B.}\ \bibnamefont {Li}}, \bibinfo {author} {\bibfnamefont {X.-L.}\ \bibnamefont {Hei}}, \bibinfo {author} {\bibfnamefont {X.}~\bibnamefont {Zhang}}, \bibinfo {author} {\bibfnamefont {M.}~\bibnamefont {Mochizuki}}, \bibinfo {author} {\bibfnamefont {F.-L.}\ \bibnamefont {Li}},\ and\ \bibinfo {author} {\bibfnamefont {F.}~\bibnamefont {Nori}},\ }\bibfield  {title} {\bibinfo {title} {Magnon-skyrmion hybrid quantum systems: Tailoring interactions via magnons},\ }\href@noop {} {\bibfield  {journal} {\bibinfo  {journal} {Phys. Rev. Lett.}\ }\textbf {\bibinfo {volume} {132}},\ \bibinfo {pages} {193601} (\bibinfo {year} {2024})}\BibitemShut {NoStop}%
    \bibitem [{\citenamefont {Chen}\ \emph {et~al.}(2021)\citenamefont {Chen}, \citenamefont {Hu},\ and\ \citenamefont {Yu}}]{Chen2021ChiralSkyrmions}%
      \BibitemOpen
      \bibfield  {author} {\bibinfo {author} {\bibfnamefont {J.}~\bibnamefont {Chen}}, \bibinfo {author} {\bibfnamefont {J.}~\bibnamefont {Hu}},\ and\ \bibinfo {author} {\bibfnamefont {H.}~\bibnamefont {Yu}},\ }\bibfield  {title} {\bibinfo {title} {{Chiral emission of exchange spin waves by magnetic skyrmions}},\ }\href {https://doi.org/10.1021/acsnano.0c07805} {\bibfield  {journal} {\bibinfo  {journal} {ACS Nano}\ }\textbf {\bibinfo {volume} {15}},\ \bibinfo {pages} {4372} (\bibinfo {year} {2021})}\BibitemShut {NoStop}%
    \bibitem [{\citenamefont {Zelent}\ \emph {et~al.}(2023)\citenamefont {Zelent}, \citenamefont {Moalic}, \citenamefont {Mruczkiewicz}, \citenamefont {Li}, \citenamefont {Zhou},\ and\ \citenamefont {Krawczyk}}]{Zelent2023}%
      \BibitemOpen
      \bibfield  {author} {\bibinfo {author} {\bibfnamefont {M.}~\bibnamefont {Zelent}}, \bibinfo {author} {\bibfnamefont {M.}~\bibnamefont {Moalic}}, \bibinfo {author} {\bibfnamefont {M.}~\bibnamefont {Mruczkiewicz}}, \bibinfo {author} {\bibfnamefont {X.}~\bibnamefont {Li}}, \bibinfo {author} {\bibfnamefont {Y.}~\bibnamefont {Zhou}},\ and\ \bibinfo {author} {\bibfnamefont {M.}~\bibnamefont {Krawczyk}},\ }\bibfield  {title} {\bibinfo {title} {{Stabilization and racetrack application of asymmetric N{\'e}el skyrmions in hybrid nanostructures}},\ }\href@noop {} {\bibfield  {journal} {\bibinfo  {journal} {Sci. Rep.}\ }\textbf {\bibinfo {volume} {13}},\ \bibinfo {pages} {13572} (\bibinfo {year} {2023})}\BibitemShut {NoStop}%
    \bibitem [{\citenamefont {Hsu}\ \emph {et~al.}(2016)\citenamefont {Hsu}, \citenamefont {Zhen}, \citenamefont {Stone}, \citenamefont {Joannopoulos},\ and\ \citenamefont {Soljacic}}]{Hsu2016}%
      \BibitemOpen
      \bibfield  {author} {\bibinfo {author} {\bibfnamefont {C.~W.}\ \bibnamefont {Hsu}}, \bibinfo {author} {\bibfnamefont {B.}~\bibnamefont {Zhen}}, \bibinfo {author} {\bibfnamefont {A.~D.}\ \bibnamefont {Stone}}, \bibinfo {author} {\bibfnamefont {J.~D.}\ \bibnamefont {Joannopoulos}},\ and\ \bibinfo {author} {\bibfnamefont {M.}~\bibnamefont {Soljacic}},\ }\bibfield  {title} {\bibinfo {title} {Bound states in the continuum},\ }\href {https://doi.org/10.1038/natrevmats.2016.48} {\bibfield  {journal} {\bibinfo  {journal} {Nat. Rev. Mater.}\ }\textbf {\bibinfo {volume} {1}},\ \bibinfo {pages} {16048} (\bibinfo {year} {2016})}\BibitemShut {NoStop}%
    \bibitem [{\citenamefont {Gallardo}\ \emph {et~al.}(2019)\citenamefont {Gallardo}, \citenamefont {Cort\'es-Ortu\~no}, \citenamefont {Schneider}, \citenamefont {Rold\'an-Molina}, \citenamefont {Ma}, \citenamefont {Troncoso}, \citenamefont {Lenz}, \citenamefont {Fangohr}, \citenamefont {Lindner},\ and\ \citenamefont {Landeros}}]{Gallardo2019}%
      \BibitemOpen
      \bibfield  {author} {\bibinfo {author} {\bibfnamefont {R.~A.}\ \bibnamefont {Gallardo}}, \bibinfo {author} {\bibfnamefont {D.}~\bibnamefont {Cort\'es-Ortu\~no}}, \bibinfo {author} {\bibfnamefont {T.}~\bibnamefont {Schneider}}, \bibinfo {author} {\bibfnamefont {A.}~\bibnamefont {Rold\'an-Molina}}, \bibinfo {author} {\bibfnamefont {F.}~\bibnamefont {Ma}}, \bibinfo {author} {\bibfnamefont {R.~E.}\ \bibnamefont {Troncoso}}, \bibinfo {author} {\bibfnamefont {K.}~\bibnamefont {Lenz}}, \bibinfo {author} {\bibfnamefont {H.}~\bibnamefont {Fangohr}}, \bibinfo {author} {\bibfnamefont {J.}~\bibnamefont {Lindner}},\ and\ \bibinfo {author} {\bibfnamefont {P.}~\bibnamefont {Landeros}},\ }\bibfield  {title} {\bibinfo {title} {Flat bands, indirect gaps, and unconventional spin-wave behavior induced by a periodic {Dzyaloshinskii-Moriya} interaction},\ }\href {https://doi.org/10.1103/PhysRevLett.122.067204} {\bibfield  {journal} {\bibinfo  {journal} {Phys. Rev. Lett.}\ }\textbf {\bibinfo {volume} {122}},\ \bibinfo {pages}
      {067204} (\bibinfo {year} {2019})}\BibitemShut {NoStop}%
    \bibitem [{\citenamefont {Awschalom}\ \emph {et~al.}(2021)\citenamefont {Awschalom}, \citenamefont {Du}, \citenamefont {He}, \citenamefont {Heremans}, \citenamefont {Hoffmann}, \citenamefont {Hou}, \citenamefont {Kurebayashi}, \citenamefont {Li}, \citenamefont {Liu}, \citenamefont {Novosad}, \citenamefont {Sklenar}, \citenamefont {Sullivan}, \citenamefont {Sun}, \citenamefont {Tang}, \citenamefont {Tyberkevych}, \citenamefont {Trevillian}, \citenamefont {Tsen}, \citenamefont {Weiss}, \citenamefont {Zhang}, \citenamefont {Zhang}, \citenamefont {Zhao},\ and\ \citenamefont {Zollitsch}}]{Awschalom2021}%
      \BibitemOpen
      \bibfield  {author} {\bibinfo {author} {\bibfnamefont {D.~D.}\ \bibnamefont {Awschalom}}, \bibinfo {author} {\bibfnamefont {C.~R.}\ \bibnamefont {Du}}, \bibinfo {author} {\bibfnamefont {R.}~\bibnamefont {He}}, \bibinfo {author} {\bibfnamefont {F.~J.}\ \bibnamefont {Heremans}}, \bibinfo {author} {\bibfnamefont {A.}~\bibnamefont {Hoffmann}}, \bibinfo {author} {\bibfnamefont {J.}~\bibnamefont {Hou}}, \bibinfo {author} {\bibfnamefont {H.}~\bibnamefont {Kurebayashi}}, \bibinfo {author} {\bibfnamefont {Y.}~\bibnamefont {Li}}, \bibinfo {author} {\bibfnamefont {L.}~\bibnamefont {Liu}}, \bibinfo {author} {\bibfnamefont {V.}~\bibnamefont {Novosad}}, \bibinfo {author} {\bibfnamefont {J.}~\bibnamefont {Sklenar}}, \bibinfo {author} {\bibfnamefont {S.~E.}\ \bibnamefont {Sullivan}}, \bibinfo {author} {\bibfnamefont {D.}~\bibnamefont {Sun}}, \bibinfo {author} {\bibfnamefont {H.}~\bibnamefont {Tang}}, \bibinfo {author} {\bibfnamefont {V.}~\bibnamefont {Tyberkevych}}, \bibinfo {author} {\bibfnamefont {C.}~\bibnamefont
      {Trevillian}}, \bibinfo {author} {\bibfnamefont {A.~W.}\ \bibnamefont {Tsen}}, \bibinfo {author} {\bibfnamefont {L.~R.}\ \bibnamefont {Weiss}}, \bibinfo {author} {\bibfnamefont {W.}~\bibnamefont {Zhang}}, \bibinfo {author} {\bibfnamefont {X.}~\bibnamefont {Zhang}}, \bibinfo {author} {\bibfnamefont {L.}~\bibnamefont {Zhao}},\ and\ \bibinfo {author} {\bibfnamefont {C.~W.}\ \bibnamefont {Zollitsch}},\ }\bibfield  {title} {\bibinfo {title} {Quantum engineering with hybrid magnonic systems and materials},\ }\href {https://doi.org/10.1109/TQE.2021.3057799} {\bibfield  {journal} {\bibinfo  {journal} {IEEE Trans. Quantum Eng.}\ }\textbf {\bibinfo {volume} {2}},\ \bibinfo {pages} {1} (\bibinfo {year} {2021})}\BibitemShut {NoStop}%
    \bibitem [{\citenamefont {Li}\ \emph {et~al.}(2020)\citenamefont {Li}, \citenamefont {Zhang}, \citenamefont {Tyberkevych}, \citenamefont {Kwok}, \citenamefont {Hoffmann},\ and\ \citenamefont {Novosad}}]{Li2020}%
      \BibitemOpen
      \bibfield  {author} {\bibinfo {author} {\bibfnamefont {Y.}~\bibnamefont {Li}}, \bibinfo {author} {\bibfnamefont {W.}~\bibnamefont {Zhang}}, \bibinfo {author} {\bibfnamefont {V.}~\bibnamefont {Tyberkevych}}, \bibinfo {author} {\bibfnamefont {W.-K.}\ \bibnamefont {Kwok}}, \bibinfo {author} {\bibfnamefont {A.}~\bibnamefont {Hoffmann}},\ and\ \bibinfo {author} {\bibfnamefont {V.}~\bibnamefont {Novosad}},\ }\bibfield  {title} {\bibinfo {title} {{Hybrid magnonics: Physics, circuits, and applications for coherent information processing}},\ }\href {https://doi.org/10.1063/5.0020277} {\bibfield  {journal} {\bibinfo  {journal} {J. Appl. Phys.}\ }\textbf {\bibinfo {volume} {128}},\ \bibinfo {pages} {130902} (\bibinfo {year} {2020})}\BibitemShut {NoStop}%
    \bibitem [{\citenamefont {Papp}\ \emph {et~al.}(2021)\citenamefont {Papp}, \citenamefont {Porod},\ and\ \citenamefont {Csaba}}]{Papp2021}%
      \BibitemOpen
      \bibfield  {author} {\bibinfo {author} {\bibfnamefont {{\'A}.}~\bibnamefont {Papp}}, \bibinfo {author} {\bibfnamefont {W.}~\bibnamefont {Porod}},\ and\ \bibinfo {author} {\bibfnamefont {G.}~\bibnamefont {Csaba}},\ }\bibfield  {title} {\bibinfo {title} {Nanoscale neural network using non-linear spin-wave interference},\ }\href {https://doi.org/10.1038/s41467-021-26711-z} {\bibfield  {journal} {\bibinfo  {journal} {Nat. Commun.}\ }\textbf {\bibinfo {volume} {12}},\ \bibinfo {pages} {6422} (\bibinfo {year} {2021})}\BibitemShut {NoStop}%
    \bibitem [{\citenamefont {Fripp}\ \emph {et~al.}(2023)\citenamefont {Fripp}, \citenamefont {Au}, \citenamefont {Shytov},\ and\ \citenamefont {Kruglyak}}]{Fripp2023NonlinearNeurons}%
      \BibitemOpen
      \bibfield  {author} {\bibinfo {author} {\bibfnamefont {K.~G.}\ \bibnamefont {Fripp}}, \bibinfo {author} {\bibfnamefont {Y.}~\bibnamefont {Au}}, \bibinfo {author} {\bibfnamefont {A.~V.}\ \bibnamefont {Shytov}},\ and\ \bibinfo {author} {\bibfnamefont {V.~V.}\ \bibnamefont {Kruglyak}},\ }\bibfield  {title} {\bibinfo {title} {{{Nonlinear chiral magnonic resonators: Toward magnonic neurons}}},\ }\href {https://doi.org/10.1063/5.0149466} {\bibfield  {journal} {\bibinfo  {journal} {Appl. Phys. Lett.}\ }\textbf {\bibinfo {volume} {122}},\ \bibinfo {pages} {172403} (\bibinfo {year} {2023})}\BibitemShut {NoStop}%
    \bibitem [{\citenamefont {Szulc}\ \emph {et~al.}(2022)\citenamefont {Szulc}, \citenamefont {Tacchi}, \citenamefont {Hierro-Rodríguez}, \citenamefont {Díaz}, \citenamefont {Gruszecki}, \citenamefont {Graczyk}, \citenamefont {Quirós}, \citenamefont {Markó}, \citenamefont {Martín}, \citenamefont {Vélez}, \citenamefont {Schmool}, \citenamefont {Carlotti}, \citenamefont {Krawczyk},\ and\ \citenamefont {Álvarez Prado}}]{Szulc2022}%
      \BibitemOpen
      \bibfield  {author} {\bibinfo {author} {\bibfnamefont {K.}~\bibnamefont {Szulc}}, \bibinfo {author} {\bibfnamefont {S.}~\bibnamefont {Tacchi}}, \bibinfo {author} {\bibfnamefont {A.}~\bibnamefont {Hierro-Rodríguez}}, \bibinfo {author} {\bibfnamefont {J.}~\bibnamefont {Díaz}}, \bibinfo {author} {\bibfnamefont {P.}~\bibnamefont {Gruszecki}}, \bibinfo {author} {\bibfnamefont {P.}~\bibnamefont {Graczyk}}, \bibinfo {author} {\bibfnamefont {C.}~\bibnamefont {Quirós}}, \bibinfo {author} {\bibfnamefont {D.}~\bibnamefont {Markó}}, \bibinfo {author} {\bibfnamefont {J.~I.}\ \bibnamefont {Martín}}, \bibinfo {author} {\bibfnamefont {M.}~\bibnamefont {Vélez}}, \bibinfo {author} {\bibfnamefont {D.~S.}\ \bibnamefont {Schmool}}, \bibinfo {author} {\bibfnamefont {G.}~\bibnamefont {Carlotti}}, \bibinfo {author} {\bibfnamefont {M.}~\bibnamefont {Krawczyk}},\ and\ \bibinfo {author} {\bibfnamefont {L.~M.}\ \bibnamefont {Álvarez Prado}},\ }\bibfield  {title} {\bibinfo {title} {Reconfigurable magnonic crystals based on
      imprinted magnetization textures in hard and soft dipolar-coupled bilayers},\ }\href {https://doi.org/10.1021/acsnano.2c04256} {\bibfield  {journal} {\bibinfo  {journal} {ACS Nano}\ }\textbf {\bibinfo {volume} {16}},\ \bibinfo {pages} {14168} (\bibinfo {year} {2022})}\BibitemShut {NoStop}%
    \bibitem [{\citenamefont {Guslienko}\ and\ \citenamefont {Gareeva}(2017{\natexlab{b}})}]{Guslienko2017}%
      \BibitemOpen
      \bibfield  {author} {\bibinfo {author} {\bibfnamefont {K.}~\bibnamefont {Guslienko}}\ and\ \bibinfo {author} {\bibfnamefont {Z.}~\bibnamefont {Gareeva}},\ }\bibfield  {title} {\bibinfo {title} {Magnetic skyrmion low frequency dynamics in thin circular dots},\ }\href {https://doi.org/https://doi.org/10.1016/j.jmmm.2017.06.094} {\bibfield  {journal} {\bibinfo  {journal} {J. Magn. Magn. Mater.}\ }\textbf {\bibinfo {volume} {442}},\ \bibinfo {pages} {176} (\bibinfo {year} {2017}{\natexlab{b}})}\BibitemShut {NoStop}%
    \bibitem [{\citenamefont {Zelent}\ \emph {et~al.}(2017)\citenamefont {Zelent}, \citenamefont {Tóbik}, \citenamefont {Krawczyk}, \citenamefont {Guslienko},\ and\ \citenamefont {Mruczkiewicz}}]{PSSR:PSSR201700259}%
      \BibitemOpen
      \bibfield  {author} {\bibinfo {author} {\bibfnamefont {M.}~\bibnamefont {Zelent}}, \bibinfo {author} {\bibfnamefont {J.}~\bibnamefont {Tóbik}}, \bibinfo {author} {\bibfnamefont {M.}~\bibnamefont {Krawczyk}}, \bibinfo {author} {\bibfnamefont {K.~Y.}\ \bibnamefont {Guslienko}},\ and\ \bibinfo {author} {\bibfnamefont {M.}~\bibnamefont {Mruczkiewicz}},\ }\bibfield  {title} {\bibinfo {title} {Bi-stability of magnetic skyrmions in ultrathin multilayer nanodots induced by magnetostatic interaction},\ }\href {https://doi.org/https://doi.org/10.1002/pssr.201700259} {\bibfield  {journal} {\bibinfo  {journal} {Phys. Status Solidi RRL}\ }\textbf {\bibinfo {volume} {11}},\ \bibinfo {pages} {1700259} (\bibinfo {year} {2017})}\BibitemShut {NoStop}%
    \bibitem [{\citenamefont {Lemesh}\ and\ \citenamefont {Beach}(2018)}]{lemesh2018TwistedMultilayers}%
      \BibitemOpen
      \bibfield  {author} {\bibinfo {author} {\bibfnamefont {I.}~\bibnamefont {Lemesh}}\ and\ \bibinfo {author} {\bibfnamefont {G.~S.~D.}\ \bibnamefont {Beach}},\ }\bibfield  {title} {\bibinfo {title} {Twisted domain walls and skyrmions in perpendicularly magnetized multilayers},\ }\href {https://doi.org/10.1103/PhysRevB.98.104402} {\bibfield  {journal} {\bibinfo  {journal} {Phys. Rev. B}\ }\textbf {\bibinfo {volume} {98}},\ \bibinfo {pages} {104402} (\bibinfo {year} {2018})}\BibitemShut {NoStop}%
    \bibitem [{\citenamefont {Woo}\ \emph {et~al.}(2016)\citenamefont {Woo}, \citenamefont {Litzius}, \citenamefont {Kr{\"{u}}ger}, \citenamefont {Im}, \citenamefont {Caretta}, \citenamefont {Richter}, \citenamefont {Mann}, \citenamefont {Krone}, \citenamefont {Reeve}, \citenamefont {Weigand}, \citenamefont {Agrawal}, \citenamefont {Lemesh}, \citenamefont {Mawass}, \citenamefont {Fischer}, \citenamefont {Kl{\"{a}}ui},\ and\ \citenamefont {Beach}}]{woo2016observation}%
      \BibitemOpen
      \bibfield  {author} {\bibinfo {author} {\bibfnamefont {S.}~\bibnamefont {Woo}}, \bibinfo {author} {\bibfnamefont {K.}~\bibnamefont {Litzius}}, \bibinfo {author} {\bibfnamefont {B.}~\bibnamefont {Kr{\"{u}}ger}}, \bibinfo {author} {\bibfnamefont {M.-Y.}\ \bibnamefont {Im}}, \bibinfo {author} {\bibfnamefont {L.}~\bibnamefont {Caretta}}, \bibinfo {author} {\bibfnamefont {K.}~\bibnamefont {Richter}}, \bibinfo {author} {\bibfnamefont {M.}~\bibnamefont {Mann}}, \bibinfo {author} {\bibfnamefont {A.}~\bibnamefont {Krone}}, \bibinfo {author} {\bibfnamefont {R.~M.}\ \bibnamefont {Reeve}}, \bibinfo {author} {\bibfnamefont {M.}~\bibnamefont {Weigand}}, \bibinfo {author} {\bibfnamefont {P.}~\bibnamefont {Agrawal}}, \bibinfo {author} {\bibfnamefont {I.}~\bibnamefont {Lemesh}}, \bibinfo {author} {\bibfnamefont {M.-A.}\ \bibnamefont {Mawass}}, \bibinfo {author} {\bibfnamefont {P.}~\bibnamefont {Fischer}}, \bibinfo {author} {\bibfnamefont {M.}~\bibnamefont {Kl{\"{a}}ui}},\ and\ \bibinfo {author} {\bibfnamefont {G.~S.~D.}\
      \bibnamefont {Beach}},\ }\bibfield  {title} {\bibinfo {title} {{Observation of room-temperature magnetic skyrmions and their current-driven dynamics in ultrathin metallic ferromagnets}},\ }\href {https://doi.org/10.1038/nmat4593} {\bibfield  {journal} {\bibinfo  {journal} {Nat. Mater.}\ }\textbf {\bibinfo {volume} {15}},\ \bibinfo {pages} {501} (\bibinfo {year} {2016})}\BibitemShut {NoStop}%
    \bibitem [{\citenamefont {Suna}(1986)}]{Suna1986PerpendicularFilm}%
      \BibitemOpen
      \bibfield  {author} {\bibinfo {author} {\bibfnamefont {A.}~\bibnamefont {Suna}},\ }\bibfield  {title} {\bibinfo {title} {{Perpendicular magnetic ground state of a multilayer film}},\ }\href {https://doi.org/10.1063/1.336684} {\bibfield  {journal} {\bibinfo  {journal} {J. Appl. Phys.}\ }\textbf {\bibinfo {volume} {59}},\ \bibinfo {pages} {313} (\bibinfo {year} {1986})}\BibitemShut {NoStop}%
    \bibitem [{\citenamefont {Vetrova}\ \emph {et~al.}(2021)\citenamefont {Vetrova}, \citenamefont {Zelent}, \citenamefont {{\v{S}}olt{\'{y}}s}, \citenamefont {Gubanov}, \citenamefont {Sadovnikov}, \citenamefont {{\v{S}}cepka}, \citenamefont {D{\'{e}}rer}, \citenamefont {Stoklas}, \citenamefont {Cambel},\ and\ \citenamefont {Mruczkiewicz}}]{Vetrova2021InvestigationDot}%
      \BibitemOpen
      \bibfield  {author} {\bibinfo {author} {\bibfnamefont {I.~V.}\ \bibnamefont {Vetrova}}, \bibinfo {author} {\bibfnamefont {M.}~\bibnamefont {Zelent}}, \bibinfo {author} {\bibfnamefont {J.}~\bibnamefont {{\v{S}}olt{\'{y}}s}}, \bibinfo {author} {\bibfnamefont {V.~A.}\ \bibnamefont {Gubanov}}, \bibinfo {author} {\bibfnamefont {A.~V.}\ \bibnamefont {Sadovnikov}}, \bibinfo {author} {\bibfnamefont {T.}~\bibnamefont {{\v{S}}cepka}}, \bibinfo {author} {\bibfnamefont {J.}~\bibnamefont {D{\'{e}}rer}}, \bibinfo {author} {\bibfnamefont {R.}~\bibnamefont {Stoklas}}, \bibinfo {author} {\bibfnamefont {V.}~\bibnamefont {Cambel}},\ and\ \bibinfo {author} {\bibfnamefont {M.}~\bibnamefont {Mruczkiewicz}},\ }\bibfield  {title} {\bibinfo {title} {{Investigation of self-nucleated skyrmion states in the ferromagnetic/nonmagnetic multilayer dot}},\ }\href {https://doi.org/10.1063/5.0045835} {\bibfield  {journal} {\bibinfo  {journal} {Appl. Phys. Lett.}\ }\textbf {\bibinfo {volume} {118}},\ \bibinfo {pages} {212409} (\bibinfo {year}
      {2021})}\BibitemShut {NoStop}%
    \bibitem [{\citenamefont {Riveros}\ \emph {et~al.}(2021)\citenamefont {Riveros}, \citenamefont {Tejo}, \citenamefont {Escrig}, \citenamefont {Guslienko},\ and\ \citenamefont {Chubykalo-Fesenko}}]{Riveros2021Field-dependentNanodots}%
      \BibitemOpen
      \bibfield  {author} {\bibinfo {author} {\bibfnamefont {A.}~\bibnamefont {Riveros}}, \bibinfo {author} {\bibfnamefont {F.}~\bibnamefont {Tejo}}, \bibinfo {author} {\bibfnamefont {J.}~\bibnamefont {Escrig}}, \bibinfo {author} {\bibfnamefont {K.}~\bibnamefont {Guslienko}},\ and\ \bibinfo {author} {\bibfnamefont {O.}~\bibnamefont {Chubykalo-Fesenko}},\ }\bibfield  {title} {\bibinfo {title} {Field-dependent energy barriers of magnetic {N\'eel} skyrmions in ultrathin circular nanodots},\ }\href {https://doi.org/10.1103/PhysRevApplied.16.014068} {\bibfield  {journal} {\bibinfo  {journal} {Phys. Rev. Appl.}\ }\textbf {\bibinfo {volume} {16}},\ \bibinfo {pages} {014068} (\bibinfo {year} {2021})}\BibitemShut {NoStop}%
    \bibitem [{\citenamefont {Gubbiotti}\ \emph {et~al.}(2018)\citenamefont {Gubbiotti}, \citenamefont {Xiong}, \citenamefont {Montoncello}, \citenamefont {Giovannini},\ and\ \citenamefont {Adeyeye}}]{Gubbiotti2018SpinStudy}%
      \BibitemOpen
      \bibfield  {author} {\bibinfo {author} {\bibfnamefont {G.}~\bibnamefont {Gubbiotti}}, \bibinfo {author} {\bibfnamefont {L.~L.}\ \bibnamefont {Xiong}}, \bibinfo {author} {\bibfnamefont {F.}~\bibnamefont {Montoncello}}, \bibinfo {author} {\bibfnamefont {L.}~\bibnamefont {Giovannini}},\ and\ \bibinfo {author} {\bibfnamefont {A.~O.}\ \bibnamefont {Adeyeye}},\ }\bibfield  {title} {\bibinfo {title} {{Spin wave dispersion and intensity correlation in width-modulated nanowire arrays: A Brillouin light scattering study}},\ }\href {https://doi.org/10.1063/1.5047393} {\bibfield  {journal} {\bibinfo  {journal} {J. Appl. Phys.}\ }\textbf {\bibinfo {volume} {124}},\ \bibinfo {pages} {083903} (\bibinfo {year} {2018})}\BibitemShut {NoStop}%
    \bibitem [{\citenamefont {Mruczkiewicz}\ \emph {et~al.}(2017)\citenamefont {Mruczkiewicz}, \citenamefont {Graczyk}, \citenamefont {Lupo}, \citenamefont {Adeyeye}, \citenamefont {Gubbiotti},\ and\ \citenamefont {Krawczyk}}]{Mruczkiewicz2017}%
      \BibitemOpen
      \bibfield  {author} {\bibinfo {author} {\bibfnamefont {M.}~\bibnamefont {Mruczkiewicz}}, \bibinfo {author} {\bibfnamefont {P.}~\bibnamefont {Graczyk}}, \bibinfo {author} {\bibfnamefont {P.}~\bibnamefont {Lupo}}, \bibinfo {author} {\bibfnamefont {A.}~\bibnamefont {Adeyeye}}, \bibinfo {author} {\bibfnamefont {G.}~\bibnamefont {Gubbiotti}},\ and\ \bibinfo {author} {\bibfnamefont {M.}~\bibnamefont {Krawczyk}},\ }\bibfield  {title} {\bibinfo {title} {{Spin-wave nonreciprocity and magnonic band structure in a thin permalloy film induced by dynamical coupling with an array of Ni stripes}},\ }\href {https://doi.org/10.1103/PhysRevB.96.104411} {\bibfield  {journal} {\bibinfo  {journal} {Phys. Rev. B}\ }\textbf {\bibinfo {volume} {96}},\ \bibinfo {pages} {104411} (\bibinfo {year} {2017})}\BibitemShut {NoStop}%
    \bibitem [{\citenamefont {Graczyk}\ \emph {et~al.}(2018)\citenamefont {Graczyk}, \citenamefont {Krawczyk}, \citenamefont {Dhuey}, \citenamefont {Yang}, \citenamefont {Schmidt},\ and\ \citenamefont {Gubbiotti}}]{Graczyk2018}%
      \BibitemOpen
      \bibfield  {author} {\bibinfo {author} {\bibfnamefont {P.}~\bibnamefont {Graczyk}}, \bibinfo {author} {\bibfnamefont {M.}~\bibnamefont {Krawczyk}}, \bibinfo {author} {\bibfnamefont {S.}~\bibnamefont {Dhuey}}, \bibinfo {author} {\bibfnamefont {W.-G.}\ \bibnamefont {Yang}}, \bibinfo {author} {\bibfnamefont {H.}~\bibnamefont {Schmidt}},\ and\ \bibinfo {author} {\bibfnamefont {G.}~\bibnamefont {Gubbiotti}},\ }\bibfield  {title} {\bibinfo {title} {Magnonic band gap and mode hybridization in continuous permalloy films induced by vertical dynamic coupling with an array of permalloy ellipses},\ }\href {https://doi.org/10.1103/PhysRevB.98.174420} {\bibfield  {journal} {\bibinfo  {journal} {Phys. Rev. B}\ }\textbf {\bibinfo {volume} {98}},\ \bibinfo {pages} {174420} (\bibinfo {year} {2018})}\BibitemShut {NoStop}%
    \bibitem [{\citenamefont {Qin}\ \emph {et~al.}(2021)\citenamefont {Qin}, \citenamefont {Holl{\"a}nder}, \citenamefont {Flajsman}, \citenamefont {Hermann}, \citenamefont {Dreyer}, \citenamefont {Woltersdorf},\ and\ \citenamefont {van Dijken}}]{Qin2021}%
      \BibitemOpen
      \bibfield  {author} {\bibinfo {author} {\bibfnamefont {H.}~\bibnamefont {Qin}}, \bibinfo {author} {\bibfnamefont {R.~B.}\ \bibnamefont {Holl{\"a}nder}}, \bibinfo {author} {\bibfnamefont {L.}~\bibnamefont {Flajsman}}, \bibinfo {author} {\bibfnamefont {F.}~\bibnamefont {Hermann}}, \bibinfo {author} {\bibfnamefont {R.}~\bibnamefont {Dreyer}}, \bibinfo {author} {\bibfnamefont {G.}~\bibnamefont {Woltersdorf}},\ and\ \bibinfo {author} {\bibfnamefont {S.}~\bibnamefont {van Dijken}},\ }\bibfield  {title} {\bibinfo {title} {{Nanoscale magnonic Fabry-Parot resonator for low-loss spin-wave manipulation}},\ }\href {https://doi.org/10.1038/s41467-021-22520-6} {\bibfield  {journal} {\bibinfo  {journal} {Nat. Commun.}\ }\textbf {\bibinfo {volume} {12}},\ \bibinfo {pages} {2293} (\bibinfo {year} {2021})}\BibitemShut {NoStop}%
    \bibitem [{\citenamefont {Wang}\ \emph {et~al.}(2024{\natexlab{a}})\citenamefont {Wang}, \citenamefont {Yan}, \citenamefont {Kuznetsov}, \citenamefont {Flajsman}, \citenamefont {Qin},\ and\ \citenamefont {van Dijken}}]{Wang2024c}%
      \BibitemOpen
      \bibfield  {author} {\bibinfo {author} {\bibfnamefont {Y.}~\bibnamefont {Wang}}, \bibinfo {author} {\bibfnamefont {W.}~\bibnamefont {Yan}}, \bibinfo {author} {\bibfnamefont {N.}~\bibnamefont {Kuznetsov}}, \bibinfo {author} {\bibfnamefont {L.}~\bibnamefont {Flajsman}}, \bibinfo {author} {\bibfnamefont {H.}~\bibnamefont {Qin}},\ and\ \bibinfo {author} {\bibfnamefont {S.}~\bibnamefont {van Dijken}},\ }\bibfield  {title} {\bibinfo {title} {{Spin-wave diffraction, caustic beam emission, and Talbot carpets in a yttrium iron garnet film with magnonic Fabry-Perot resonator gratings}},\ }\href {https://doi.org/10.1103/PhysRevApplied.22.014038} {\bibfield  {journal} {\bibinfo  {journal} {Phys. Rev. Appl.}\ }\textbf {\bibinfo {volume} {22}},\ \bibinfo {pages} {014038} (\bibinfo {year} {2024}{\natexlab{a}})}\BibitemShut {NoStop}%
    \bibitem [{\citenamefont {Tacchi}\ \emph {et~al.}(2023)\citenamefont {Tacchi}, \citenamefont {Flores-Far{\'i}as}, \citenamefont {Petti}, \citenamefont {Brevis}, \citenamefont {Cattoni}, \citenamefont {Scaramuzzi}, \citenamefont {Girardi}, \citenamefont {Cort{\'e}s-Ortuno}, \citenamefont {Gallardo}, \citenamefont {Albisetti}, \citenamefont {Carlotti},\ and\ \citenamefont {Landeros}}]{Tacchi2023}%
      \BibitemOpen
      \bibfield  {author} {\bibinfo {author} {\bibfnamefont {S.}~\bibnamefont {Tacchi}}, \bibinfo {author} {\bibfnamefont {J.}~\bibnamefont {Flores-Far{\'i}as}}, \bibinfo {author} {\bibfnamefont {D.}~\bibnamefont {Petti}}, \bibinfo {author} {\bibfnamefont {F.}~\bibnamefont {Brevis}}, \bibinfo {author} {\bibfnamefont {A.}~\bibnamefont {Cattoni}}, \bibinfo {author} {\bibfnamefont {G.}~\bibnamefont {Scaramuzzi}}, \bibinfo {author} {\bibfnamefont {D.}~\bibnamefont {Girardi}}, \bibinfo {author} {\bibfnamefont {D.}~\bibnamefont {Cort{\'e}s-Ortuno}}, \bibinfo {author} {\bibfnamefont {R.~A.}\ \bibnamefont {Gallardo}}, \bibinfo {author} {\bibfnamefont {E.}~\bibnamefont {Albisetti}}, \bibinfo {author} {\bibfnamefont {G.}~\bibnamefont {Carlotti}},\ and\ \bibinfo {author} {\bibfnamefont {P.}~\bibnamefont {Landeros}},\ }\bibfield  {title} {\bibinfo {title} {Experimental observation of flat bands in one-dimensional chiral magnonic crystals},\ }\href {https://doi.org/10.1021/acs.nanolett.2c04215} {\bibfield  {journal} {\bibinfo
       {journal} {Nano Lett.}\ }\textbf {\bibinfo {volume} {23}},\ \bibinfo {pages} {6776} (\bibinfo {year} {2023})}\BibitemShut {NoStop}%
    \bibitem [{\citenamefont {Centa{\l}a}\ and\ \citenamefont {K{\l}os}(2023)}]{Centala2023}%
      \BibitemOpen
      \bibfield  {author} {\bibinfo {author} {\bibfnamefont {G.}~\bibnamefont {Centa{\l}a}}\ and\ \bibinfo {author} {\bibfnamefont {J.~W.}\ \bibnamefont {K{\l}os}},\ }\bibfield  {title} {\bibinfo {title} {{Compact localized states in magnonic Lieb lattices}},\ }\href {https://doi.org/10.1038/s41598-023-39816-w} {\bibfield  {journal} {\bibinfo  {journal} {Sci. Rep.}\ }\textbf {\bibinfo {volume} {13}},\ \bibinfo {pages} {12676} (\bibinfo {year} {2023})}\BibitemShut {NoStop}%
    \bibitem [{\citenamefont {Wang}\ \emph {et~al.}(2024{\natexlab{b}})\citenamefont {Wang}, \citenamefont {Wang}, \citenamefont {Chen}, \citenamefont {Legrand}, \citenamefont {Chen}, \citenamefont {Sheng}, \citenamefont {Xia}, \citenamefont {Lan}, \citenamefont {Zhang}, \citenamefont {Yuan}, \citenamefont {Dong}, \citenamefont {Han}, \citenamefont {Ansermet},\ and\ \citenamefont {Yu}}]{Wang2024b}%
      \BibitemOpen
      \bibfield  {author} {\bibinfo {author} {\bibfnamefont {J.}~\bibnamefont {Wang}}, \bibinfo {author} {\bibfnamefont {H.}~\bibnamefont {Wang}}, \bibinfo {author} {\bibfnamefont {J.}~\bibnamefont {Chen}}, \bibinfo {author} {\bibfnamefont {W.}~\bibnamefont {Legrand}}, \bibinfo {author} {\bibfnamefont {P.}~\bibnamefont {Chen}}, \bibinfo {author} {\bibfnamefont {L.}~\bibnamefont {Sheng}}, \bibinfo {author} {\bibfnamefont {J.}~\bibnamefont {Xia}}, \bibinfo {author} {\bibfnamefont {G.}~\bibnamefont {Lan}}, \bibinfo {author} {\bibfnamefont {Y.}~\bibnamefont {Zhang}}, \bibinfo {author} {\bibfnamefont {R.}~\bibnamefont {Yuan}}, \bibinfo {author} {\bibfnamefont {J.}~\bibnamefont {Dong}}, \bibinfo {author} {\bibfnamefont {X.}~\bibnamefont {Han}}, \bibinfo {author} {\bibfnamefont {J.-P.}\ \bibnamefont {Ansermet}},\ and\ \bibinfo {author} {\bibfnamefont {H.}~\bibnamefont {Yu}},\ }\bibfield  {title} {\bibinfo {title} {Broad-wave-vector spin pumping of flat-band magnons},\ }\href
      {https://doi.org/10.1103/PhysRevApplied.21.044024} {\bibfield  {journal} {\bibinfo  {journal} {Phys. Rev. Appl.}\ }\textbf {\bibinfo {volume} {21}},\ \bibinfo {pages} {044024} (\bibinfo {year} {2024}{\natexlab{b}})}\BibitemShut {NoStop}%
    \bibitem [{\citenamefont {Chen}\ \emph {et~al.}(2018)\citenamefont {Chen}, \citenamefont {Liu}, \citenamefont {Liu}, \citenamefont {Xiao}, \citenamefont {Xia}, \citenamefont {Bauer}, \citenamefont {Wu},\ and\ \citenamefont {Yu}}]{Chen2018}%
      \BibitemOpen
      \bibfield  {author} {\bibinfo {author} {\bibfnamefont {J.}~\bibnamefont {Chen}}, \bibinfo {author} {\bibfnamefont {C.}~\bibnamefont {Liu}}, \bibinfo {author} {\bibfnamefont {T.}~\bibnamefont {Liu}}, \bibinfo {author} {\bibfnamefont {Y.}~\bibnamefont {Xiao}}, \bibinfo {author} {\bibfnamefont {K.}~\bibnamefont {Xia}}, \bibinfo {author} {\bibfnamefont {G.~E.~W.}\ \bibnamefont {Bauer}}, \bibinfo {author} {\bibfnamefont {M.}~\bibnamefont {Wu}},\ and\ \bibinfo {author} {\bibfnamefont {H.}~\bibnamefont {Yu}},\ }\bibfield  {title} {\bibinfo {title} {Strong interlayer magnon-magnon coupling in magnetic metal-insulator hybrid nanostructures},\ }\href {https://doi.org/10.1103/PhysRevLett.120.217202} {\bibfield  {journal} {\bibinfo  {journal} {Phys. Rev. Lett.}\ }\textbf {\bibinfo {volume} {120}},\ \bibinfo {pages} {217202} (\bibinfo {year} {2018})}\BibitemShut {NoStop}%
    \bibitem [{\citenamefont {Moalic}\ \emph {et~al.}(2024)\citenamefont {Moalic}, \citenamefont {Zelent}, \citenamefont {Szulc},\ and\ \citenamefont {Krawczyk}}]{Moalic2024}%
      \BibitemOpen
      \bibfield  {author} {\bibinfo {author} {\bibfnamefont {M.}~\bibnamefont {Moalic}}, \bibinfo {author} {\bibfnamefont {M.}~\bibnamefont {Zelent}}, \bibinfo {author} {\bibfnamefont {K.}~\bibnamefont {Szulc}},\ and\ \bibinfo {author} {\bibfnamefont {M.}~\bibnamefont {Krawczyk}},\ }\bibfield  {title} {\bibinfo {title} {The role of non-uniform magnetization texture for magnon{--}magnon coupling in an antidot lattice},\ }\href {https://doi.org/10.1038/s41598-024-61246-5} {\bibfield  {journal} {\bibinfo  {journal} {Sci. Rep.}\ }\textbf {\bibinfo {volume} {14}},\ \bibinfo {pages} {11501} (\bibinfo {year} {2024})}\BibitemShut {NoStop}%
    \bibitem [{\citenamefont {Pal}\ \emph {et~al.}(2023)\citenamefont {Pal}, \citenamefont {Majumder}, \citenamefont {Otani},\ and\ \citenamefont {Barman}}]{Kumar2023}%
      \BibitemOpen
      \bibfield  {author} {\bibinfo {author} {\bibfnamefont {P.~K.}\ \bibnamefont {Pal}}, \bibinfo {author} {\bibfnamefont {S.}~\bibnamefont {Majumder}}, \bibinfo {author} {\bibfnamefont {Y.}~\bibnamefont {Otani}},\ and\ \bibinfo {author} {\bibfnamefont {A.}~\bibnamefont {Barman}},\ }\bibfield  {title} {\bibinfo {title} {{Bias-Field Tunable Magnon-Magnon Coupling in Ni80Fe20 Nanocross Array}},\ }\href {https://doi.org/https://doi.org/10.1002/qute.202300003} {\bibfield  {journal} {\bibinfo  {journal} {Adv. Quantum Technol.}\ }\textbf {\bibinfo {volume} {6}},\ \bibinfo {pages} {2300003} (\bibinfo {year} {2023})}\BibitemShut {NoStop}%
    \bibitem [{\citenamefont {Hu}\ \emph {et~al.}(2024)\citenamefont {Hu}, \citenamefont {Xie}, \citenamefont {Lu},\ and\ \citenamefont {He}}]{Bo2024}%
      \BibitemOpen
      \bibfield  {author} {\bibinfo {author} {\bibfnamefont {B.}~\bibnamefont {Hu}}, \bibinfo {author} {\bibfnamefont {Z.-K.}\ \bibnamefont {Xie}}, \bibinfo {author} {\bibfnamefont {J.}~\bibnamefont {Lu}},\ and\ \bibinfo {author} {\bibfnamefont {W.}~\bibnamefont {He}},\ }\bibfield  {title} {\bibinfo {title} {{Mapping the magnon--magnon hybrid state onto the Bloch sphere}},\ }\href {https://doi.org/10.1063/5.0202372} {\bibfield  {journal} {\bibinfo  {journal} {Appl. Phys. Lett.}\ }\textbf {\bibinfo {volume} {124}},\ \bibinfo {pages} {232402} (\bibinfo {year} {2024})}\BibitemShut {NoStop}%
    \bibitem [{\citenamefont {Qin}\ \emph {et~al.}(2018)\citenamefont {Qin}, \citenamefont {Both}, \citenamefont {H{\"a}m{\"a}l{\"a}inen}, \citenamefont {Yao},\ and\ \citenamefont {van Dijken}}]{Qin2018}%
      \BibitemOpen
      \bibfield  {author} {\bibinfo {author} {\bibfnamefont {H.}~\bibnamefont {Qin}}, \bibinfo {author} {\bibfnamefont {G.-J.}\ \bibnamefont {Both}}, \bibinfo {author} {\bibfnamefont {S.~J.}\ \bibnamefont {H{\"a}m{\"a}l{\"a}inen}}, \bibinfo {author} {\bibfnamefont {L.}~\bibnamefont {Yao}},\ and\ \bibinfo {author} {\bibfnamefont {S.}~\bibnamefont {van Dijken}},\ }\bibfield  {title} {\bibinfo {title} {{Low-loss YIG-based magnonic crystals with large tunable bandgaps}},\ }\href@noop {} {\bibfield  {journal} {\bibinfo  {journal} {Nat. Commun.}\ }\textbf {\bibinfo {volume} {9}},\ \bibinfo {pages} {5445} (\bibinfo {year} {2018})}\BibitemShut {NoStop}%
    \bibitem [{\citenamefont {Adhikari}\ \emph {et~al.}(2021)\citenamefont {Adhikari}, \citenamefont {Choudhury}, \citenamefont {Barman}, \citenamefont {Otani},\ and\ \citenamefont {Barman}}]{Adhikari_2021}%
      \BibitemOpen
      \bibfield  {author} {\bibinfo {author} {\bibfnamefont {K.}~\bibnamefont {Adhikari}}, \bibinfo {author} {\bibfnamefont {S.}~\bibnamefont {Choudhury}}, \bibinfo {author} {\bibfnamefont {S.}~\bibnamefont {Barman}}, \bibinfo {author} {\bibfnamefont {Y.}~\bibnamefont {Otani}},\ and\ \bibinfo {author} {\bibfnamefont {A.}~\bibnamefont {Barman}},\ }\bibfield  {title} {\bibinfo {title} {Observation of magnon-magnon coupling with high cooperativity in {Ni$_{80}$Fe$_{20}$} cross-shaped nanoring array},\ }\href {https://doi.org/10.1088/1361-6528/ac0ddc} {\bibfield  {journal} {\bibinfo  {journal} {Nanotechnology}\ }\textbf {\bibinfo {volume} {32}},\ \bibinfo {pages} {395706} (\bibinfo {year} {2021})}\BibitemShut {NoStop}%
    \bibitem [{\citenamefont {Gartside}\ \emph {et~al.}(2021)\citenamefont {Gartside}, \citenamefont {Vanstone}, \citenamefont {Dion}, \citenamefont {Stenning}, \citenamefont {Arroo}, \citenamefont {Kurebayashi},\ and\ \citenamefont {Branford}}]{Gartside2021}%
      \BibitemOpen
      \bibfield  {author} {\bibinfo {author} {\bibfnamefont {J.~C.}\ \bibnamefont {Gartside}}, \bibinfo {author} {\bibfnamefont {A.}~\bibnamefont {Vanstone}}, \bibinfo {author} {\bibfnamefont {T.}~\bibnamefont {Dion}}, \bibinfo {author} {\bibfnamefont {K.~D.}\ \bibnamefont {Stenning}}, \bibinfo {author} {\bibfnamefont {D.~M.}\ \bibnamefont {Arroo}}, \bibinfo {author} {\bibfnamefont {H.}~\bibnamefont {Kurebayashi}},\ and\ \bibinfo {author} {\bibfnamefont {W.~R.}\ \bibnamefont {Branford}},\ }\bibfield  {title} {\bibinfo {title} {Reconfigurable magnonic mode-hybridisation and spectral control in a bicomponent artificial spin ice},\ }\href {https://doi.org/10.1038/s41467-021-22723-x} {\bibfield  {journal} {\bibinfo  {journal} {Nat. Commun.}\ }\textbf {\bibinfo {volume} {12}},\ \bibinfo {pages} {2488} (\bibinfo {year} {2021})}\BibitemShut {NoStop}%
    \bibitem [{\citenamefont {Wang}\ \emph {et~al.}(2024{\natexlab{c}})\citenamefont {Wang}, \citenamefont {Zhang}, \citenamefont {Li}, \citenamefont {Wei}, \citenamefont {He}, \citenamefont {Xu}, \citenamefont {Xia}, \citenamefont {Luo}, \citenamefont {Li}, \citenamefont {Dong}, \citenamefont {He}, \citenamefont {Yan}, \citenamefont {Yang}, \citenamefont {Ma}, \citenamefont {Chai}, \citenamefont {Yan}, \citenamefont {Wan}, \citenamefont {Han},\ and\ \citenamefont {Yu}}]{Wang2024}%
      \BibitemOpen
      \bibfield  {author} {\bibinfo {author} {\bibfnamefont {Y.}~\bibnamefont {Wang}}, \bibinfo {author} {\bibfnamefont {Y.}~\bibnamefont {Zhang}}, \bibinfo {author} {\bibfnamefont {C.}~\bibnamefont {Li}}, \bibinfo {author} {\bibfnamefont {J.}~\bibnamefont {Wei}}, \bibinfo {author} {\bibfnamefont {B.}~\bibnamefont {He}}, \bibinfo {author} {\bibfnamefont {H.}~\bibnamefont {Xu}}, \bibinfo {author} {\bibfnamefont {J.}~\bibnamefont {Xia}}, \bibinfo {author} {\bibfnamefont {X.}~\bibnamefont {Luo}}, \bibinfo {author} {\bibfnamefont {J.}~\bibnamefont {Li}}, \bibinfo {author} {\bibfnamefont {J.}~\bibnamefont {Dong}}, \bibinfo {author} {\bibfnamefont {W.}~\bibnamefont {He}}, \bibinfo {author} {\bibfnamefont {Z.}~\bibnamefont {Yan}}, \bibinfo {author} {\bibfnamefont {W.}~\bibnamefont {Yang}}, \bibinfo {author} {\bibfnamefont {F.}~\bibnamefont {Ma}}, \bibinfo {author} {\bibfnamefont {G.}~\bibnamefont {Chai}}, \bibinfo {author} {\bibfnamefont {P.}~\bibnamefont {Yan}}, \bibinfo {author} {\bibfnamefont {C.}~\bibnamefont
      {Wan}}, \bibinfo {author} {\bibfnamefont {X.}~\bibnamefont {Han}},\ and\ \bibinfo {author} {\bibfnamefont {G.}~\bibnamefont {Yu}},\ }\bibfield  {title} {\bibinfo {title} {Ultrastrong to nearly deep-strong magnon-magnon coupling with a high degree of freedom in synthetic antiferromagnets},\ }\href {https://doi.org/10.1038/s41467-024-46474-7} {\bibfield  {journal} {\bibinfo  {journal} {Nat. Commun.}\ }\textbf {\bibinfo {volume} {15}},\ \bibinfo {pages} {2077} (\bibinfo {year} {2024}{\natexlab{c}})}\BibitemShut {NoStop}%
    \bibitem [{\citenamefont {Krivosik}\ and\ \citenamefont {Patton}(2010)}]{Krivosik2010}%
      \BibitemOpen
      \bibfield  {author} {\bibinfo {author} {\bibfnamefont {P.}~\bibnamefont {Krivosik}}\ and\ \bibinfo {author} {\bibfnamefont {C.~E.}\ \bibnamefont {Patton}},\ }\bibfield  {title} {\bibinfo {title} {Hamiltonian formulation of nonlinear spin-wave dynamics: Theory and applications},\ }\href {https://doi.org/10.1103/PhysRevB.82.184428} {\bibfield  {journal} {\bibinfo  {journal} {Phys. Rev. B}\ }\textbf {\bibinfo {volume} {82}},\ \bibinfo {pages} {184428} (\bibinfo {year} {2010})}\BibitemShut {NoStop}%
    \bibitem [{\citenamefont {Merbouche}\ \emph {et~al.}(2022)\citenamefont {Merbouche}, \citenamefont {Divinskiy}, \citenamefont {Nikolaev}, \citenamefont {Kaspar}, \citenamefont {Pernice}, \citenamefont {Gouerne}, \citenamefont {Lebrun}, \citenamefont {Cros}, \citenamefont {Ben~Youssef}, \citenamefont {Bortolotti}, \citenamefont {Anane}, \citenamefont {Demokritov},\ and\ \citenamefont {Demidov}}]{Merbouche2022}%
      \BibitemOpen
      \bibfield  {author} {\bibinfo {author} {\bibfnamefont {H.}~\bibnamefont {Merbouche}}, \bibinfo {author} {\bibfnamefont {B.}~\bibnamefont {Divinskiy}}, \bibinfo {author} {\bibfnamefont {K.~O.}\ \bibnamefont {Nikolaev}}, \bibinfo {author} {\bibfnamefont {C.}~\bibnamefont {Kaspar}}, \bibinfo {author} {\bibfnamefont {W.~H.~P.}\ \bibnamefont {Pernice}}, \bibinfo {author} {\bibfnamefont {D.}~\bibnamefont {Gouerne}}, \bibinfo {author} {\bibfnamefont {R.}~\bibnamefont {Lebrun}}, \bibinfo {author} {\bibfnamefont {V.}~\bibnamefont {Cros}}, \bibinfo {author} {\bibfnamefont {J.}~\bibnamefont {Ben~Youssef}}, \bibinfo {author} {\bibfnamefont {P.}~\bibnamefont {Bortolotti}}, \bibinfo {author} {\bibfnamefont {A.}~\bibnamefont {Anane}}, \bibinfo {author} {\bibfnamefont {S.~O.}\ \bibnamefont {Demokritov}},\ and\ \bibinfo {author} {\bibfnamefont {V.~E.}\ \bibnamefont {Demidov}},\ }\bibfield  {title} {\bibinfo {title} {{Giant nonlinear self-phase modulation of large-amplitude spin waves in microscopic YIG waveguides}},\ }\href
      {https://doi.org/10.1038/s41598-022-10822-8} {\bibfield  {journal} {\bibinfo  {journal} {Sci. Rep.}\ }\textbf {\bibinfo {volume} {12}},\ \bibinfo {pages} {7246} (\bibinfo {year} {2022})}\BibitemShut {NoStop}%
    \bibitem [{\citenamefont {Wang}\ \emph {et~al.}(2023{\natexlab{b}})\citenamefont {Wang}, \citenamefont {Verba}, \citenamefont {Heinz}, \citenamefont {Schneider}, \citenamefont {Wojewoda}, \citenamefont {Davidkova}, \citenamefont {Levchenko}, \citenamefont {Dubs}, \citenamefont {Mauser}, \citenamefont {Urb{\'a}nek}, \citenamefont {Pirro},\ and\ \citenamefont {Chumak}}]{Qi2023}%
      \BibitemOpen
      \bibfield  {author} {\bibinfo {author} {\bibfnamefont {Q.}~\bibnamefont {Wang}}, \bibinfo {author} {\bibfnamefont {R.}~\bibnamefont {Verba}}, \bibinfo {author} {\bibfnamefont {B.}~\bibnamefont {Heinz}}, \bibinfo {author} {\bibfnamefont {M.}~\bibnamefont {Schneider}}, \bibinfo {author} {\bibfnamefont {O.}~\bibnamefont {Wojewoda}}, \bibinfo {author} {\bibfnamefont {K.}~\bibnamefont {Davidkova}}, \bibinfo {author} {\bibfnamefont {K.}~\bibnamefont {Levchenko}}, \bibinfo {author} {\bibfnamefont {C.}~\bibnamefont {Dubs}}, \bibinfo {author} {\bibfnamefont {N.~J.}\ \bibnamefont {Mauser}}, \bibinfo {author} {\bibfnamefont {M.}~\bibnamefont {Urb{\'a}nek}}, \bibinfo {author} {\bibfnamefont {P.}~\bibnamefont {Pirro}},\ and\ \bibinfo {author} {\bibfnamefont {A.~V.}\ \bibnamefont {Chumak}},\ }\bibfield  {title} {\bibinfo {title} {Deeply nonlinear excitation of self-normalized short spin waves},\ }\href {https://doi.org/10.1126/sciadv.adg4609} {\bibfield  {journal} {\bibinfo  {journal} {Sci. Adv.}\ }\textbf {\bibinfo
      {volume} {9}},\ \bibinfo {pages} {eadg4609} (\bibinfo {year} {2023}{\natexlab{b}})}\BibitemShut {NoStop}%
    \bibitem [{\citenamefont {Yu}\ \emph {et~al.}(2016)\citenamefont {Yu}, \citenamefont {d’Allivy Kelly}, \citenamefont {Cros}, \citenamefont {Bernard}, \citenamefont {Bortolotti}, \citenamefont {Anane}, \citenamefont {Brandl}, \citenamefont {Heimbach},\ and\ \citenamefont {Grundler}}]{Yu2016}%
      \BibitemOpen
      \bibfield  {author} {\bibinfo {author} {\bibfnamefont {H.}~\bibnamefont {Yu}}, \bibinfo {author} {\bibfnamefont {O.}~\bibnamefont {d’Allivy Kelly}}, \bibinfo {author} {\bibfnamefont {V.}~\bibnamefont {Cros}}, \bibinfo {author} {\bibfnamefont {R.}~\bibnamefont {Bernard}}, \bibinfo {author} {\bibfnamefont {P.}~\bibnamefont {Bortolotti}}, \bibinfo {author} {\bibfnamefont {A.}~\bibnamefont {Anane}}, \bibinfo {author} {\bibfnamefont {F.}~\bibnamefont {Brandl}}, \bibinfo {author} {\bibfnamefont {F.}~\bibnamefont {Heimbach}},\ and\ \bibinfo {author} {\bibfnamefont {D.}~\bibnamefont {Grundler}},\ }\bibfield  {title} {\bibinfo {title} {{Approaching soft X-ray wavelengths in nanomagnet-based microwave technology}},\ }\href@noop {} {\bibfield  {journal} {\bibinfo  {journal} {Nat. Commun.}\ }\textbf {\bibinfo {volume} {7}},\ \bibinfo {pages} {11255} (\bibinfo {year} {2016})}\BibitemShut {NoStop}%
    \bibitem [{\citenamefont {Szulc}(2024)}]{data}%
      \BibitemOpen
      \bibfield  {author} {\bibinfo {author} {\bibfnamefont {K.}~\bibnamefont {Szulc}},\ }\href {https://doi.org/10.5281/zenodo.10964849} {\bibinfo {title} {{Dataset for "Reconfigurable spin-wave platform based on interplay between nanodots and waveguide in hybrid magnonic crystal"}}},\ \bibinfo {howpublished} {\url{https://doi.org/10.5281/zenodo.10964849}} (\bibinfo {year} {2024})\BibitemShut {NoStop}%
    \end{thebibliography}
\end{document}



\title{\textsc{Supplementary Materials}\\Reconfigurable spin-wave platform based on interplay between nanodots and waveguide in hybrid magnonic crystal}

 
\author{Krzysztof~Szulc}
\email{szulc@ifmpan.poznan.pl}
\affiliation{%
Institute of Molecular Physics, Polish Academy of Sciences, M. Smoluchowskiego 17, 60-179 Pozna\'{n}, Poland
}%
\affiliation{%
Institute of Spintronics and Quantum Information, Faculty of Physics and Astronomy, Adam Mickiewicz University, Pozna\'{n}, Uniwersytetu Pozna\'{n}skiego 2, 61-614 Pozna\'{n}, Poland 
}%
\author{Mateusz~Zelent}%
\affiliation{%
Institute of Spintronics and Quantum Information, Faculty of Physics and Astronomy, Adam Mickiewicz University, Pozna\'{n}, Uniwersytetu Pozna\'{n}skiego 2, 61-614 Pozna\'{n}, Poland 
}%
\author{Maciej~Krawczyk}%
\affiliation{%
Institute of Spintronics and Quantum Information, Faculty of Physics and Astronomy, Adam Mickiewicz University, Pozna\'{n}, Uniwersytetu Pozna\'{n}skiego 2, 61-614 Pozna\'{n}, Poland 
}%


\begin{abstract}
We present a hybrid magnonic crystal composed of a chain of nanodots with strong perpendicular magnetic anisotropy and Dzyaloshinskii--Moriya interaction, positioned above a permalloy waveguide. The micromagnetic study examines two different magnetization states in the nanodots: a single-domain state and an egg-shaped skyrmion state. Due to the dipolar coupling between the dot and the waveguide, a strongly bound hybrid magnetization texture is formed in the system. Our results show complex spin-wave spectra, combining the effects of periodicity, magnetization texture, and hybridization of the propagating waves in the waveguide with the dot/skyrmion modes. The dynamics of the systems are characterized by several key features which include differences in band-gap sizes, the presence of flat bands in the skyrmion state that can form both bound and hybridized states, the latter sometimes leading to the presence of additional non-Bragg band gaps, and a broad frequency range of only waveguide-dominated modes in the single-domain state. Thus, the study shows that the proposed hybrid magnonic crystals have many distinct functionalities, highlighting their reconfigurable potential, magnon--magnon couplings, mode localization, and bound states overlapping with the propagating waves. This opens up potential applications in analog and quantum magnonics, spin-wave filtering, and the establishment of magnonic neural networks.
\end{abstract}

\maketitle


\section{Description of mode localization}

\begin{figure*}[!t]
    \includegraphics{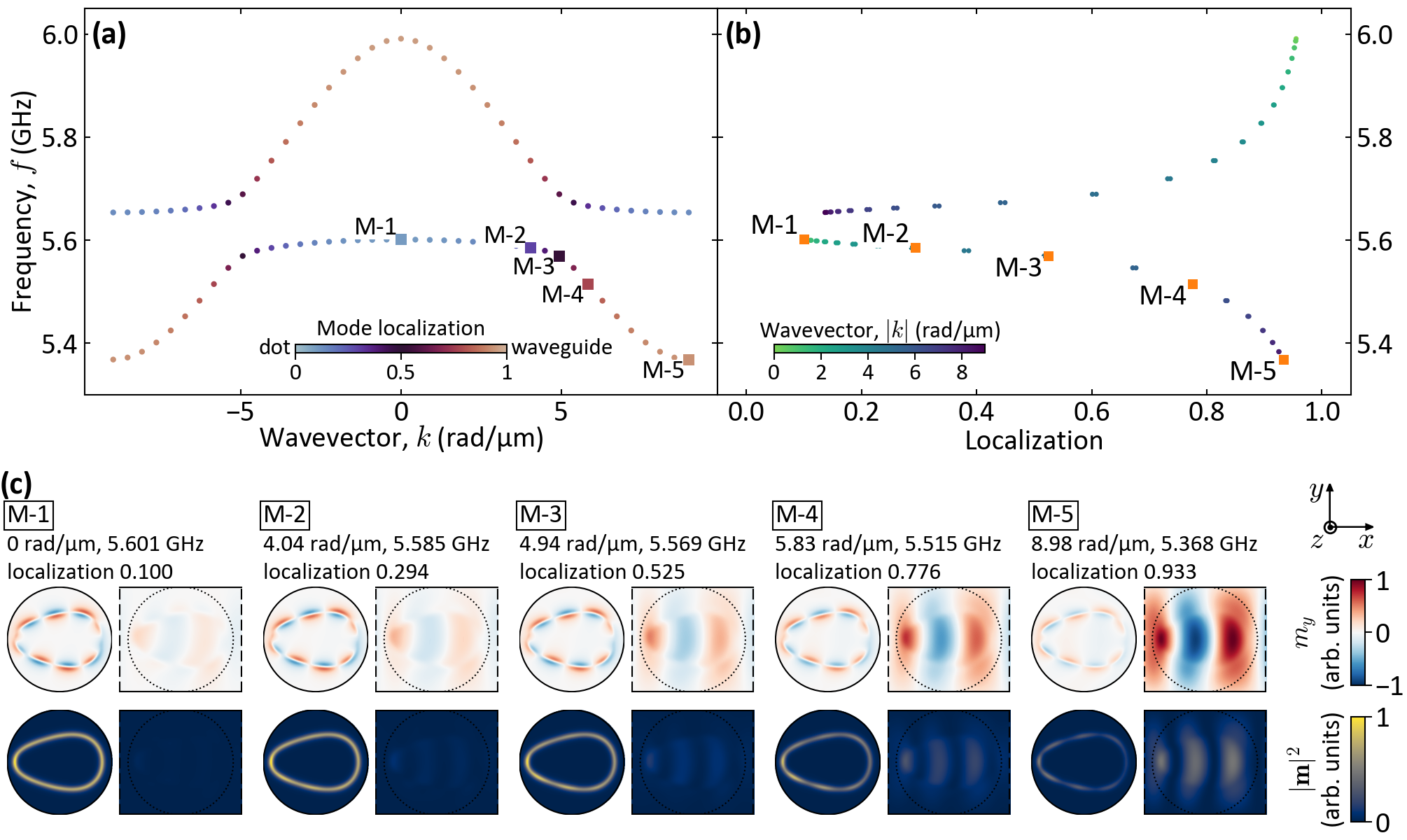}
    \caption{(a-b) The zoom-in of Figures~4(c-d) showing the dispersion relation in the first Brillouin zone and localization in the W/Sk system. (c) The SW mode profiles for 5 modes marked from M-1 to M-5 in (a) and (b). The modes are marked on the dispersion relations with a square point and a label. In the top row, each mode profile displays the $m_y$ magnetization component in the $xy$-plane at the center of the dot (left side) and at the center of the waveguide (right side). The intensity is normalized so that the maximum value of $|m_y|$ is 1 for each of the mode profiles. In the bottom row, analogically, the local intensity is shown calculated as $|\mathbf{m}|^2$.}
    \label{fig:localization}
\end{figure*}

The parameter of the mode localization introduced in Eq.~3 and shown in Fig.~4(a-d) and its connection with the mode profiles shown in Fig.~4(e) are not straightforward. In this section, we extend the description of mode localization basing on two hybridized branches---one dot-dominated and second waveguide-dominated---at around 5.6 GHz in the W/Sk system. They are shown in Fig.~\ref{fig:localization}. Figs.~\ref{fig:localization}(a-b) show the zoom-in of Figs.~4(c-d) from the manuscript. From the bottom of these two branches, we have chosen 5 modes of different wavevector with different values of localization. Their mode profiles are shown in Fig.~\ref{fig:localization}(c).

The top row of Fig.~\ref{fig:localization}(c) shows the $m_y$ component of the magnetization, the same as in Fig.~4(e). In the bottom row, the spatial distribution of the $|\mathbf{m}|^2$ is shown. This value is under the integral in Eq.~4, which is used to calculate the mode localization, therefore, assuming almost uniform distribution of $\mathbf{m}$ across the thickness of the dot and waveguide, these graphs are graphical representations of mode localization. This is apparent from the comparison of the top and bottom row of Fig.~\ref{fig:localization}(c) that the distribution of $m_y$ is not equivalent to $|\mathbf{m}|^2$. The area of the distribution of both parameters matches for the waveguide but it is different for dot as the nodal line for $m_y$ component along the skyrmion circumference matches the maximum of $|\mathbf{m}|^2$. Also, the relative amplitude is different, e.g., while $m_y$ distribution of the mode M-3 gives similar maximum values, for $|\mathbf{m}|^2$ distribution the maximum is significantly higher in the dot. It is because of the difference in the contribution of each magnetization component to the mode localization. In the waveguide, the $m_y$ component shown in the mode profiles is strongly dominant. In the dot in skyrmion state, the highest amplitude component is $m_z$, but is less dominant than $m_y$ component in the waveguide. Because of this, the distributions of $m_y$ and $|\mathbf{m}|^2$ matches only to some extent. It is also worth to look more careful on the distribution of $|\mathbf{m}|^2$. The mode M-3 has the localization close to 0.5 -- the situation where it is equally distributed between the dot and the waveguide. Despite that, the maximum amplitude is significantly larger in the dot. One can notice that the amplitude in the dot is distributed over smaller area than in the waveguide but, except of that, the contribution of the thickness should not be forgotten---the thickness of the waveguide is three times larger than of the dot, so the total intensity of a cross-section should be approximately three times smaller in the waveguide to lead to mode localization of 0.5.

\section{The dynamics of isolated subsystems}

\begin{figure*}[!t]
    \includegraphics{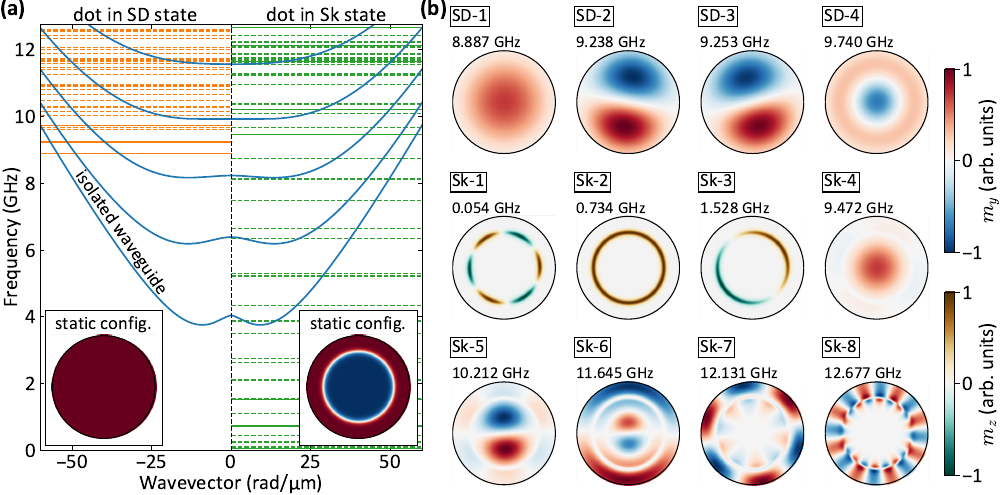}
    \caption{(a) Dispersion relation of an isolated waveguide (solid blue lines) and frequencies of the resonant modes of a dot in a single-domain state (horizontal dashed orange lines) and in a skyrmion state (horizontal dashed green lines). Solid orange and green lines correspond to the modes which profiles are shown in (b). Please note that the resonant modes are characterized solely by their frequencies and they are not connected with the wavevector presented on the horizontal axis. The division of the single-domain state and skyrmion modes on the negative and positive wavevector sides is made solely for the sake of presentation clarity. (b) Resonant mode profiles of the dot of four modes in the SD state and eight modes in the Sk state. Please note that all SD modes and Sk-4 -- Sk-8 modes are presented with $m_y$ component (top color bar), while modes Sk-1 -- Sk-3 are presented with $m_z$ component (bottom color bar). All modes are normalized to the maximum absolute value of the mode. The animated version of this figure is available in Supporting Information.}
    \label{fig:disp}
\end{figure*}

Figure~\ref{fig:disp}(a) shows the comparison between the dynamics of the isolated subsystems. The dispersion relation of an isolated waveguide is shown with solid blue lines. The resonant modes of a dot are shown with horizontal dashed lines: orange line for an isolated dot in the single-domain state and green line for an isolated dot in the skyrmion state.

The lowest zero-$k$ frequency of the waveguide is 4.04 GHz and it reaches a minimum of 3.76 GHz for $k = \SI{9.0}{\radian/\micro\meter}$. Higher-order modes have their minima at 6.19 GHz, 8.17 GHz, 9.92 GHz, and 11.57 GHz, respectively. First four modes exhibit a backward-wave regime at small wavevectors, while higher modes can only propagate forward.

In case of the dot, its static configuration has a very large impact on the frequency of resonant modes. In a single-domain state, the lowest mode has a frequency of 8.89 GHz and it is a fundamental mode (SD-1). Modes with higher frequencies are clockwise (CW) [e.g. SD-2] and counterclockwise (CCW) [e.g. SD-3] azimuthal modes, as well as higher-order radial modes (e.g. SD-4). 

The skyrmion state exhibits numerous low-frequency modes, which are all CW (e.g. Sk-1) and CCW (e.g. Sk-3) azimuthal modes in the skyrmion domain wall, except of one skyrmion breathing mode (Sk-2) (which can be considered the 0th order azimuthal mode). The first mode not associated with the skyrmion domain wall is the fundamental mode of a skyrmion core (Sk-4) at the frequency 9.47 GHz. The higher-frequency modes includes higher-order azimuthal and radial modes, which can be localized either in the skyrmion core (e.g. Sk-5), outside the skyrmion (e.g. Sk-7) or in both the core and outside (e.g. Sk-6). Interestingly, some of the skyrmion domain wall modes in this range can also be strongly excited outside the skyrmion (e.g. Sk-8).

\section{Comparison between W/Sk system and a dot chain}

\begin{table*}[!ht]
    \centering
    \caption{The comparison of the skyrmion domain wall modes in three different systems: single dot, dot chain, and W/Sk system as defined in the main manuscript. For a single dot, we present the value of the mode frequency, while for dot chain and W/Sk system, we show the lowest and the highest frequency of the band and the bandwidth. Please note that the frequencies are in MHz, while bandwidths are in kHz.}
    \begin{tabular}{c|r|rrr|rrr}
        & \multicolumn{1}{c}{Single dot} & \multicolumn{3}{c}{Dot chain} & \multicolumn{3}{c}{W/Sk system} \\ \hline
        & \multicolumn{1}{c}{$f$} & \multicolumn{1}{c}{$f_{\rm min}$} & \multicolumn{1}{c}{$f_{\rm max}$}& \multicolumn{1}{c}{Bandwidth}& \multicolumn{1}{c}{$f_{\rm min}$}& \multicolumn{1}{c}{$f_{\rm max}$}& \multicolumn{1}{c}{Bandwidth} \\
        Mode & \multicolumn{1}{c}{(MHz)} & \multicolumn{1}{c}{(MHz)} & \multicolumn{1}{c}{(MHz)}& \multicolumn{1}{c}{(kHz)}& \multicolumn{1}{c}{(MHz)}& \multicolumn{1}{c}{(MHz)}& \multicolumn{1}{c}{(kHz)} \\ \hline
        CW 2 & 2628 & 2627 & 2628 & 1309.06 & 3065 & 3078 & 13085.05  \\ 
        CW 1 & 1528 & 1542 & 1549 & 7533.08 & 2023 & 2044 & 21187.33  \\ 
        breathing & 734 & 747 & 780 & 33223.62 & 1192 & 1226 & 34143.46  \\ 
        CCW 1 & 266 & 290 & 296 & 5996.20 & 482 & 490 & 8453.12  \\ 
        CCW 2 & 91 & 106 & 107 & 354.05 & 238 & 239 & 1061.96  \\ 
        CCW 3 & 54 & 64 & 64 & 42.51 & 107 & 108 & 533.27  \\ 
        CCW 4 & 106 & 117 & 117 & 264.79 & 132 & 133 & 178.41  \\ 
        CCW 5 & 235 & 237 & 237 & 13.78 & 303 & 304 & 728.22  \\ 
        CCW 6 & 441 & 438 & 438 & 2.25 & 505 & 505 & 208.15  \\ 
        CCW 7 & 729 & 719 & 719 & 1.99 & 780 & 781 & 265.48  \\ 
        CCW 8 & 1101 & 1083 & 1083 & 0.96 & 1140 & 1140 & 26.74  \\ 
        CCW 9 & 1561 & 1534 & 1534 & 8.47 & 1586 & 1586 & 36.19  \\ 
        CCW 10 & 2113 & 2075 & 2075 & 0.10 & 2122 & 2122 & 9.06  \\ 
        CCW 11 & 2760 & 2710 & 2710 & 0.08 & 2750 & 2750 & 11.54  \\ 
        CCW 12 & 3504 & 3440 & 3440 & 0.01 & 3474 & 3474 & 54.62  \\ 
    \end{tabular}
    \label{tab:tab}
\end{table*}

\begin{figure*}[!t]
    \includegraphics{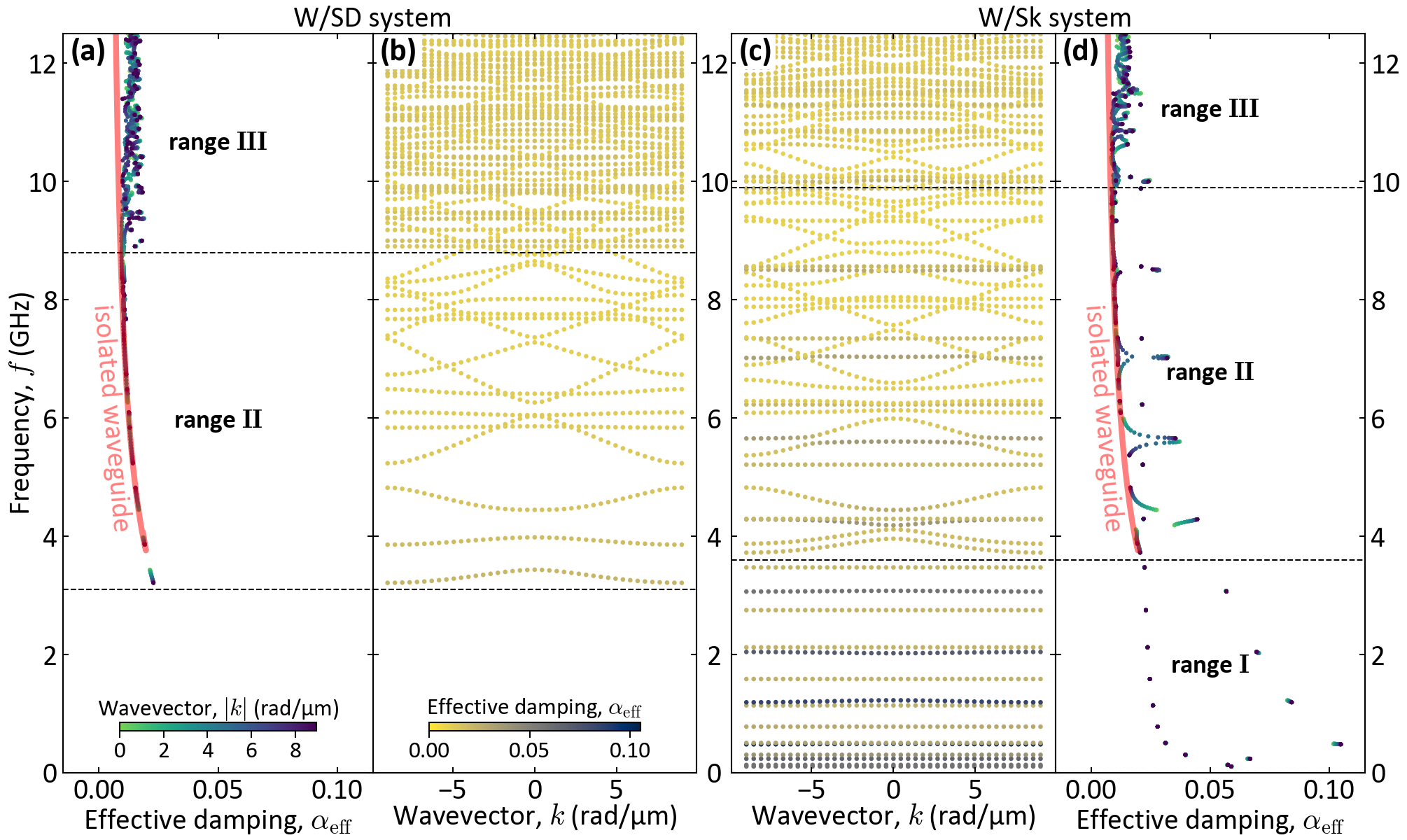}
    \caption{The dispersion relation in the first Brillouin zone presents effective damping $\alpha_{\rm eff}$ in (b) the W/SD system and (c) W/Sk system. The effective damping of each mode is indicated by the color of the point on the dispersion. The corresponding plots with the effective damping are shown in (a) for the W/SD system and in (d) for the W/Sk system. Here, the color of the point marks the absolute value of the wavevector. Dashed black vertical lines mark the limits of ranges. For the reference, the solid red line in (a) and (d) indicates the effective damping for the isolated waveguide.}
    \label{fig:imag}
\end{figure*}

In order to investigate the contribution of the dipolar interaction between the dot and the waveguide to the bandwidth of the skyrmion domain wall modes, we studied a one-dimensional chain of dots in skyrmion state, which is a subsystem of the W/Sk system. Additionally, as a reference, we studied a single dot in a skyrmion state.

Table~\ref{tab:tab} shows the simulation data for the modes within the frequency range I of the W/Sk system, as depicted in Fig.~4(c,d) in the main manuscript. This range contains 15 modes, ranging from the 2nd CW to the 12th CCW mode. Modes at higher frequencies may be significantly impacted by interaction with waveguide modes and are therefore not included in Table~\ref{tab:tab}. 

First of all, it is important to note that the static configurations of these systems are different. In a single dot, the skyrmion is round. In a dot chain, the dipolar interaction between the dots is very small so the skyrmion remains round. In the W/Sk system, the skyrmion changes its shape to an egg-like shape, as shown in Fig.~2(c) in the main manuscript. This change in shape significantly impacts the mode frequencies. As shown in Table~\ref{tab:tab}, the frequencies of modes in a single dot and an array of dots are very similar, differing by no more than 65 MHz. On the other hand, modes in the W/Sk system can differ from a single dot modes as much as 516 MHz for the 1st CW mode. However, for the higher-order CCW modes, this difference is strongly reduced.

When comparing the bandwidths of the same modes in different systems, it is clear that the dipolar interaction between the dot and the waveguide significantly contributes to this value. The bandwidths of all modes, except of the 4th CCW mode, are larger in W/Sk system, indicating that the presence of the waveguide enhance the interaction between the skyrmions. This effect is particularly noticeable for higher-order CCW modes (5th order and higher), whose bandwidths are orders of magnitude larger in the W/Sk system. However, it is difficult to distinguish the contribution of modified static configuration of a skyrmion and dynamic dipolar interaction through the waveguide.

\section{Impact of the damping}

\begin{figure*}[!t]
    \includegraphics{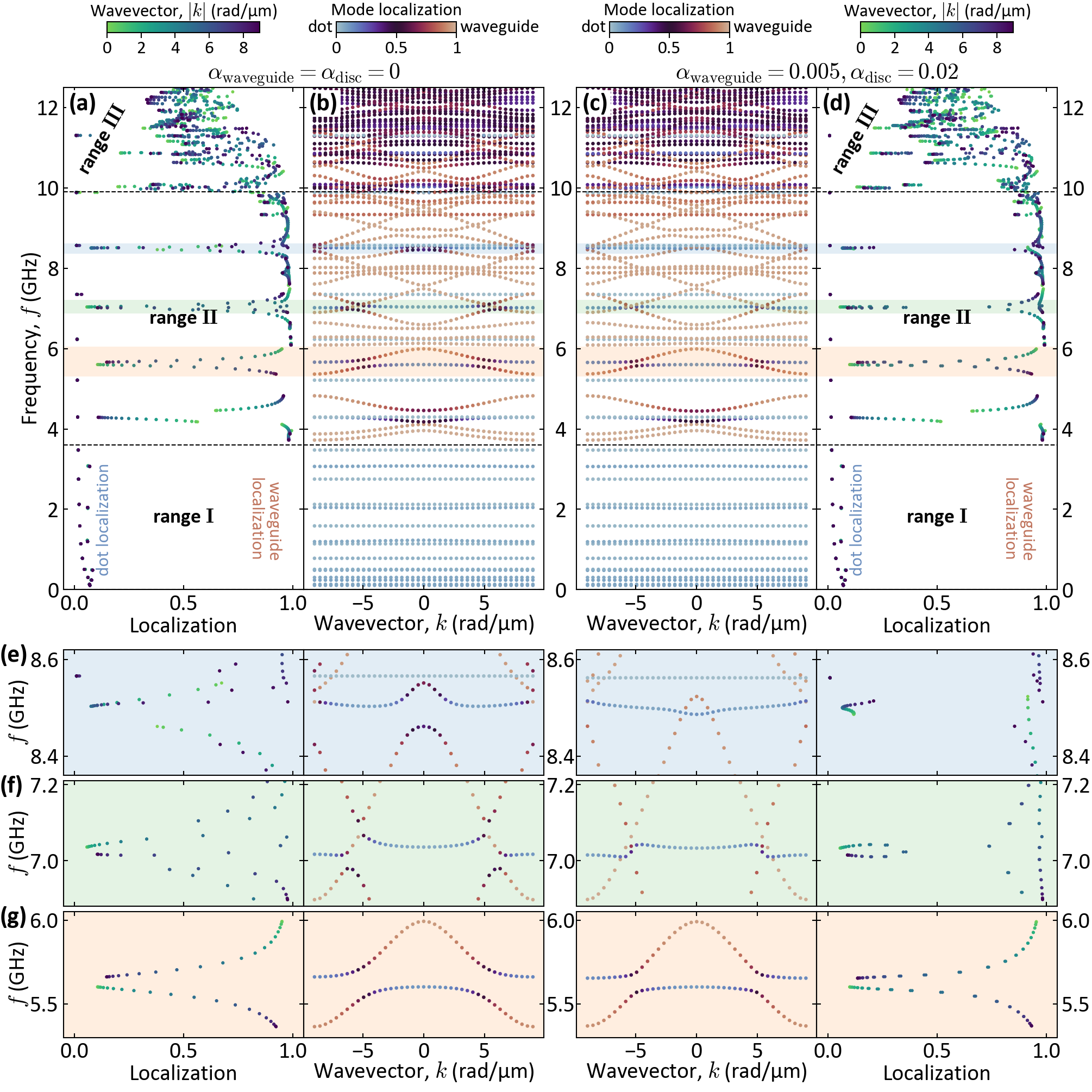}
    \caption{The dispersion relation in the first Brillouin zone presents the localization of modes in the W/Sk system (b) when the damping is neglected and (c) when it is taken into account. Each mode localization value is indicated by the color of the point on the dispersion. The corresponding plots with the localization value are shown in the case of (a) neglected damping and (d) included damping. Here, the color of the points mark the absolute value of the wavevector. Dashed black vertical lines mark the limits of the characteristic ranges. (e-g) Zoom-ins of figures (a-d) marked with (e) blue, (f) green, and (g) orange color, respectively.}
    \label{fig:loc+damping}
\end{figure*}

In the structures in which the damping varies between subsystems (just as here where the damping of Py waveguide is 0.005 while of the dots is 0.02), the excitation lifetime can give additional information about the system. In the results of the simulations, this information can be found in the frequencies which assume complex values. The ratio of imaginary part to real part of the frequency gives the value of the effective damping $\alpha_{\rm eff} = \mathrm{Im}(f)/\mathrm{Re}(f)$. This parameter is directly connected with the excitation lifetime since $\mathbf{m}(\mathbf{r},t) = \Tilde{\mathbf{m}}(\mathbf{r})e^{i \omega t} = \Tilde{\mathbf{m}}(\mathbf{r})e^{i (\omega'+i\omega'') t} = \Tilde{\mathbf{m}}(\mathbf{r})e^{i \omega' t}e^{-\omega'' t} = \Tilde{\mathbf{m}}(\mathbf{r})e^{i \omega' t}e^{-\alpha_{\rm eff} \omega' t} = \Tilde{\mathbf{m}}(\mathbf{r})e^{i \omega' t}e^{-t/T}$ so $T=1/\alpha_{\rm eff} \omega'$. This value depends on the frequency since the contribution of dipolar interactions to the effective damping is nonlinear. It is connected with the Gilbert damping in such a way that $\alpha_{\rm eff} \geq \alpha$.

The effect of the effective damping on the dispersion relation is shown in Fig.~\ref{fig:imag}. Subfigures (b) and (c) show the dispersion relation for the W/SD system and W/Sk system, respectively, like in Fig.~4(b) and (c) of the main text, but here the color of points marks the value of the effective damping instead of the mode localization. In addition, the value of the effective damping is plotted in subfigures (a) and (d), with the color indicating the wavenumber value from the first Brillouin zone. For reference, the solid red line in (a) and (d) indicates the effective damping for the isolated waveguide.

In the W/SD system, the effective damping is relatively low. In range II, where the modes are localized almost exclusively in the waveguide, the points follow the case of the isolated waveguide. In range III, where the modes are coupled with the dots, the points no longer follow the case of the isolated waveguide, with the effective damping increasing in value. It comes from the fact that due to the coupling with dots, the excitation inside the dot is attenuated faster, effectively decreasing the mode lifetime in whole system.

In the W/Sk system the situation is similar. In range I, the modes are strongly localized in the dots (see, Fig.~4(d) in the main text) and their effective damping is high. The CW modes have higher $\alpha_{\rm eff}$ than CCW modes but the maximum is at the 1st CCW mode. In range II, the modes localized in dots have higher effective damping than modes localized in the waveguide. In the areas where these modes hybridize, the effective damping is gradually changing. This effect resembles the gradual change of the localization shown in Fig.~4(d) in the main text. It shows a clear correlation between the mode localization and the effective damping. The situation in range III is analogical to range III in the W/SD system.

In addition to providing information about the mode lifetime, the damping has a significant effect on the spin-wave dynamics, affecting the dispersion relation, especially in the regions of the hybridization between the modes.

Such an effect is shown in Fig.~\ref{fig:loc+damping}. It compares dispersion relations of the W/Sk system in the case of no damping in both subsystems [Fig.~\ref{fig:loc+damping}(b)] with the case of realistic values of damping [Fig.~\ref{fig:loc+damping}(c)]. Most of the modes are not affected by the damping. However, at the hybridizations between the waveguide-dominated and dot-dominated modes, the dispersion of the involved modes can be significantly modulated. We marked three most interesting frequency ranges of dispersion relations with blue, green, and orange rectangles. Their zoom-ins are shown in subfigures (e), (f), and (g), respectively. In (g), around 5.5 GHz, the interaction between the modes in both cases leads to the opening of the band gap. However, the gap is slightly larger for the case without the damping (60 MHz) than with the damping (53 MHz). A stronger effect is visible in subfigure (e) around 7 GHz. The presence of damping in the system leads to the closure of the band gap. However, despite the gap being closed, the modes are still interacting, since the dot-localized mode is slightly dispersive, and its localization increases up to 0.36. Even more pronounced influence of the damping is shown in subfigure (f) around 8.5 GHz, where the dot mode opens two band gaps in the case of no damping, while the presence of damping leads to the closure of these gaps. Also, the coupling is significantly diminished resulting in the separation of modes localized in the dots and the waveguide.

\bibliography{main}